%% file: thesis.tex
\documentclass{eethesis}
\usepackage{graphicx}
\usepackage{latexsym}

\renewcommand{\type}{Dissertation Proposal}
\renewcommand{\type}{Dissertation}

\renewcommand{\degree}{DOCTOR OF PHILOSOPHY}    

\renewcommand{\major}{Physics}  

\def\dalemb#1#2{{\vbox{\hrule height .#2pt
        \hbox{\vrule width.#2pt height#1pt \kern#1pt
                \vrule width.#2pt}
        \hrule height.#2pt}}}

\def\cF{{\cal F}}
\def\cA{{\cal A}}

\def\cL{{\cal L}}
\def\0{{\sst{(0)}}}
\def\1{{\sst{(1)}}}
\def\2{{\sst{(2)}}}
\def\3{{\sst{(3)}}}
\def\4{{\sst{(4)}}}
\def\5{{\sst{(5)}}}
\def\6{{\sst{(6)}}}
\def\7{{\sst{(7)}}}
\def\8{{\sst{(8)}}}
\def\n{{\sst{(n)}}}

\def\tD{\widetilde {\cal D}}

\def\Z{\rlap{\sf Z}\mkern3mu{\sf Z}}
\def\R{\rlap{\rm I}\mkern3mu{\rm R}}
\def\cD{{\cal D}}
\def\cN{{\cal N}}

\def\ep{\epsilon}
\def\td{\tilde}
\def\wtd{\widetilde}
\def\half{{\textstyle{1\over2}}}

\let\a=\alpha \let\b=\beta \let\g=\gamma \let\d=\delta \let\e=\epsilon
    \let\k=\kappa
\let\l=\lambda \let\m=\mu \let\n=\nu   \let\r=\rho
\let\s=\sigma \let\t=\tau  \let\f=\phi \let\c=\chi 
\let\w=\omega  \let\D=\Delta  
   \let\F=\Phi 
 \let\W=\Omega \let\G=\Gamma
\let\la=\label  
  
\def\nn{\nonumber} \def\bd{\begin{document}} \def\ed{\end{document}}
\def\ds{\documentstyle} \let\fr=\frac \let\bl=\bigl \let\br=\bigr
\let\Br=\Bigr \let\Bl=\Bigl
\let\bm=\bibitem
\let\na=\nabla
\let\pa=\partial \let\ov=\overline
\def\vp{\varphi}
\def\vep{\varepsilon}
\def\ve{\varepsilon}
\def\vf{\varphi}
\def\wg{\wedge}
\def\bD{\bar{D}}
\def\bcD{\bar{\cal D}}
\newcommand{\be}{\begin{equation}}
\newcommand{\ee}{\end{equation}}
\def\ba{\begin{eqnarray}}
\def\ea{\end{eqnarray}}
\def\ft#1#2{{\textstyle{{\scriptstyle #1}\over {\scriptstyle #2}}}}
\def\fft#1#2{{#1 \over #2}}
\def\del{\partial}
\def\sst#1{{\scriptscriptstyle #1}}
\def\oneone{\rlap 1\mkern4mu{\rm l}}
\def\ie{{\it i.e.\ }}
\def\via{{\it via}}
\def\semi{{\ltimes}}
\def\str{{\rm str}}
\def\jm{{\rm j}}
\def\im{{\rm i}}
\def\mapright#1{\smash{\mathop{-\!\!\!-\!\!\!-\!\!\!-\!\!\!-\!\!\!
             \longrightarrow}\limits^{#1}}}
\def\maprightt#1#2{\smash{\mathop{-\!\!\!-\!\!\!-\!\!\!-\!\!\!-\!\!\!
             \longrightarrow}\limits^{#1}_{#2}}}

\newcommand{\ho}[1]{$\, ^{#1}$}
\newcommand{\hoch}[1]{$\, ^{#1}$}
\newcommand{\bea}{\begin{eqnarray}}
\newcommand{\eea}{\end{eqnarray}}
\newcommand{\ra}{\rightarrow}
\newcommand{\lra}{\longrightarrow}
\newcommand{\Lra}{\Leftrightarrow}
\newcommand{\ap}{\alpha^\prime}
\newcommand{\bp}{\tilde \beta^\prime}
\newcommand{\tr}{{\rm tr} }
\newcommand{\Tr}{{\rm Tr} }
\newcommand{\NP}{Nucl. Phys. }

\def\del{{\partial}}
\def\vev#1{\left\langle #1 \right\rangle}
\def\cn{{\cal N}}
\def\co{{\cal O}}
\newcommand{\mathbb}[1]{\mbox{\Bbb #1}}
\def\IC{{\mathbb C}}
\def\IR{{\mathbb R}}
\def\IZ{{\mathbb Z}}
\def\RP{{\bf RP}}
\def\CP{{\bf CP}}
\def\Poincare{{Poincar\'e }}
\def\tr{{\rm tr}}
\def\tp{{\tilde \Phi}}

\def\TL{\hfil$\displaystyle{##}$}
\def\TR{$\displaystyle{{}##}$\hfil}
\def\TC{\hfil$\displaystyle{##}$\hfil}
\def\TT{\hbox{##}}
\def\HLINE{\noalign{\vskip1\jot}\hline\noalign{\vskip1\jot}} 
\def\seqalign#1#2{\vcenter{\openup1\jot
  \halign{\strut #1\cr #2 \cr}}}
\def\lbldef#1#2{\expandafter\gdef\csname #1\endcsname {#2}}
\def\eqn#1#2{\lbldef{#1}{(\ref{#1})}%
\begin{equation} #2 \label{#1} \end{equation}}
\def\eqalign#1{\vcenter{\openup1\jot
    \halign{\strut\span\TL & \span\TR\cr #1 \cr
   }}}
\def\eno#1{(\ref{#1})}
\def\href#1#2{#2}
\def\half{{1 \over 2}}

\def\ads{{\it AdS}}
\def\adsp{{\it AdS}$_{p+2}$}
\def\cft{{\it CFT}}

\newcommand{\beq}{\begin{equation}}
\newcommand{\eeq}{\end{equation}}
\newcommand{\ber}{\begin{eqnarray}}
\newcommand{\eer}{\end{eqnarray}}
\newcommand{\beqar}{\begin{eqnarray}}
\newcommand{\eeqar}{\end{eqnarray}}

\newcommand{\nonu}{\nonumber}
\newcommand{\oh}{\displaystyle{\frac{1}{2}}}
\newcommand{\dsl}
  {\kern.06em\hbox{\raise.15ex\hbox{$/$}\kern-.56em\hbox{$\partial$}}}
\newcommand{\id}{i\!\!\not\!\partial}
\newcommand{\as}{\not\!\! A}
\newcommand{\ps}{\not\! p}
\newcommand{\ks}{\not\! k}
\newcommand{\dv}{d^2x}
\newcommand{\Dsl}{\not\!\! D}
\newcommand{\Bsl}{\not\!\! B}
\newcommand{\Psl}{\not\!\! P}
\newcommand{\eeqarr}{\end{eqnarray}}
\newcommand{\ZZ}{{\rm \kern 0.275em Z \kern -0.92em Z}\;}
\def\tilg{\tilde{g}}
\def\tilF{\tilde{F}}
\def\tilA{\tilde{A}}
\def\varf{\varphi}
\def\tilf{\tilde{\phi}}
\def\tilh{\tilde{h}}
\def\rme{{\rm e}}

\begin{document}
\pagenumbering{roman}
\include{title}
\include{approval}
\include{abstract}
\include{ded}
\include{ack}
\include{lists}
\pagenumbering{arabic}
\setlength{\headheight}{12pt}
\pagestyle{myheadings}
\include{ch1}

\include{ch2}

\include{ch3}

\include{ch4}
\include{ch5}

\include{refs}
\include{appendA}
\include{appendB}

\include{appendC}

\include{appendD}

\include{appendE}
\include{appendF}

\include{vita}
\end{document}

%% file: title.tex
\maketitlepage
{GAUGED SUPERGRAVITIES FROM SPHERICAL REDUCTIONS}   
{TUAN ANH TRAN} 
{DOCTOR OF PHILOSOPHY}   
{December 2001}
{Physics}         

%% file: approval.tex

\approvaltwo
{GAUGED SUPERGRAVITIES FROM SPHERICAL REDUCTIONS}
{TUAN ANH TRAN}
{Christopher N. Pope}
{Richard L. Arnowitt}
{Robert C. Webb}
{Stephen A. Fulling}
{Thomas W. Adair III}
{Member's name}
{December 2001}



%% file: abstract.tex
\absone
{Gauged Supergravities from Spherical Reductions}
{December 2001}
{Tuan Anh Tran}
{B.S., Hanoi University, Vietnam}  
{Dr. Christopher N. Pope}
{This dissertation is devoted to deriving the bosonic sectors of certain 
gauged supergravities in various dimensions from reducing eleven-dimensional 
supergravity, type IIA and type IIB supergravities in
ten dimensions on certain spherical spaces. Explicit non-linear
Kaluza-Klein ans\"atze for reductions of eleven-dimensional
supergravity and of type IIA and type IIB supergravities on $S^n$ and 
$S^n\times T^m$ are presented. Knowing explicit non-linear ans\"atze
is proven to be very useful in finding super Yang-Mills operators of 
gauge theories via AdS/CFT correspondence. We present a sample
calculation which
allows us to find a super Yang-Mills operator using a non-linear
ansatz. Knowing non-linear ans\"atze is also useful for finding
supergravity duals to certain twisted supersymmetric gauge theories. 
These supergravity solutions are branes wrapped on certain
supersymmetric cycles. Some solutions, which are dual to gauge
theories in three and five dimensions, are presented. 
}


%% file: ded.tex
\dedicate{my parents for their love and their support}

%% file: ack.tex
\acknow{I am very grateful to my advisor, Christopher N. Pope,
for not only teaching me Physics, but also for helping me with various 
issues in life. I am also grateful to Mirjam Cveti\v{c} and Hong L\"u 
for the research we did together.

 I am thankful to Glenn Agnolet, Richard Arnowitt, Michael Duff and
 Ergin Sezgin for teaching me various subjects of physics.

 I am also thankful to Igor Lavrinenko, Mihail Mihailescu, Carlos 
N\'u\~nez, Inyong Park,
Arta Sadrzadeh and Martin Schvellinger for the ideas they shared with
me and for the work we did together.

 Many stimulating discussions with Karim Benakli, Sadik Deger, 
Bashkar Dutta, Jianxin Lu, Sudipta Mukherji, Igor Rudychev, Hisham
Sati, Per Sundell and Kaiwen Xu are acknowledged. 

 I would like to thank the people in the HEP section at ICTP, Trieste,
Italy for teaching me HEP while I was in the Diploma Program. 
Particularly, I would 
like to thank Faheem Hussain and Seif Randjbar-Daemi for their help while 
I was at ICTP. Many conversations with K.S. Narain about 
aspects of string theory partially influenced my 
decision to work on supergravity and string theory. I would like to
thank Goran Senjanovi\'c for his supervision while I was there.

 I am indebted to Robert Foot and to Hoang Ngoc Long (my
undergraduate advisor) for introducing me to particle physics.

 I would like to thank Angela, Bob, Christoph, Francesca, Irina, Karen, Lucia,
Nandita, Natasha and Sante for many enjoyable hours throughout the
course of my study.

 
}

%% file: lists.tex
\pagestyle{headings}
\setlength{\headheight}{36pt}
\tableofcontents
\listoftables
\listoffigures
\clearpage

%% file: ch1.tex
\chapter{Introduction}
\body
\vspace{1cm}
Finding an ultimate quantum field theory that describes a
unification of four known, fundamental interactions, namely
electromagnetic, gravitational, strong and weak interactions, has
been a dream for many generations of physicists, and it might
still be a dream for many others yet to come. Over the years, 
there have been many attempts to achieve this dream. Presently,
ten-dimensional string theory and eleven-dimensional
M-theory~\cite{Green:1987sp,Polchinski:1998rq}, whose low-energy
limits are ten-dimensional and eleven-dimensional
supergravities~\cite{Cremmer:1978km,Schwarz:1983qr} respectively, are
our best candidates for the final theory. To describe our 
four-dimensional world, the extra dimensions of string theory and
M-theory are assumed to be very
small and compact, \`a la the Kaluza-Klein idea~\cite{Duff:1986hr}. By
small, we mean that it is of the order of the Planck scale.
One reason the Kaluza-Klein idea is very attractive is that it unifies
gravitational and gauge interactions. A satisfactory by-product of
learning how to perform dimensional reduction via Kaluza-Klein
reduction is that we find many of the lower-dimensional theories 
we wish to consider are derivable from a single and unique theory,
{\it the Theory of Everything}.

In the 1980s, many different supergravity theories
in various dimensions were constructed~\cite{Salam:1989fm}. Ungauged
maximal supergravities in various dimensions from $D=10$ to $D=3$ can
be obtained by Kaluza-Klein reductions of eleven-dimensional 
supergravity on a $T^n$ ($ 1 \leq n
\leq 8$)~\cite{Cremmer:1998ct}. Reducing these ungauged maximal supergravities
on $T^{1,m}$ torus, which includes a time-like direction, gives rise to
Euclidean supergravities in various dimensions~\cite{Cremmer:1998em,Hull:1998br}.
It has long been believed that many gauged supergravities arise from
the Kaluza-Klein reduction of eleven-dimensional supergravity or the type
IIA and IIB supergravities on certain spherical internal spaces. It
was shown long ago that at the level of linearized fluctuations,
the massless spectrum of the maximal gauged supergravities in
$D=4$ and $D=7$ can be obtained from appropriate truncations of those
coming from the $S^7$ and $S^4$ reductions of $D=11$ supergravity
\cite{Duff:1983gq,Pilch:1984xy}.  Likewise, the spectrum of maximal 
gauged supergravity in $D=5$ is known to arise
from a truncation of the $S^5$ reduction from type IIB supergravity
\cite{Gunaydin:1985fk,Kim:1985ez}. Although 
these linearized results are rather easily established, it is much 
harder to determine whether the truncations to the massless
supermultiplets are consistent at the full non-linear level.
The key point of concern here is that once the non-linear interactions are
included, the possibility exists that in a full, untruncated reduction,
source terms built purely from the fields of the massless multiplet
might arise in the field equations for the lower-dimensional massive
multiplets, making it inconsistent to set the massive fields to zero.
For the Kaluza-Klein reductions on $S^1$, and $n$-dimensional torus
$T^n$, which can be viewed as a sequences of $S^1$ reductions, the
truncations to the massless sector are always consistent. The reason
for this consistency can be understood in the following way: The
massless fields are singlets under the $U(1)^n$ isometry group
of the $n$-torus, while the massive modes are non-singlets. The
products of singlets under the group action can never generate
non-singlets, so there are no couplings between massless and
massive modes. Thus, no matter how non-linear the higher-dimensional
theory, the Kaluza-Klein reduction on $T^n$ will be a consistent one.
However, the story is significantly different when one performs the
Kaluza-Klein reductions on arbitrary spaces, and in particular, on spherical
spaces. Indeed, it is not hard to establish that in a spherical reduction of a
generic, higher-dimensional theory, there will definitely be such
couplings, making a consistent truncation to the massless sector
impossible.  Thus, if the spherical reductions (with truncation to the
massless supermultiplet) in supergravities are to be consistent, it
must be because of remarkable special properties of these particular
theories.  General arguments suggesting that reductions to the
supergravity multiplet should always be consistent were constructed in
\cite{Pope:1987ad}.
Despite all the apparent obstacles, it was shown~\cite{deWit:1987iy} 
that the four-dimensional, $\cN=8$ gauged $SO(8)$ supergravity can
be embedded at the full non-linear level in eleven-dimensional
supergravity. However, the result is rather implicit, so practically
speaking, it
is hard to use. Furthermore, there is no demonstration at the non-linear
level for the consistency of the complete $S^5$ reduction from type IIB
supergravity. Therefore, finding explicit, non-linear reduction ans\"atze 
remains a challenging problem.
 
A few years ago, a new duality was
conjectured~\cite{Maldacena:1998re,Gubser:1998bc,Witten:1998qj}, 
which relates supergravities in anti-de Sitter backgrounds to
conformal field theories on their boundaries (AdS/CFT). 
Since the anti-de Sitter 
backgrounds arise as solutions of lower-dimensional gauged
supergravities, the conjectured AdS/CFT correspondence has led to a 
revival of interest in deriving these gauged supergravities by
Kaluza-Klein reduction from the ten-dimensional fundamental string 
theories and eleven-dimensional M-theory. 

In the last few years, much progress has been achieved in understanding the full
non-linear structure of certain Kaluza-Klein sphere reductions. The
starting point of the whole program of finding the non-linear
reduction ans\"atze was an observation that the rotating
D3-brane should have a ``decoupling limit'' similar to that of the 
non-rotating
D3-brane.~\footnote{The decoupling limit is a point where the space-time
of the non-rotating D3-brane becomes a product space $M_5\times
S^5$. If the non-rotating D3-brane is extremal, $M_5$ is
five-dimensional anti-de Sitter space-time ($AdS_5$).} After taking the
decoupling limit, the rotating D3-brane metric has the form of an
$AdS_5$ black hole times $S^5$. From this observation, we have learned
a great deal about the structure of the non-linear reduction
ans\"atze, and we were able to conjecture, and then verify, the
correct reduction ansatz. The ansatz describes the exact embedding of
the five-dimensional, $\cN=2$ gauged $U(1)^3$ supergravity into type
IIB supergravity. Similarly, the ans\"atze which allow the embedding
of the $U(1)^2$ and $U(1)^4$ truncations of $D=7$ and $D=4$ gauged
supergravities into eleven-dimensional supergravity were
constructed~\cite{Cvetic:1999xp}. Not long after our
work~\cite{Cvetic:1999xp}, using a formalism based on an
analysis of the supersymmetry transformation rules, the complete
ansatz for the $S^4$ reduction of eleven-dimensional supergravity
giving rise to the seven-dimensional gauged $SO(5)$ supergravity was
obtained~\cite{Nastase:1999cb,Nastase:1999kf}. The consistent
reduction of massive type IIA supergravity on a locally $S^4$ space,
to give $\cN=2$ gauged $SU(2)$ supergravity in $D=6$, has also been 
obtained~\cite{Cvetic:1999un}. Further examples of AdS with warped
space-times were constructed in~\cite{Cvetic:2000cj,Cvetic:2000yp}.

Studying the singular limit of the known $S^4$ reduction of
eleven-dimensional supergravity~\cite{Nastase:1999kf}, which gives
rise to the standard $SO(5)$-gauged maximal supergravity in $D=7$, we 
constructed a consistent reduction of type IIA on
$S^3$~\cite{Cvetic:2000ah}, leading to a maximal gauged supergravity in seven dimensions
with the full set of massless $SO(4)$ Yang-Mills fields. We also
obtained the consistent reduction of type IIA supergravity on $S^4$,
giving rise to an $SO(5)$-gauged maximal supergravity in
$D=6$. Details of these constructions will be presented in chapter
II. Also in chapter II, a consistent reduction ansatz, which allows
us to embed any solution of six-dimensional Romans'
theory~\cite{Romans:1986tw} without mass parameter into
ten-dimensional supergravity is presented. As will be seen in chapter II,
this six-dimensional Romans' theory can be obtained from a reduction
of type IIA supergravity on $S^1\times S^3$. In addition, a reduction
ansatz of type IIA on $S^1\times S^3\times S^1$, giving rise to Romans'
theory in $D=5$ without the $U(1)$ gauge-coupling, will be also
presented in chapter II.

In addition to maximal Abelian
truncations in $D=4, 5$ and 7~\cite{Cvetic:1999xp}, certain reductions to
truncations of the maximal gauged supergravities in various dimensions
have been constructed. These have the advantage of being considerably
simpler than the maximal theories, allowing the reduction ansatz to be
presented in a more explicit form. Cases that have been constructed
include the $\cN=2$ gauged $SU(2)$ supergravity in
$D=7$~\cite{Lu:1999bc}; the $\cN=4$ gauged $SU(2)\times U(1)$ in
$D=5$~\cite{Lu:1999bw}; the $\cN=4$ gauged $SO(4)$ supergravity in
$D=4$~\cite{Cvetic:1999au}. Using these results, certain ungauged
supergravities in $D=4$, $D=5$ and $D=6$ can be embedded into type
IIA, IIB and eleven-dimensional supergravities~\cite{Lu:2000xc,Cvetic:2000gj}.
One can also consider non-supersymmetric
truncations of the maximal gauged supergravities. The consistent
truncations of the $D=7$, $D=5$ and $D=4$ supergravities to the
graviton plus scalar subsectors comprising only the diagonal scalars
in the $SL(N,\R)/SO(N)$ scalar sub-manifolds were considered (with
$N=5, 6$ and 8, respectively), and the consistent embeddings in $D=11$
and $D=10$ were
constructed~\cite{Cvetic:1999xx,Cvetic:2000eb,Cvetic:2000tb}. Another complete
ansatz for the consistent embedding of a subsector, which consists of
20 scalars and 15 Yang-Mills fields, of the five-dimensional $\cN=8$
gauged $SO(6)$ supergravity into type IIB was
obtained~\cite{Cvetic:2000nc}. Various ans\"atze of type IIB 
supergravity on $S^5$ and on $S^3\times T^2$ will also be presented in
chapter III.

In chapter IV, certain applications of non-linear Kaluza-Klein
ans\"atze will be presented. In particular, we will
show~\cite{Park:2000du} that with the
full non-linear reduction ansatz the calculation of $n$-point
correlation functions in super Yang-Mills (SYM) theory via AdS/CFT
correspondence appears to be much simpler than those previously
appearing in the literature, which use the linearized ansatz followed
by non-linear field redefinitions~\cite{Lee:1998bx}. 

Recently, non-linear Kaluza-Klein
reductions ans\"atze have been extensively used in finding
supergravity solutions corresponding to certain twisted field
theories~\cite{Bershadsky:1996qy}. This searching provides some new
examples of AdS/CFT correspondence. Wrapping M5-branes and D3-branes
on Riemann-surfaces that are holomorphically embedded in Calabi-Yau 2-
or 3-folds, giving rise to four-dimensional field theories with $\cN=2,
1$ supersymmetry and two-dimensional field theories with (4, 4) and
(2, 2) supersymmetry respectively, were
analyzed~\cite{Maldacena:2000mw}. Type IIB NS5-branes wrapped on
$S^2$ leading to four-dimensional $\cN=1$ super Yang-Mills
theory was studied~\cite{Maldacena:2000yy}. Some related work
associated with Dp-branes and M5-branes wrapping Riemann surfaces in
Calabi-Yau spaces had studied earlier
in~\cite{Alishahiha:1999ds,Cvetic:2000cj,Fayyazuddin:1999zu,Fayyazuddin:2000em,Brinne:2000fh,Klebanov:2000nc,Klebanov:2000hb}.
These studies have been extended for the cases of M5-branes
and NS5-branes wrapping on associative 3-cycles in seven-dimensional 
manifolds with $G_2$ holonomy~\cite{Acharya:2000mu}; D3-branes~\cite{Nieder:2000kc}
wrapped on associative 3-cycles; M5-branes wrapping on K\"ahler
4-cycles, special Lagrangian 3-, 4-, and 5-cycles, co-associative
4-cycles and Caley 4-cycles~\cite{Gauntlett:2000ng}; D4-branes as well
as NS5-branes wrapping on 2- and 3-cycles~\cite{Nunez:2001pt};
D6-branes wrapping on $S^2$ and $S^3$~\cite{Edelstein:2001pu}; 
NS5-branes wrapping on 
$S^3\times T^2$~\cite{Schvellinger:2001ib,Maldacena:2001pb}; M2-branes
wrapped on 2-cycles in Calabi-Yau 2-, 3-, 4- and
5-folds~\cite{Gauntlett:2001qs}; D6-branes
wrapped on co-associative 4-cycles with constant curvature of seven
manifold of $G_2$ holonomy~\cite{Hernandez:2001bh}; 5-branes wrapping
on $S^2$ in a Calabi-Yau 2-folds~\cite{Gauntlett:2001ps}; D6-branes
wrapping on K\"{a}hler 4-cycles~\cite{Gomis:2001vk,Gomis:2001vg}. 
We will also present some supergravity
solutions~\cite{Nunez:2001pt,Schvellinger:2001ib} which are dual to 
certain supersymmetric gauge theories via AdS/CFT correspondence in
chapter IV. 

Chapter V is devoted to conclusions. Brief reviews of eleven-dimensional
supergravity and type IIA, IIB and massive type IIA in $D=10$ are
presented in appendices A, B, C, and D, respectively. Appendix E is 
devoted to a review of
six-dimensional, $SU(2)$-gauged supergravity and the $\cN=4$, 
$SU(2)\times U(1)$-gauged supergravity in $D=5$ is presented in
appendix F.

%% file: ch2.tex
\chapter{Type IIA supergravity on $S^{\lowercase{m}}\times 
T^{\lowercase{n}}$}
\vspace{1cm}
\section{The $S^4$ reduction of eleven-dimensional supergravity}

The complete ansatz for the $S^4$ reduction of eleven-dimensional
supergravity was obtained in \cite{Nastase:1999kf}, using a formalism based
on an analysis of the supersymmetry transformation rules.  One may
also study the reduction from a purely bosonic standpoint, by
verifying that if the ansatz is substituted into the
eleven-dimensional equations of motion, it consistently yields the
equations of motion of the seven-dimensional gauged $SO(5)$
supergravity.  We shall carry out this procedure here, in order to
establish notation, and to obtain the complete system of
seven-dimensional bosonic equations of motion, which we shall need
in the later part.

After some manipulation, the Kaluza-Klein $S^4$ reduction ansatz
obtained in \cite{Nastase:1999kf} for eleven-dimensional supergravity can
be expressed as follows:
\bea 
d\hat s_{11}^2 &=& \Delta^{1/3}\, ds_{7}^2 +
\fr1{g^2}\Delta^{-2/3}\, T^{-1}_{ij}\, \cD\mu^i\,
\cD\mu^j\;,
\label{11d-metric-ansatz-s4} 
\eea
\bea 
\hat F_\4 &=& \fr1{4!}\, \ep_{i_1\cdots i_5}\, \Big[ -
\fr1{g^3} U\, \Delta^{-2} \mu^{i_1}\cD\mu^{i_2}\wedge \cdots \wedge
\cD\mu^{i_5}\nn\\
&& + \fr4{g^3} \Delta^{-2}\, T^{i_1 m}\, \cD T^{i_2 n}\, \mu^m\,
\mu^n\, \cD\mu^{i_3}
\wedge \cdots \wedge \cD\mu^{i_5}\nn\\
&& + \fr6{g^2} \Delta^{-1} F_\2^{i_1 i_2} \wedge \cD\mu^{i_3}\wedge
\cD\mu^{i_4}\, T^{i_5 j}\, \mu^j \Big] - T_{ij}\, {*S_\3^i}\, \mu^j
+ \fr1{g}\, S_\3^i \wedge \cD\mu^i\;,\label{11d-4form-ansatz-s4}\\
{{\hat *}\hat F_\4} &=& - g U \ep_\7 - \fr1{g} T^{-1}_{ij}
{*\cD}T^{ik} \mu_k\wedge \cD\mu^j +\fr1{2g^2} T^{-1}_{ik}
T^{-1}_{j\ell}\,
{*F_\2^{ij}} \wedge \cD\mu^k \wedge \cD\mu^\ell \nn\\
&& - \fr1{6 g^3} \, \Delta^{-1}\, \ep_{ij \ell_1\ell_2
\ell_3} {*S_\3^m}\, T_{im}\,  T_{jk}\, \mu^k \wedge
\cD\mu^{\ell_1}\wedge \cD\mu^{\ell_2} \wedge \cD\mu^{\ell_3}\nn\\
& & + \fr1{g^4} \Delta^{-1}\, T_{ij}\, S_\3^i\, \mu^j\wedge W\;,
\label{11d-hodge4form-ansatz-s4} 
\eea
where
\bea 
U &\equiv& 2 T_{ij}\, T_{jk}\, \mu^i\, \mu^k - \Delta\,
T_{ii}\;,\;\;
\Delta \equiv T_{ij}\, \mu^i\, \mu^j\;,\;\;
F_\2^{ij} =\cD A^{ij}_\1 \equiv 
dA_\1^{ij} + g A_\1^{ik}\wedge A_\1^{kj}\;,\nn\\
\cD\mu^i &\equiv& d\mu^i + g A_\1^{ij}\,\mu^j\;,\;\;
\cD T_{ij} \equiv dT_{ij} + g A_\1^{ik}\, T_{kj} + g A_\1^{jk}\,
T_{ik}\;,\;\; \mu^i\, \mu^i \equiv 1\;,\nn\\
W&\equiv& \fr1{24}\, \ep_{i_1\cdots i_5}\, \mu^{i_1}\,
\cD\mu^{i_2}\wedge \cdots \wedge \cD\mu^{i_5}\;,
\eea
where the symmetric matrix $T_{ij}$ is unimodular and hat denotes
eleven-dimensional fields. Furthermore, $T_{ij}$ parameterizes the 
scalar coset $SL(6,\R)/SO(6)$. 

To obtain seven-dimensional equations of motion, we first consider the 
Bianchi identity $d\hat F_\4 = 0$. Substituting
(\ref{11d-4form-ansatz-s4}) into this, we obtain the following equations:
\bea
\cD(T_{ij}\, {* S_\3^j}) &=& F_\2^{ij}\wedge S_\3^j\;,
\label{7d-bianchi-id}\\
H_\4^i &=& g T_{ij}\, {* S_\3^j} +  \fr1{8}  \ep_{i {j_1}\cdots
{j_4}} F_\2^{{j_1} {j_2}}\wedge\, F_\2^{{j_3} {j_4}}\;,
\label{7d-h4-eq} 
\eea
where we define
\be 
H_\4^i \equiv \cD S_\3^i = dS_\3^i + g\, A_\1^{ij}\wedge
S_\3^j\;. 
\label{7d-h4-def} 
\ee
Next, we substitute the ansatz into the $D=11$ field
equation~(\ref{11d-4form-eq}) for $\hat F_\4$ and we get
\bea 
{\cD\Big(T^{-1}_{ik} T^{-1}_{j\ell} {*F_\2^{ij}}\Big)} &=& -2
g\, T^{-1}_{i[k} {*\cD T_{\ell] i}} - \fr1{2g}\, \ep_{i_1 i_2 i_3 k
\ell}\, F_2^{i_1 i_2}\, H_\4^{i_3}
\nn\\
&& + \fr3{2g} \delta_{i_1 i_2 k\ell}^{j_1 j_2 j_3 j_4}\, F_\2^{i_1
i_2}\wedge F_\2^{j_1 j_2}\wedge  F_\2^{j_3 j_4} -
 S_\3^k\wedge S_\3^\ell\;.
\label{7d-gauge-eq}\\
\cD\Big(\, T^{-1}_{ik} *\cD T_{kj}\Big) &=& 2 g^2 (2 T_{ik}\,
T_{kj} - T_{kk}\, T_{ij})\ep_\7 + T^{-1}_{im}\, T^{-1}_{k\ell}\,
{*F_\2^{m\ell}}\wedge F_\2^{kj}\nn\\
&& + T_{jk}\, {*S_\3^k} \wedge S_\3^i - \fr15 \delta_{ij}
\Big[ 2 g^2 \Big(2T_{ik} T_{ik} - 2 (T_{ii})^2 \Big) \ep_\7  \nn\\
&& + T^{-1}_{nm} T^{-1}_{k\ell}\, {*F_\2^{m\ell}} \wedge F_\2^{kn}
+ T_{k\ell } \, {*S_\3^k} \wedge S_\3^\ell \Big]\;,
\label{7d-scalars-eq} 
\eea
for the Yang-Mills and scalar equations of motion in
$D=7$. Note from (\ref{7d-gauge-eq}) that it would be
inconsistent to set the Yang-Mills fields to zero while retaining
the scalars $T_{ij}$, since the currents $T^{-1}_{i[k} {*\cD T_{\ell]
i}}$ act as sources for them.  A truncation where the Yang-Mills
fields are set to zero is consistent, however, if the
scalars are also truncated to the diagonal subsector $T_{ij}={\rm
diag}(X_1,X_2,\ldots, X_6)$, as in the consistent reductions
constructed in \cite{Cvetic:1999xx,Cvetic:2000eb}.

We find that all the  equations of motion can be derived from the
following seven-dimensional Lagrangian
\bea 
{\cal L}_7 &=& R\, {*\oneone} - \fr14 T^{-1}_{ij}\, {*\cD
T_{jk}}\wedge T^{-1}_{k\ell}\, \cD T_{\ell i} -\fr1{4}\,
T^{-1}_{ik}\, T^{-1}_{j\ell}\, {* F_\2^{ij}}\wedge F_\2^{k\ell}
-\fr12 T_{ij}\, {*S_\3^i}\wedge S_\3^j \nn\\
&&+ \fr1{2g} S_\3^i\wedge H_\4^i - \fr1{8g}  \ep_{i j_1\cdots
j_4}\, S_\3^i\wedge F_\2^{j_1 j_2}\wedge F_\2^{j_3 j_4} + \fr1g
\Omega_\7 - V\, {*\oneone}\;,
\label{7d-lag} 
\eea
where $H_\4^i$ are given by (\ref{7d-h4-def}) and the potential $V$ is
given by
\be 
V = \fr12  g^2 \Big(2 T_{ij}\, T_{ij} - (T_{ii})^2 \Big)\;,
\ee
and $\Omega_\7$ is a Chern-Simons type of term built from the
Yang-Mills fields, which has the property that its variation with
respect to $A_\1^{ij}$ gives
\be 
\delta \Omega_\7 = \fr34 \delta_{i_1 i_2 k\ell}^{j_1 j_2 j_3
j_4}\, F_\2^{i_1 i_2}\wedge F_\2^{j_1 j_2}\wedge  F_\2^{j_3
j_4}\wedge \delta A_\1^{k\ell}\;. 
\ee
Note that the $S_\3^i$ are viewed as fundamental fields in the
Lagrangian, and that (\ref{7d-h4-eq}) is their first-order equation.
In fact (\ref{7d-lag}) is precisely the bosonic sector of the
Lagrangian describing maximal gauged seven-dimensional
supergravity that was derived in \cite{Pernici:1984xx}.  An explicit
expression for the 7-form $\Omega_\7$ can be found there.

Although we have fully checked the eleven-dimensional Bianchi
identity and field equation for $\hat F_\4$ here, we have not
completed the task of substituting the ansatz into the
eleven-dimensional Einstein equations.  This would be an extremely
complicated calculation, on account of the Yang-Mills gauge
fields. However, various complete consistency checks, including
the higher-dimensional Einstein equation, have been performed in
various truncations of the full $\cN=4$ maximal supergravity
embedding, including the $\cN=2$ gauged theory in \cite{Lu:1999bc},
and the non-supersymmetric truncation in \cite{Cvetic:2000eb} where the
gauge fields are set to zero and only the diagonal scalars in
$T_{ij}$ are retained.  All the evidence points to the full
consistency of the reduction.\footnote{The original demonstration
in \cite{Nastase:1999kf}, based on the reduction of the eleven-dimensional
supersymmetry transformation rules, also provides extremely
compelling evidence. Strictly speaking, the arguments presented
there also fall short of a complete and rigorous proof, since they
involve an approximation in which the quartic fermion terms in the
theory are neglected.}

    It is perhaps worth making a few further remarks on the nature of
the reduction ansatz.  One might wonder whether the ansatz
(\ref{11d-4form-ansatz-s4}) on the 4-form field strength $\hat F_\4$ could be
re-expressed as an ansatz on its potential $\hat A_\3$.  As it
stands, (\ref{11d-4form-ansatz-s4}) only satisfies the Bianchi identity $d\hat
F_\4=0$ by virtue of the lower-dimensional equations
(\ref{7d-bianchi-id}) and (\ref{7d-h4-eq}).  However, if (\ref{7d-h4-eq}) is
substituted into (\ref{11d-4form-ansatz-s4}), we obtain an expression that
satisfies $d\hat F_\4=0$ without the use of any lower-dimensional
equations.  However, one does still have to make use of the fact
that the $\mu^i$ coordinates satisfy the constraint $\mu^i\,
\mu^i=1$, and this prevents one from writing an explicit ansatz
for $\hat A_\3$ that has a manifest $SO(5)$ symmetry.  One could
solve for one of the $\mu^i$ in terms of the others, but this
would break the manifest local symmetry from $SO(5)$ to
$SO(4)$.  In principle though, this could be done, and then one
could presumably substitute the resulting ansatz directly into the
eleven-dimensional Lagrangian.  After integrating out the internal
4-sphere directions, one could then in principle obtain a
seven-dimensional Lagrangian in which, after re-organizing terms,
the local $SO(5)$ symmetry could again become manifest.

    It should, of course, be emphasized that merely substituting an
ansatz into a Lagrangian and integrating out the internal
directions to obtain a lower-dimensional Lagrangian is justifiable
only if one already has an independent proof of the consistency of
the proposed reduction ansatz.\footnote{A classic illustration is
provided by the example of five-dimensional pure gravity with an
(inconsistent) Kaluza-Klein reduction in which the scalar dilaton
is omitted. Substituting this into the five-dimensional
Einstein-Hilbert action yields the perfectly self-consistent
Einstein-Maxwell action in $D=4$, but fails to reveal that setting
the scalar to zero is inconsistent with the internal component of
the five-dimensional Einstein equation.} If one is in any case going
to work with the higher-dimensional field equations in order to
prove the consistency, it is not clear that there would be any
significant benefit to be derived from then re-expressing the
ansatz in a form where it could be substituted into the
Lagrangian.

   It is interesting to observe that one cannot take the limit
$g\rightarrow 0$ in the Lagrangian (\ref{7d-lag}), on account of the
terms proportional to $g^{-1}$ in the second line.  We know, on
the other hand, that it must be possible to recover the ungauged
$D=7$ theory by turning off the gauge coupling constant.  In fact
the problem is associated with a pathology in taking the limit at
the level of the Lagrangian, rather than in the equations of
motion.  This can be seen by looking instead at the
seven-dimensional equations of motion, which were given earlier.
The only apparent obstacle to taking the limit $g\rightarrow 0$ is
in the Yang-Mills equations (\ref{7d-gauge-eq}), but in fact this
illusory.  If we substitute the first-order equation (\ref{7d-h4-eq})
into (\ref{7d-gauge-eq}) it gives
\be 
{\cD\Big(T^{-1}_{ik} T^{-1}_{j\ell} {*F_\2^{ij}}\Big)} = -2 g
T^{-1}_{i[k} {*\cD T_{\ell] i}} - \fr1{2}\, \ep_{i_1 i_2 i_3 k
\ell}\, F_2^{i_1 i_2}\wedge T_{ij}\, {*S_\3^j} -S_\3^k\wedge
S_\3^\ell\;, 
\label{gaugev2} 
\ee
which has a perfectly smooth $g\rightarrow 0$ limit.  It is clear
that equations of motion (\ref{7d-h4-eq}) and (\ref{7d-scalars-eq}) and the
Einstein equations of motion also have a smooth limit.  (The
reason why the Einstein equations have the smooth limit is because
the $1/g$ terms in the Lagrangian (\ref{7d-lag}) do not involve the
metric, and thus they give no contribution.)

    Unlike in the gauged theory, we should not treat the $S^i_\3$
fields as fundamental variables in a Lagrangian formulation in the
ungauged limit.  This is because once the gauge coupling $g$ is
sent to zero, the fields $S^i_\3$ behave like 3-form field
strengths.  This can be seen from the first-order equation of
motion (\ref{7d-h4-eq}), which in the limit $g\rightarrow 0$ becomes
\be 
dS_\3^i = \fr18 \ep_{ij_1\cdots j_4}\, dA_\1^{j_1j_2}\wedge
dA_\1^{j_3j_4}\;,
\label{s3ibianchi} 
\ee
and should now be interpreted as a Bianchi identity.  This can be
solved by introducing 2-form gauge potentials $A_\2^i$, with the
$S_\3^i$ given by
\be 
S_\3^i = dA_\2^i + \fr18 \ep_{ij_1\cdots j_4}\,
A_\1^{j_1j_2}\wedge dA_\1^{j_3j_4}\;.
\label{s3isol} 
\ee
In terms of these 2-form potentials, the equations of motion can
now be obtained from the Lagrangian
\bea 
{\cal L}^0_7 &=& R\, {*\oneone} - \fr14 T^{-1}_{ij}\, {*d
T_{jk}}\wedge T^{-1}_{k\ell}\, d T_{\ell i} -\fr1{4}\,
T^{-1}_{ik}\, T^{-1}_{j\ell}\, {* F_\2^{ij}}\wedge
F_\2^{k\ell}-\fr12 T_{ij}\, {*S_\3^i}\wedge S_\3^j\nn\\
&&+ \fr12 A_\1^{ij}\wedge S_\3^i\wedge S_\3^j -2 S_\3^i\wedge
A_\2^j \wedge dA_\1^{ij} \;,
\label{d7-ungauge-lag} 
\eea
where $S_\3^i$ is given by (\ref{s3isol}).  This is precisely the
bosonic Lagrangian of the ungauged maximal supergravity in $D=7$.

       It is worth exploring in a little more detail why it is
possible to take a smooth $g\rightarrow 0$ limit in the
seven-dimensional equations of motion, but not in the Lagrangian.
We note that in this limit the Lagrangian (\ref{7d-lag}) can be
expressed as
\be 
{\cal L}_7 = \fr1{g}\, L + {\cal O}(1)\;, 
\ee
where
\be 
L = \fr12 S_\3^i dS_\3^i -\fr18 \ep_{i j_1\cdots j_4}\,
S_\3^i\wedge F_\2^{j_1 j_2}\wedge F_\2^{j_3 j_4} +
\Omega_\7\;.
\label{Lterms} 
\ee
The term $L/g$, which diverges in the $g\rightarrow 0$ limit,
clearly emphasizes that the Lagrangian (\ref{d7-ungauge-lag}) is not
merely the $g\rightarrow 0$ limit of (\ref{7d-lag}).  However if we
make use of the equations of motion, we find that in the
$g\rightarrow 0$ limit the $S_\3^i$ can be solved by
(\ref{s3isol}).  Substituting this into (\ref{Lterms}), we find
that in this limit it becomes
\be 
L=\fr1{16} \epsilon_{ij_1\cdots j_4} dA_\2^i\wedge
dA_\1^{j_1j_2}\wedge dA_\1^{j_3j_4} + {\cal O}(g)\;, 
\ee
and so the singular terms in $L/g$ form a total derivative and
hence can be subtracted from the Lagrangian.  This analysis
explains why it is possible to take a smooth $g\rightarrow 0$
limit in the equations of motion, but not in the Lagrangian.

\section{Type IIA on $S^3$}

    Here we examine, at the level of the seven-dimensional theory
itself, how to take a limit in which the $SO(5)$-gauged sector is
broken down to $SO(4)$.  In a later section, we shall show how
this can be interpreted as an $S^3$ reduction of type IIA
supergravity.  We shall do that by showing how to take a limit in
which the internal $S^4$ in the original reduction from $D=11$
becomes $\R\times S^3$. For now, however, we shall examine the
$SO(4)$-gauged limit entirely from the perspective of the
seven-dimensional theory itself.

    To take the limit, we break the $SO(5)$ covariance by splitting
the $\underline 5$ index $i$ as
\be 
i = (0,\a)\;, 
\ee
where $1\le \a\le 4$.  We also introduce a constant parameter
$\lambda$, which will be sent to zero as the limit is taken.  We
find that the various seven-dimensional fields, and the $SO(5)$
gauge-coupling  constant, should be scaled as follows:
\bea 
&&g= \lambda^2\, \td g\;,\qquad A_\1^{0\a} = \lambda^3\, \wtd
A_\1^{0\a}\;,\qquad A_\1^{\a\b} = \lambda^{-2}\, \wtd
A_\1^{\a\b}\;,\nn\\
&&S_\3^0 = \lambda^{-4}\, \wtd S_\3^0\;,\qquad
S_\3^\a = \lambda\, \wtd S_\3^\a\;.\label{s3scal}\\
&&\nn\\
&&T^{-1}_{ij} = \pmatrix{\lambda^{-8}\, \Phi & \lambda^{-3}\,
            \Phi\, \chi_\a\cr
            \lambda^{-3}\, \Phi\, \chi_\a & \lambda^2\,
M^{-1}_{\a\b} + \lambda^2\, \Phi\, \chi_\a\, \chi_\beta }\;.\nn
\eea
As we show in the next section, this rescaling corresponds to a
degeneration of $S^4$ to $R\times S^3$. Note that in this
rescaling, we have also performed a decomposition of the scalar
matrix $T^{-1}_{ij}$ that is of the form of a Kaluza-Klein metric
decomposition. It is useful also to present the consequent
decomposition for $T_{ij}$, which turns out to be
\be 
T_{ij} =\pmatrix{ \lambda^8\, \Phi^{-1} + \lambda^8\,
\chi_\gamma\, \chi^\gamma &  - \lambda^3\, \chi^\a \cr
 -\lambda^3\, \chi^\a & \lambda^{-2}\, M_{\a\beta} }\;.
\ee
Calculating the determinant, we get
\be 
\det(T_{ij}) = \Phi^{-1}\, \det(M_{\a\beta}) \;. 
\ee
Since we know that $\det(T_{ij})=1$, it follows that
\be 
\Phi = \det(M_{\a\beta})\;.
\label{phim} 
\ee
The fields $\chi_\a$ are ``axionic'' scalars.  Note that we shall
also have
\bea 
&&H_\4^0 = \lambda^{-4}\, \wtd H_\4^0\;,\qquad H_\4^\a =
\lambda\,
\wtd H_\4^\a\;,\nn\\
&&\wtd H_\4^0 = d\wtd S_\3^0\;,\qquad \wtd H_\4^\a = \tD \wtd
S_\3^\a - \td g\, \wtd A_\1^{0\a}\wedge \wtd S_\3^0\;. \eea
We have defined an $SO(4)$-covariant exterior derivative $\tD$,
which acts on quantities with $SO(4)$ indexes $\a,\beta,\ldots$ in
the obvious way:
\be \tD\, X_\a = d X_\a + \td g\, \wtd A_\1^{a\beta}\,
X_\beta\;, \ee
etc. It is helpful also to make the following further field
redefinitions:
\bea G_\2^\a &\equiv& \wtd F_\2^{0\a} + \chi_\beta\, \wtd
F_\2^{\beta\a}\;,\nn\\
G_\3^\a &\equiv& \wtd S_\3^\a - \chi_\a\, \wtd S_\3^0\;,
\label{fieldredefs}\\
G_\1^\a &\equiv& \tD\chi_\a - \td g\, \wtd A_\1^{0\a}\;,\nn
\eea
where $\wtd F_\2^{0\a}\equiv \tD \wtd A_\1^{0\a}$.

    We may now substitute these redefined fields into the
seven-dimensional equations of motion.  We find that a smooth
limit in which $\lambda$ is sent to zero exists, leading to an
$SO(4)$-gauged seven-dimensional theory.  Our results for the
seven-dimensional equations of motion are as follows.  The fields
$H_\4^i$ become
\be \wtd H_\4^0 = d\wtd S_\3^0\;,\qquad \wtd H_\4^\a = \tD
G_\3^\a + G_\1^\a\wedge \wtd S_\3^0  +\chi_\a\, d
S_\3^0\;.\label{h0ha} \ee
The first-order equations (\ref{7d-h4-def}) give
\bea \wtd H_\4^0 &=& \fr18 \ep_{\a_1\cdots \a_4}\, \wtd
F_\2^{\a_1\a_2}\wedge
                     \wtd F_\2^{\a_3\a_4}\;,\nn\\
\wtd F_\4^\a &=& \td g\, M_{\a\beta}\, {*G_\3^\beta} -\fr12
\ep_{\a\beta\gamma\delta}\,G_\2^{\beta}\wedge \wtd
F_\2^{\gamma\delta} - G_\1^\a\wedge \wtd S_\3^0\;,
\label{s3firstorder} \eea
where we have defined
\be 
F_\4^\a\equiv \tD G_\3^\a\;. 
\ee

   The second-order equations (\ref{7d-bianchi-id}) (which are nothing but
Bianchi identities following from (\ref{7d-h4-def})) become
\bea d(\Phi^{-1}\, {*\wtd S_\3^0}) &=& M_{\a\beta}\,
{*G_\3^\a}\wedge G_\1^\beta
            + G_\2^\a\wedge G_\3^\a\;,\nn\\
\tD(M_{\a\beta}\, {*G_\3^\beta}) &=& \wtd F_\2^{\a\beta}\wedge
G_\3^\beta - G_\2^\a\wedge \wtd S_\3^0\;. \eea

    The Yang-Mills equations (\ref{7d-gauge-eq}) become
\bea 
\tD(\Phi\, M^{-1}_{\alpha\beta}\, {*G_\2^\beta}) &=& \td
g\, \Phi\, M_{\a\beta}\, {*G_\1^\beta}  - \wtd S_\3^0\wedge
G_\3^\a - \fr12 \ep_{\a\beta_1\beta_2\beta_3}\,
M_{\beta_3\gamma}\, \wtd
F_\2^{\beta_1\beta_2}\wedge {*G_\3^\gamma} \;,\nn\\
\tD\, \Big[ M^{-1}_{\gamma\a}\, M^{-1}_{\delta\beta} \, {*\wtd
F_\2^{\gamma\delta}}\Big]&=& -2\td g\, M^{-1}_{\gamma[\a}\, 
{*\tD M_{\beta]\gamma}} - G_\3^\a\wedge G_\3^\beta +
\Phi\, M^{-1}_{\a\gamma}\, G_\1^\beta\wedge {*G_\2^\gamma}\nn\\
&&-\Phi\, M^{-1}_{\b\gamma}\,
G_\1^\a\wedge {*G_\2^\gamma}
 -\ep_{\a\beta\gamma\delta}\, M_{\delta\lambda}\, G_\2^\gamma\wedge
{*G_\3^\lambda} \nn\\
& & - \fr12 \Phi^{-1}\, \ep_{\a\beta\gamma\delta}\,
\wtd F_\2^{\gamma\delta} \wedge {* \wtd S_\3^0}\;. 
\label{s3ym}
\eea

    Finally, the scalar field equations (\ref{7d-scalars-eq}) give the
following:
\bea d(\Phi^{-1}\, {*d\Phi}) &=& \Phi\, M_{\a\beta}\,
{*G_1^\a}\wedge G_\1^\beta + \Phi\, M^{-1}_{\a\beta}\,
{*G_\2^\a}\wedge G_\2^\beta\nn\\
&&-\Phi^{-1}\, {*\wtd S_\3^0}\wedge \wtd S_\3^0 +\fr15 Q\;,\nn\\
\tD(\Phi\, M_{\a\beta}\, {*G_\1^\beta}) &=&\Phi\,
M^{-1}_{\beta\gamma}\, {*G_\2^\gamma}\wedge \wtd F_\2^{\a\beta}
- M_{\a\beta}\, {*G_\3^\beta}\wedge \wtd S_\3^0\;,\nn\\
\tD(M^{-1}_{\a\gamma}\, {*\tD M_{\gamma\beta}}) &=& \Phi\,
M_{\beta\gamma} {*G_\1^\gamma}\wedge G_\1^\a + M_{\beta\gamma}\,
{*G_\3^\gamma}\wedge G_\3^\a \nn\\
& & - \Phi\, M^{-1}_{\a\gamma}\,
{*G_\2^\gamma} \wedge G_\2^\beta + M^{-1}_{\a\gamma}\, 
M^{-1}_{\lambda\delta}\, {*\wtd
F_\2^{\gamma\delta}} \wedge \wtd F_\2^{\lambda\beta}\nn\\
& & + 2\td g^2(
2M_{\a\gamma}\, M_{\gamma\beta} - M_{\gamma\gamma}\,
M_{\a\beta})\, \ep_\7 - \fr15 \delta_{\a\beta}\, Q\;.
\label{s3scalar}
\eea
In these equations, the quantity $Q$ is the limit of the trace
term multiplying $\delta_{ij}$ in (\ref{7d-scalars-eq}), and is given
by
\bea
 Q&=&2\td g^2 \, \Big(2 M_{\a\beta}\, M_{\a\beta} -
(M_{\a\a})^2\Big)\, \ep_\7 - M^{-1}_{\a\gamma}\,
M^{-1}_{\beta\delta}\, {*\wtd
F_\2^{\a\beta}}\wedge \wtd F_\2^{\gamma\delta} \nn\\
&&+ \Phi^{-1}\, {*\wtd S_\3^0}\wedge \wtd S_\3^0
 -2 \Phi\, M^{-1}_{\a\beta}\, {*G_\2^{\a}}
\wedge G_\2^{\beta} + M_{\a\beta}\, {*G_\3^\a}\wedge
G_\3^\beta\;. \label{newtraceterm} \eea

   Having obtained the seven-dimensional equations of motion for the
$SO(4)$-gauged limit, we can now seek a Lagrangian from which they
can be generated.  A crucial point is that the equations involving
$\wtd H_\4^0$ in (\ref{h0ha}) and (\ref{s3firstorder}) give
\be d\wtd S_\3^0 =  \fr18 \ep_{\a_1\cdots \a_4}\, \wtd
F_\2^{\a_1\a_2}\wedge
                     \wtd F_\2^{\a_3\a_4}\;,
\ee
which allows us to strip off the exterior derivative by writing
\be \wtd S_\3^0 = dA_\2 + \omega_\3\;, \ee
where $\wtd S_\3^0$ is now viewed as a field strength with 2-form
potential $A_\2$, and
\be \omega_\3 \equiv \fr18 \ep_{\a_1\cdots \a_4}\, (\wtd
F_\2^{\a_1\a_2}\wedge \wtd A_\1^{\a_3\a_4} - \fr13 \td g\, \wtd
A_\1^{\a_1\a_2}\wedge \wtd A_\1^{\a_3\beta}\wedge \wtd
A_\1^{\beta\a_4})\;. \ee
We can now see that the equations of motion can be derived from
the following seven-dimensional Lagrangian, in which $A_\2$, and
not its field strength $\wtd S_\3^0\equiv dA_\2 +\omega_\3$, is
viewed as a fundamental field:
\bea {\cal L}_7 &=& R\, {*\oneone} - \fr{1}{4}\, \Phi^{-2}\,
{*d\Phi}\wedge d\Phi - \fr14 \wtd M^{-1}_{\a\beta}\,
{*\tD}\wtd M_{\beta\gamma}\wedge \wtd M^{-1}_{\gamma\delta}\, \tD \wtd
M_{\delta\a}
-\fr12 \Phi^{-1}\, {*\wtd S_\3^0}\wedge \wtd S_\3^0\nn\\
&& -\fr14 M^{-1}_{\a\gamma}\, M^{-1}_{\beta\delta}\, {*\wtd
F_\2^{\a\beta}}\wedge \wtd F_\2^{\gamma\delta} - \fr12 \Phi\,
M^{-1}_{\a\beta}\, {*G_\2^\a}\wedge G_\2^\beta -\fr12\Phi\,
M_{\a\beta}\, {*G_\1^\a}\wedge G_\1^\beta \nn\\
&&- \fr12 M_{\a\beta}\,{*G_\3^\a}\wedge G_\3^\beta -\wtd V\,
{*\oneone} +\fr1{2\td g} \tD \wtd S_\3^\a\wedge \wtd S_\3^\a
+\wtd S_\3^\a\wedge \wtd S_\3^0\wedge A_\1^{0\a}+\fr1{\td g}\wtd
\Omega_\7\nn\\
&&+\fr1{2\td g}\, \ep_{\a\beta\gamma\delta}\, \wtd S_\3^\a \wedge
\wtd F_\2^{0\beta}\wedge \wtd F_\2^{\gamma\delta} + \fr14
\ep_{\a_1\cdots\a_4}\, \wtd S_\3^0\wedge \wtd
F_\2^{\a_1\a_2}\wedge \wtd A_\1^{0\a_3}\wedge \wtd A_\1^{0\a_4}\;,
\label{d7lag0}
\eea
where $\wtd \Omega_\7$ is built purely from $\wtd A_\1^{\a\beta}$
and $\wtd A_\1^{0\a}$. It is defined by the requirement that its
variations with respect to $\wtd A_\1^{\a\beta}$ and $\wtd
A_\1^{0\a}$ should produce the necessary terms in the equations of
motion for these fields.  Since it has a rather complicated
structure, we shall not present it here. Note that $\wtd M_{\a\beta} \equiv 
\Phi^{-1/4}\, M_{\a\beta}$, where $\wtd M_{\a\b}$ is the unimodular matrix. 

Using the above scaling limit of the gauged
$SO(5)$ theory in seven dimensions, in which an $SO(4)$ gauging
survives, we show that it leads to a
degeneration in which the 4-sphere becomes $\R\times S^3$.  We can
then re-interpret the reduction from $D=11$ as an initial
``ordinary'' Kaluza-Klein reduction step from $D=11$ to give the
type IIA supergravity in $D=10$, followed by a non-trivial
reduction of the type IIA theory on $S^3$, in which the entire
$SO(4)$ isometry group is gauged. Note that the $S^3$ reduction of
type IIA supergravity discussed in \cite{Chamseddine:1999uy}, giving a
seven-dimensional theory with just an $SU(2)$ gauging, was
re-derived in \cite{Nastase:2000tu} as a singular limit of the $S^4$
reduction of $D=11$ supergravity that was obtained in
\cite{Nastase:1999kf}.  Since the $S^3$ reduction in \cite{Chamseddine:1999uy} retains
only the left-acting $SU(2)$ of the $SO(4)\sim SU(2)_L\times
SU(2)_R$ of gauge fields, the consistency of that reduction is
guaranteed by group-theoretic arguments, based on the fact that
all the retained fields are singlets under the right-acting
$SU(2)_R$.  The subtleties of the consistency of the $S^4$
reduction in \cite{Nastase:1999kf} are therefore lost in the singular limit
to $\R\times S^3$ discussed in \cite{Nastase:2000tu}, since a truncation
to the $SU(2)_L$ subgroup of the $SO(4)$ gauge group is made.  By
contrast, the $\R\times S^3$ singular limit that we consider here
retains all the fields of the $S^4$ reduction in \cite{Nastase:1999kf}, and
the proof of the consistency of the resulting $S^3$ reduction of
the type IIA theory follows from the non-trivial consistency of
the reduction in \cite{Nastase:1999kf}, and has no simple group-theoretic
explanation.

    To take this limit, we combine the scalings of seven-dimensional
quantities derived in the previous section with an
appropriately-matched rescaling of the coordinates $\mu^i$ defined
on the internal 4-sphere.  As in \cite{Cvetic:1999pu}, we see that after
splitting the $\mu^i$ into $\mu^0$ and $\mu^\a$, these additional
scalings should take the form
\be \mu^0 = \lambda^5\, \td\mu^0\;,\qquad \mu^\a
=\td\mu^\a\;.\label{mulim} \ee
In the limit where $\lambda$ goes to zero, we see that the
original constraint $\mu^i\, \mu^i=1$ becomes
\be \td\mu^\a\, \td\mu^\a=1\;, \ee
implying that the $\td\mu^\a$ coordinates define a 3-sphere, while
the coordinate $\td\mu^0$ is now unconstrained and ranges over the
real line $\R$.

    Combining this with the rescalings of the previous section, we
find that the $S^4$ metric reduction ansatz~(\ref{11d-metric-ansatz-s4}) 
becomes
\bea 
d\hat s_{11}^2 &=& \lambda^{-2/3}\, \Big[ \wtd \Delta^{1/3}\,
ds_7^2 + \fr1{\td g^2}\, \wtd\Delta^{-2/3}\, M^{-1}_{\a\b}\,
\tD\td\mu^\a \,\tD\td\mu^\b \nn\\
&& + \fr1{\td g^2}\,
\wtd\Delta^{-2/3}\, \Phi\, (d\td\mu_0 + \td g\, \wtd A_\1^{0\a}\,
\td \mu^\a + \chi_\a\, \tD\td\mu^\a)^2 \Big]\;, \label{r1s3met}
\eea
where
\be \wtd\Delta \equiv M_{\a\beta}\, \td\mu^\a\, \td\mu^\beta\;.
\ee
Thus $\td\mu_0$ can be interpreted as the ``extra'' coordinate of
a standard type of Kaluza-Klein reduction from $D=11$ to $D=10$,
with
\be d\hat s_{11}^2 = \rme^{-\fr16\phi}\, ds_{10}^2 + \rme^{\fr43
\phi}\, (d\td\mu_0 + \cA_\1)^2\;.\label{1step} \ee

    By comparing (\ref{1step}) with (\ref{r1s3met}), we can read off
the $S^3$ reduction ansatz for the ten-dimensional fields.  Thus
we find that the ten-dimensional metric is reduced according to
\be ds_{10}^2 = \Phi^{1/8}\, \Big[ \wtd\Delta^{1/4}\, ds_7^2
+\fr1{\td g^2}\, \wtd\Delta^{-3/4}\,   M^{-1}_{\a\b}\,
\tD\td\mu^\a \, \tD\td\mu^\b\Big]\;, \ee
while the ansatz for the dilaton $\phi$ of the ten-dimensional
theory is
\be \rme^{2\phi} = \wtd\Delta^{-1}\, \Phi^{3/2}\;. \ee
Finally, the reduction ansatz for the ten-dimensional Kaluza-Klein
vector is
\be \cA_\1 =  \td g\, \wtd A_\1^{0\a}\, \td \mu^\a + \chi_\a\,
\tD\td\mu^\a\;.\label{1formans} \ee
These results for the $S^3$ reduction of the ten-dimensional
metric and dilaton agree precisely with the results obtained in
\cite{Cvetic:2000dm}. (Note that the field $\Phi$ is called $Y$ there, and
our $M_{\a\b}$ is called $T_{ij}$ there.)   Note that the field
strength $\cF_\2=d\cA_\1$ following from (\ref{1formans}) has the
simple expression
\be \cF_\2 = \td g\, G_\2^\a\, \td\mu^\a + G_\1^\a\wedge 
\tD\td\mu^\a\;. \ee

    So far, we have read off the reduction ans\"atze for those fields
of ten-dimensional type IIA supergravity that come from the
reduction of the eleven-dimensional metric.  The remaining type
IIA fields come from the reduction of the eleven-dimensional
4-form.  Under the standard Kaluza-Klein procedure, this reduces
as follows:
\be \hat F_\4 = F_\4 + F_\3\wedge (d\td\mu^0 +
\cA_\1)\;.\label{4fs1} \ee
By applying the $\lambda$-rescaling derived previously to the
$S^4$ reduction ansatz (\ref{11d-4form-ansatz-s4}) for the eleven-dimensional
4-form, and comparing with (\ref{4fs1}), we obtain the following
expressions for the $S^3$ reduction ans\"atze for the
ten-dimensional 4-form and 3-form fields:
\bea F_\4 &=&
\fr{\wtd{\D}^{-1}}{\td{g}^3}\,M_{\a\b}\,G_\1^\a\,\td{\m}^\b\wedge
\wtd{W} + \fr{\wtd{\D}^{-1}}{2\td{g}^2}\, \e_{\a_1\ldots\a_4}\,
M_{\a_4\b}\td{\m}^\b\,G_\2^{\a_1}\wedge\tD\td{\m}^{\a_2}
\wedge\tD\td{\m}^{\a_3}\nn\\
& & - M_{\a\b}{*G_\3^{\a}\td{\m}^{\b}}
+ \fr1{\td{g}}\, G_\3^\a\wedge\tD\td{\m}^\a\;,\\
F_\3 &=& -\fr{\wtd{U}\wtd{\D}^{-2}}{\td{g}^3}\, \wtd W +
\fr{\wtd\D^{-2}}{2\td{g}^3}\, \e_{\a_1\ldots\a_4}\,
M_{\a_1\b}\td{\m}^\b\, \tD M_{\a_2\g}\td{\m}^\g
\wedge\tD\td{\m}^{\a_3}\wedge\tD\td{\m}^{\a_4}\nn\\
& & +\fr{\wtd{\D}^{-1}}{2\td{g}^2}\, \e_{\a_1\ldots\a_4}\,
M_{\a_1\b}\td{\m}^\b\,\wtd{F}_\2^{\a_2\a_3}\wedge\tD\td{\m}^{\a_4}
+ \fr1{\td g}\, \wtd{S}_\3^0\;, \eea
where
\be \wtd W \equiv \fr1{6}\, \ep_{\a_1\cdots \a_4}\,
\td\mu^{\a_1}\, \tD\td\mu^{\a_2} \wedge\tD\td\mu^{\a_3}\wedge 
\tD\td\mu^{\a_4}\;. 
\ee
   It is also useful to present the expressions for the
ten-dimensional Hodge duals of the field strengths:
\bea 
 \rme^{\fr32 \phi}\bar{*}{\cal F}_\2 &=&
\frac{\tilde{\Delta}^{-1}\Phi}{\tilde{g}^5}{*G_\2^\alpha}
\tilde{\mu}^\alpha\wedge \tilde{W} +
\frac{\tilde{\Delta}^{-1}\Phi}{2\tilde{g}^4}
\epsilon_{\alpha_1\cdots \alpha_4}M_{\alpha_1\beta}\,
\tilde{\mu}^\beta M_{\alpha_2\gamma}{*G_\1^\gamma}\wedge
\tD\tilde{\mu}^{\alpha_3}\wedge\tD\tilde{\mu}^{\alpha_4}
\;,\nn\\
{\rm e}^{-\f}\, {\bar *F_\3} &=& - \td{g}\wtd{U}\e_\7 -
\fr1{\td{g}^3}
\Phi^{-1}\, {*\wtd{S}^0_\3}\wedge\wtd{W} \nn\\
& & + \fr1{2\td{g}^2} M_{\a\g}^{-1}\, M_{\b\d}^{-1}\,
{*\wtd{F}^{\a\b}_\2}\wedge\tD\td{\m}^\g\wedge\tD\td{\m}^\d
- \fr1{\td{g}}M_{\a\b}^{-1}\, {*\tD M_{\a\g}}
\td{\m}^\g\wedge\tD\,\td{\m}^{\b}\;,\nn\\
{\rm e}^{\fr12\f}\, {\bar *F_\4} &=& \fr1{\td{g}}\,\Phi\,
M_{\a\b}\, {*G_\1^\a\td{\m}^\b} - \fr1{\td{g}^2}\, \Phi
\,M_{\a\b}^{-1} \, {*G_\2^\a}\wedge\tD\td{\m}^\b +
\fr{\wtd{\D}^{-1}}{\td{g}^4}\,M_{\a\b} G_\3^\a \td{\m}^\b\wedge
\wtd{W}\nn\\
& &+ \fr{\wtd{\D}^{-1}}{2\td{g}^3}
\e_{\a_1\ldots\a_4}\,M_{\a_1\b}\td{\m}^\b\, M_{\a_2\g}\,
{*G_\3^\g}\wedge\tD\td{\m}^{\a_3}\wedge\tD\td{\m}^{\a_4}\;.
\eea
(Here we are using $\bar *$ to denote a Hodge dualization in the
ten-dimensional metric $ds_{10}^2$, to distinguish it from $*$
which denotes the seven-dimensional Hodge dual in the metric
$ds_7^2$. )

The consistency of the $S^3$ reduction of the type IIA theory
using the ansatz that we obtained in the previous subsection is
guaranteed by virtue of the consistency of the $S^4$ reduction
from $D=11$.  It is still useful, however, to examine the
reduction directly, by substituting the ansatz into the equations
of motion of type IIA supergravity~(\ref{iia-eqs}). By this means we 
can obtain an explicit verification of the validity of the limiting
procedures that we applied in obtaining the $S^3$ reduction
ansatz. Note that it is consistent to truncate the theory to the
NS-NS sector, 
namely the subsector comprising the metric, the dilaton
and the 3-form field strength.  This implies that it is possible
also to perform an $S^3$ reduction of the NS-NS sector alone,
which was indeed demonstrated in \cite{Cvetic:2000dm}. On the other hand
it is not consistent to truncate the theory to a sub-sector
comprising only the metric, the dilaton and the 4-form field
strength, which again is in agreement with the conclusion in
\cite{Cvetic:2000dm} that it is not consistent to perform an $S^4$
reduction on such a subsector.  However, as we show in the next section,
there is a consistent $S^4$ reduction if we include all the
fields of the type IIA theory.

\section{$S^4$ reduction of type IIA supergravity}

In this section, we derive the ansatz for the consistent $S^4$ reduction
of type IIA supergravity from the $S^4$ reduction ansatz of
eleven-dimensional supergravity.  In this case we do not need to
take any singular limit of the internal 4-sphere, but rather, we
extract the ``extra'' coordinate from the seven-dimensional
space-time of the original eleven-dimensional supergravity
reduction ansatz.  The resulting six-dimensional $SO(5)$-gauged
maximal supergravity can be obtained from the Kaluza-Klein
reduction of seven-dimensional gauged maximal supergravity on a
circle.

    We begin, therefore, by making a standard $S^1$ Kaluza-Klein
reduction of the seven-dimensional metric:
\be 
ds_7^2 = \rme^{-2\a\varphi}\, ds_6^2 + \rme^{8\a\varphi}\, (dz+
\bar\cA_\1)^2\;,
\label{sevensix} 
\ee
where $\a=1/\sqrt{40}$.  With this parameterization the metric
reduction preserves the Einstein frame, and the dilatonic scalar
$\varphi$ has the canonical normalization for its kinetic term in
six dimensions.\footnote{We use a bar to denote six-dimensional
fields, in cases where this is necessary to avoid an ambiguity.}
Substituting (\ref{sevensix}) into the original metric reduction
ansatz~(\ref{11d-metric-ansatz-s4}), we obtain
\be 
d\hat s_{11}^2 = \Delta^{1/3}\, \rme^{-2\a\varphi}\, ds_6^2 +
\fr1{g^2}\, \Delta^{-2/3}\, T_{ij}^{-1}\, \cD\mu^i\, \cD\mu^j +
\Delta^{1/3}\, \rme^{8\a\varphi}\, (dz+
\bar\cA_\1)^2\;.
\label{firstgo} 
\ee

    In order to extract the ansatz for the $S^4$ reduction of type IIA
supergravity, we must first rewrite (\ref{firstgo}) in the form
\be d\hat s_{11}^2 = \rme^{-\fr16 \phi}\, ds_{10}^2 +
\rme^{\fr43\phi}\, (dz + \cA_\1)^2\;,\label{delten} \ee
which is a canonical $S^1$ reduction from $D=11$ to $D=10$.  It is
not immediately obvious that this can easily be done, since the
Yang-Mills fields $A_\1^{ij}$ appearing in the covariant
differentials $\cD\mu^i$ in (\ref{firstgo}) must themselves be
reduced according to standard Kaluza-Klein rules,
\be A_\1^{ij} = \bar A_\1^{ij} + \chi^{ij}\, (dz+\bar\cA_\1)\;,
\ee
where $\bar A_\1^{ij}$ are the $SO(5)$ gauge potentials in six
dimensions, and $\chi^{ij}$ are axions in $D=6$.  Thus we
have
\be 
\cD\mu^i = \bcD\mu^i + g\, \chi^{ij}\, \mu^j\, (dz+\bar\cA_\1)\;,
\ee
where
\be \bcD\mu^i \equiv  d\mu^i + g\, \bar A_\1^{ij}\, \mu^j\;. \ee
This means that the differential $dz$ actually appears in a much
more complicated way in (\ref{firstgo}) than is apparent at first
sight. Nonetheless, we find that one can in fact ``miraculously''
complete the square, and thereby rewrite (\ref{firstgo}) in the
form of (\ref{delten}).

    To present the result, it is useful to make the following
definitions:
\bea \Omega &\equiv & \Delta^{1/3}\, \rme^{8\a\varphi} +
\Delta^{-2/3}\,
T_{ij}^{-1}\, \chi^{ik}\, \chi^{j\ell}\, \mu^k\, \mu^\ell\;,\nn\\
Z_{ij} &\equiv & T_{ij}^{-1} - \Omega^{-1}\, \Delta^{-2/3}\,
T_{ik}^{-1}\, T_{j\ell}^{-1}\, \chi^{km}\, \chi^{\ell n}\, \mu^m\,
\mu^n\;.\label{omsdef} \eea
In terms of these, we find after some algebra that we can rewrite
(\ref{firstgo}) as
\be d\hat s_{11}^2 =\Delta^{1/3}\, \rme^{-2\a\varphi}\, ds_6^2 +
\fr1{g^2}\, \Delta^{-2/3}\, Z_{ij}\, \bcD\mu^i\, \bcD\mu^j
+\Omega\, (dz+\cA_\1)^2\;,\label{second} \ee
where the ten-dimensional potential $\cA_\1$ is given in terms of
six-dimensional fields by
\be \cA_\1 = \bar \cA_\1 + \fr1{g}\, \Omega^{-1}\,
\Delta^{-2/3}\, T_{ij}^{-1}\, \chi^{jk}\, \mu^k\, \bcD\mu^i\;.
\label{d101form} \ee
This is therefore the Kaluza-Klein $S^4$ reduction ansatz for the
1-form $\cA_\1$ of the type IIA theory.  Comparing (\ref{second})
with (\ref{delten}), we see that the Kaluza-Klein reduction
ans\"atze for the metric $ds_{10}^2$ and dilaton $\phi$ of the
type IIA theory are given by
\bea 
ds_{10}^2 &=& \Omega^{1/8}\, \Delta^{1/3}\, \rme^{-2\a\varphi}\,
ds_6^2  + \fr1{g^2}\, \Omega^{1/8}\, \Delta^{-2/3}\, Z_{ij}\,
\bcD\mu^i\,\bcD\mu^j \;,\nn\\
\rme^{\fr43\phi} &=& \Omega\;.\label{d10metphi} \eea

    The $S^4$ reduction ansatz for the R-R 4-form $F_\4$ of the type
IIA theory is obtained in a similar manner, by first implementing
a standard $S^1$ Kaluza-Klein reduction on the various
seven-dimensional fields appearing in the $S^4$ reduction ansatz
(\ref{11d-4form-ansatz-s4}) for the eleven-dimensional 4-form $\hat F_\4$, and
then matching this to a standard $S^1$ reduction of $\hat F_\4$
from $D=11$ to $D=10$:
\be \hat F_\4 = F_\4 + F_\3\wedge (dz+\cA_\1)\;.\label{f4f3} \ee
Note that in doing this, it is appropriate to treat the 3-form
fields $S_\3^i$ of the seven-dimensional theory as field strengths
for the purpose of the $S^1$ reduction to $D=6$, viz.
\be S_\3^i =  \bar S_\3^i + \bar S_\2^i\wedge (dz+ \bar\cA_\1)\;.
\ee
It is worth noting also that this implies that the reduction of
the seven-dimensional Hodge duals ${*S_\3^i}$ will be given by
\be {*S_\3^i} = \rme^{4\a\varphi}\, {\bar * \bar S_\3^i}\wedge (dz+
\bar \cA_\1) + \rme^{-6\a\varphi}\, {\bar *\bar S_\2^i}\;, \ee
where $\bar *$ denotes a Hodge dualization in the six-dimensional
metric $ds_6^2$.

    With these preliminaries, it is now a mechanical, albeit somewhat
uninspiring, exercise to make the necessary substitutions into
(\ref{11d-4form-ansatz-s4}), and, by comparing with (\ref{f4f3}), read off the
expressions for $F_\4$ and $F_\3$.  These give the Kaluza-Klein
$S^4$ reductions ans\"atze for the 4-form and 3-form field
strengths of type IIA supergravity.  We shall not present the
results explicitly here, since they are rather complicated, and
are easily written down ``by inspection'' if required.  For these
purposes, the following identities are useful:
\bea 
(dz+\bar\cA_\1) &=& (dz+\cA_\1) - \fr1{g}\, \Omega^{-1}\,
\Delta^{-2/3}\,
T_{ij}^{-1}\, \chi^{jk}\, \mu^k\, \tD\mu^i\;,\nn\\
\cD X_i &=& \bcD X_i - \Omega^{-1}\, \Delta^{-2/3}\, \chi^{ij}\,
X_j\, T_{k\ell}^{-1}\, \chi^{\ell m}\, \mu^m\, \bcD\mu^k + g\,
\chi^{ij}\, (dz+\cA_\1)\;,\nn\\
\cD\mu^i &=& T_{ij}\, Z_{jk}\, \bcD\mu^k + g\, \chi^{ij}\, \mu^j\,
(dz+\cA_\1)\;,
\eea
where in the last line $X_i$ represents any six-dimensional field
in the vector representation of $SO(5)$, and the covariant
derivative generalizes to higher-rank $SO(5)$ tensors in the
obvious way.

   If we substitute the $S^4$ reduction ans\"atze given for the
ten-dimensional dilaton, metric and 1-form in (\ref{d10metphi}),
and (\ref{d101form}), together with those for $F_\4$ and $F_\3$ as
described above, into the equations of motion of type IIA
supergravity, we shall obtain a consistent reduction to six
dimensions.  This six-dimensional theory will be precisely the one
that follows by performing an ordinary $S^1$ Kaluza-Klein
reduction on the $SO(5)$-gauged maximal supergravity in $D=7$,
whose bosonic Lagrangian is given in (\ref{d7lag0}).

    It is perhaps worth remarking that the expression
(\ref{d10metphi}) for the Kaluza-Klein $S^4$ reduction of the type
IIA supergravity metric illustrates a point that has been observed
previously (for example in \cite{Cvetic:1999au,Cvetic:2000tb}), namely 
that the ansatz becomes much more complicated when axions or pseudoscalars
are involved.  Although the axions $\chi^{ij}$ would not be seen
in the metric ansatz in a linearized analysis, they make an
appearance in a rather complicated way in the full non-linear
ansatz that we have obtained here, for example in the quantities
$\Omega$ and $Z_{ij}$ defined in (\ref{omsdef}).  They will also,
of course, appear in the ans\"atze for the $F_\4$ and $F_\3$ field
strengths.  It may be that the results we are finding here could
be useful in other contexts, for providing clues as to how the
axionic scalars should appear in the Kaluza-Klein reduction
ansatz.

\section{Type IIA supergravity on $S^1\times S^3$ and $S^1\times
S^3\times S^1$}

As shown in~\cite{Cvetic:1999un} and reviewed in appendix E, the
six-dimensional $SU(2)$-gauged supergravity can be derived from
reducing massive type IIA on a local $S^4$. In this section, we
present two other reduction ans\"atze which lead to two subsectors of Romans'
theories in $D=5$ and $D=6$. 

Firstly, the dual six-dimensional Romans' theory~(\ref{6d-dual-lag})
can be obtained from a subset, which consists of a graviton, a dilaton
and a 4-form field strength, of type IIA supergravity on $S^1\times
S^3$. The idea is that we first reduce type IIA on $S^1$. After
consistently setting 4-form and one scalar to zero, we then reduce the
obtained nine-dimensional theory on $S^3$ using the formulae
in~\cite{Cvetic:2000dm}. The reduction ans\"atze are
\begin{eqnarray}
d\hat{s}_{10}^2 &=& -\frac{1}{2}{\rm e}^{\frac{\sqrt{2}}{4}\phi}\,
ds_6^2 + \frac{1}{g^2}\,{\rm e}^{-\frac{3\sqrt{2}}{4}\phi}\,
\sum_{i=1}^3\,\left(\sigma^i -
\frac{1}{\sqrt{2}}\,g\,A^i_{(1)}\right)^2 +
{\rm e}^{\frac{5\sqrt{2}}{4}\phi}\,dZ^2\;,\nonumber\\
\hat{F}_{(4)} &=& \left(\,F_{(3)} - \frac{1}{g^2}h^1\wedge
h^2\wedge h^3 +
\frac{1}{\sqrt{2}\,g}\,F^{i}_{(2)}\wedge h^i\,\right)\wedge dZ\;,\nonumber\\
\hat{\phi\baselineskip=20pt plus 1pt minus 1pt
} &=& \frac{1}{\sqrt{2}}\,\phi\;,
\label{6d-dual-iia-s1xs3}
\end{eqnarray}
where $h^i = \s^i - \frac{1}{\sqrt{2}}\,g\,A^i$. As pointed out 
in~\cite{Nunez:2001pt}, the equations of
motion produced by the ans\"atze~(\ref{6d-dual-iia-s1xs3}) are
precisely those of the ${\cal{N}}=\tilde{4}^g$
Romans' theory~(\ref{6d-dual-lag}).
 
After dualizing 3-form field following~(\ref{6d-dualization}), we
obtain the Romans' theory in $D=6$ (Appendix E). Reducing this theory on $S^1$
produces the Romans' theory in $D=5$ without $U(1)$ gauge coupling
(Appendix F). This implies that the five-dimensional Romans' theory
without the $U(1)$ gauge-coupling can also be embedded into type IIA
supergravity. The reduction ans\"atze for metric, dilaton and field
strength are
\begin{eqnarray}
d\hat{s}_{10}^2 &=& {\rm e}^{\frac{7}{8\sqrt{6}}\phi}\, ds_5^2 +
\frac{1}{4g^2}\,{\rm e}^{-\frac{9}{8\sqrt{6}}\phi}\,
\sum_{i=1}^3\,\left(\sigma^i - g\,A^i_1\right)^2 + {\rm
e}^{\frac{15}{8\sqrt{6}}\phi}\,dY^2 +
\rme^{-\frac{9}{8\sqrt{6}}\phi}\,dZ^2\;, \nonumber\\
\hat{F}_4 &=& \left(\,\rme^{\frac{4}{\sqrt{6}}\f}\,{*\cF_2} -
\frac{1}{24g^2}\epsilon_{ijk}\,h^i\wedge h^j\wedge h^k +
\frac{1}{4g}\,F^{i}_2\wedge h^i\,\right)\wedge dY\;,\nonumber\\
\hat{\phi\baselineskip=20pt plus 1pt minus 1pt } &=&
\frac{3}{4\sqrt{6}}\,\phi\;. 
\label{5d-no-g1-iia-s1xs3xs1} 
\ea

%% file: ch3.tex
\chapter{Type IIB supergravity on $S^{\lowercase{m}}\times T^{\lowercase{n}}$}
\vspace{1cm}

\section{Five-dimensional Romans' theory from type IIB on $S^5$}

To construct the reduction ans\"atze, we follow a procedure similar 
to that used in \cite{Lu:1999bc,Cvetic:1999un}.  
We take as
our starting point the $U(1)^3$ Abelian truncation obtained
in~\cite{Cvetic:1999xp}, which involved two independent scalar fields.  One can
perform a consistent truncation of this, in which two of the three $U(1)$
potentials are set equal, and at the same time one of the scalar
degrees of freedom is eliminated.  After doing so,
the metric on the internal 5-sphere takes the following form
\bea
d\omega_5^2 &=& X\, \Delta\, d\xi^2 + X^2\, s^2 ( d\tau -
g B_\1)^2\nn\\
& & + \fr14 X^{-1}\, c^2 \Big(
d\theta^2 + \sin^2\theta\, d\phi^2 + (d\psi +\cos\theta\, d\phi
- \sqrt{2} g\, A_\1)^2\Big)\;,
\label{5met1}
\eea
where $X\equiv \rme^{-\fr1{\sqrt6}\, \phi}$, and we have defined
\be
c\equiv \cos\xi\;,\qquad s\equiv\sin\xi\;.
\ee
The function $\Delta$ is given by
\be
\Delta = X^{-2}\, s^2 + X\, c^2\;.
\ee
At $X=1$, in the absence of the $U(1)$ gauge fields, the metric
(\ref{5met1}) describes a unit 5-sphere.  (The gauge field $A_\1$
appearing in (\ref{5met1}) is the one that comes from setting two
of the original $U(1)$ gauge fields in \cite{Cvetic:1999xp} equal.)

     We can now make a non-Abelian generalization of the
the 5-sphere metric ansatz, by introducing the three
left-invariant 1-forms $\sigma^i$ on the 3-sphere, which satisfy the
Cartan-Maurer equation
\be
d\sigma^i + \fr12 \ep_{ijk}\, \sigma^j\wedge \sigma^k = 0\;.
\label{cartan-maurice}
\ee
These can be written in terms of the Euler angles as 
\bea
\sigma_1 &=& \cos\psi\,d\theta + \sin\theta\sin\psi\,d\phi\;,\nn\\
\sigma_2 &=& -\sin\psi\,d\theta + \sin\theta\cos\psi\,d\phi\;,\nn\\
\sigma_3 &=& d\psi + \cos\theta\, d\phi\;.   
\label{cartan-maurice-euler}
\eea
Thus we are naturally led to generalize (\ref{5met1}) to
\be
d\omega_5^2 = X\, \Delta\, d\xi^2 + X^2\, s^2 \Big(d\tau
- g B_\1 \Big)^2 +
\fr14 X^{-1}\, c^2 \sum_{i=1}^3
(\sigma^i-\sqrt{2}g\, A_\1^i)^2\;.
\label{5met2}
\ee
The Abelian limit (\ref{5met1}) is recovered by setting the $i=1$ and
$i=2$ components of the $SU(2)$ gauge potentials to
zero.

    By proceeding along these lines, we are eventually led
to the following ans\"atze for the ten-dimensional metric,
the 5-form field strength $\hat H_\5 \equiv \hat G_\5 + {\hat *\hat
G_\5}$, and the two 2-form potentials:
\bea
d\hat s^2_{10} &=&
\Delta^{1/2}\, ds_5^2 +  g^{-2} \, X\,
\Delta^{1/2}\, d\xi^2 +
g^{-2}\Delta^{-1/2}\, X^2\, s^2\, \Big(d\tau - g\, B_\1\Big)^2 \nn\\
& &+ \fr14 g^{-2}\, \Delta^{-1/2}\, X^{-1}\,
c^2\, \sum_i (\sigma^i - \sqrt{2}g\, A_\1^i)^2\;,\nn\\
\hat{G}_\5 &=& 2g\, U \,\varepsilon_5 - \frac{3 sc}{g} X^{-1}\,
{*dX}\wedge d\xi +
\frac{c^2}{8\sqrt{2}\, g^2} X^{-2}\, {*F^i_\2}\wedge h^j\wedge h^k
\, \varepsilon_{ijk} \nn\\
& &-\frac{sc}{2\sqrt{2}\,g^2} X^{-2}\, {*F^i_\2}\wedge h^i\wedge d\xi
- \frac{sc}{g^2} X^4\, {*G_\2}\wedge d\xi\wedge (d\tau - g B_\1)\;,\nn\\
\hat{A}_\2 &\equiv& \hat A_\2^1 + \im\, \hat A_\2^2 =
 -\fr{s}{\sqrt{2}g}\, \rme^{-\im\,\tau}\, A_\2\;,\nn\\
\hat{\phi} &=& 0\;,\ \ \ \hat{\chi} = 0\;,
\label{metans}
\eea
where $h^i\equiv \sigma^i - \sqrt{2} g\, A_\1^i$, $U\equiv X^2\, c^2 +
X^{-1}\, s^2 + X^{-1}$, and $\ep_5$ is the volume form in the
five-dimensional space-time metric $ds_5^2$.
Note that we have defined the complex 2-form potential $\hat A_\2\equiv
\hat A_\2^1 + \im\, \hat A_\2^2$ in the type IIB theory.
The ten-dimensional dilaton and the axion are constants,
which without loss of generality we have set to zero.  The $SU(2)$
Yang-Mills field strengths $F_\2^i$ are given by $F_\2^i = dA_\1^i +
\frac{1}{\sqrt{2}} g\, \ep_{ijk}\, A_\1^j\wedge A_\1^k$.  It should be
emphasized that all the fields $X$, $A_\2^\a$ and $A_\1^i$, and the
metric $ds_5^2$, appearing on the right-hand sides of (\ref{metans}),
are taken to depend only on the coordinates of the five-dimensional
space-time.\footnote{Note that the metric reduction ansatz in
(\ref{metans}) has a fairly simply form, and fits in with the general
pattern of Kaluza-Klein metric ans\"atze, such as have been seen in
many previous examples.  (See
\cite{Salam:1982xd,deWit:1985nz,Nilsson:1985dj} 
for some earlier
examples, and \cite{Khavaev:1998fb,Cvetic:1999xp,Lu:1999bc,Cvetic:1999un} for
some more recent ones.)  It
is the determination of the reduction ans\"atze for the field
strengths that is the more difficult part of the problem, and since
the consistency of the reduction depends crucially on conspiracies
between the field strength and metric contributions, it is only when
the full ansatz is known that the consistency can be established.}
Note also that since the two 2-form potentials of the five-dimensional
supergravity are charged under the $U(1)$ factor in the $SU(2)\times
U(1)$ gauge group, these fields appear in the
reduction ansatz with the appropriate $\tau$ dependence for charged
$U(1)$ harmonics.

   The ten-dimensional Hodge dual of $\hat G_\5$ turns out to be
\bea
{\hat *\hat{G}_\5} &=& -\fr{sc^3}{4g^4}\, U\,
\Delta^{-2}\, d\xi\wedge \omega\wedge
\epsilon_3 + \frac{3s^2\, c^4}{8g^4} \,X^{-2}\, \Delta^{-2}\, dX\wedge
\omega\wedge\epsilon_3 \nn\\
& &-\fr{s^2\, c^2}{8\sqrt{2}\, g^3}\, X^{-2}\, \Delta^{-1}\,
F^i_\2\wedge h^j\wedge h^k\wedge
\omega\,  \epsilon_{ijk} + \fr{s\, c}{2\sqrt{2}\, g^3} \,
F^i_\2\wedge h^i\wedge
d\xi\wedge \omega\nn\\
& & - \fr{c^4}{8g^3}\, X\, \Delta^{-1}\,  G_\2\wedge\epsilon_3\;,
\label{}
\eea
where $\w\equiv  d\t - g B_\1$, and $\ep_3 \equiv h^1\wedge h^2\wedge h^3$.

   We must now verify that the ans\"atze (\ref{metans}) do indeed
yield a consistent reduction of the type IIB theory to the $\cN=4$
gauged supergravity in $D=5$.  Let us begin by noting that the Bianchi
identity for $\hat H_\5$, and the field equations for the NS-NS and
R-R 3-forms of the type IIB theory become, in the complex notation,
\be
d\hat H_\5 = \fr{\im}{2}\,
\overline{\hat{F}}_\3\wedge \hat{F}_\3\;,\qquad
d{\hat *\hat F_\3} = -\im\, \hat H_\5\wedge \hat F_\3\;.
\label{complexeom}
\ee
We now find that the Bianchi identity for $\hat H_\5$
gives rise to the following equations in $D=5$:
\bea
d(X^{-1}\, {*dX}) &=& \fr43 g^2 \, (X^{-1}-X^2) {*\oneone}
- \fr16 X^{-2}\, ({*F^i_\2}\wedge
F^i_\2 + {*{\bar A}_\2}\wedge A_\2)\nn\\
& &+ \fr13 X^4\, {*G_\2}\wedge G_\2\;,\nn\\
d(X^{-2}\, {*F^i_\2}) &=& \sqrt{2}\, g \,X^{-2}\,
\epsilon^{ijk}\, {*F^j_\2}\wedge A^k_\1 -
F^i_\2\wedge G_\2\;,\nn\\
d(X^4\, {*G_\2}) &=& -\fr12 F^i_\2\wedge F^i_\2 -
\fr12 {\bar A}_\2\wedge A_\2\;.
\label{equation}
\eea
These are precisely equations of motion of the Romans' theory.

     Plugging the ans\"atze~(\ref{metans}) into the equation of motion
for $\hat F_\3$ in (\ref{complexeom}) gives rise
to the five-dimensional first-order equation for $A_\2$ in
(\ref{5d-romans-eqs-form}), and also the second-order equation
\be
\cD(X^{2}\, {* F_\3}) = g^2\, X^{-2}\, {*A_\2}\;,\label{secondorder}
\ee
where the $U(1)$ gauge-covariant exterior derivative $\cD$ is as defined
in (\ref{5d-romans-gauge-def}).  This second order equation of motion 
is in fact implied by the first order equation. With our ans\"atze, 
the ten-dimensional equations of motion for
the dilaton and axion are automatically satisfied. Substituting the
ans\"atze into the ten-dimensional Einstein equation
(\ref{iib-einstein-eq}) does not lead to any other equations except
five-dimensional Einstein equation~(\ref{5d-romans-eqs-form}). 
Thus the full consistency of the reduction ansatz (\ref{metans}) is
established.

 From the above ans\"atze, one can perform some appropriate rescaling
 to transform them from one gauge $g$ coupling to two gauge couplings
 $g_1$ and $g_2$ which correspond to the gauge couplings of $U(1)$ and $SU(2)$ 
respectively. We then in principle can take a singular limit in which
 $g_1$ is set to zero. In this singular limit $S^5$ is deformed to
 $S^3\times T^2$. As a result of this process, we
 obtain the five-dimensional Romans' theory without the $U(1)$ 
gauge-coupling. Instead of taking the singular limit of the metric and
 field strength ans\"atze in Eq.~(\ref{metans}), we
 obtain the same theory by reducing a
 subset, which comprises a metric, a dilaton and a NS-NS 3-form, of
 type IIB supergravity on $S^3\times T^2$ together with appropriate
 dualizations~\cite{Schvellinger:2001ib}. The ans\"atze are
\ba 
d\hat{s}_{10}^2 &=& \rme^{\frac{13}{5\sqrt{6}}\f}\,ds_5^2 +
\rme^{\frac{3}{5\sqrt{6}}\f}\,(dY^2 + dZ^2) +
\frac{1}{4g^2}\,\rme^{-\frac{3}{\sqrt{6}}\f}\,\sum_{i=1}^3(\s^i -
g\,A_1^i)^2\;,\nn\\
\hat{F}_3 &=&  \rme^{\frac{4}{\sqrt{6}}\f}{*\cF_2} -
\frac{1}{24g^2}\,\epsilon_{ijk}\, h^i\wedge h^j\wedge h^k +
\frac{1}{4}\,\sum_{i=1}^3\,F_2^i\wedge h^i\;,\nn\\
\hat{\phi} &=& \sqrt{6}\,\phi\;,\;\;\;{\rm and}\;\;\;h^i = \s^i -
g\,A_1^i\;. 
\label{5d-no-g1-iib-s3xt2} 
\ea
Notice that reducing the above mentioned subset of type IIB
supergravity produces the $SU(2)$-gauged $\cN=2$ supergravity in $D=7$
without topological mass
term~\cite{Townsend:1983kk,Salam:1983fa}. Reductions of
seven-dimensional gauged supergravity on $S^1$ and on $T^2$ were done 
in~\cite{Giani:1984dw,Cowdall:1998rs,Cowdall:1998fn}.
The above mentioned seven-dimensional gauged supergravity can
also be obtained from appropriate truncation of a gauged supergravity
derived from reducing type IIA supergravity on
$S^3$~\cite{Cvetic:2000ah}. It is not clear that a supergravity
obtained from reducing type IIA on $S^3$ is dual to a theory obtained
from reducing IIB on $S^3$ in the same sense as T-duality (sometimes
it is called non-Abelian T-duality). However, reducing particular subsets
of type IIA and type IIB supergravities on $S^3$ produces the same
theory. Based on this one might say that there exists non-Abelian
T-duality. Unfortunately, an explicit construction from purely
supergravity theories has yet been done. Arguments based on worldsheet
conformal field theory were done by many
groups~\cite{Rocek:1992ps,delaOssa:1993vc,Alvarez:1994qi,Alvarez:1994zr,Alvarez:1994wj,Alvarez:1995dn,Bossard:2000xq}. 

\section{$SO(6)$ gauged supergravity in $D=5$}

In this section, we obtain the consistent reduction ansatz for a
different truncation of the maximal gauged five-dimensional supergravity.  One
can consistently set the $\bf{1}+\bf{1}+\bf{10}+\overline{\bf{10}}$ of 
scalars to zero, at the
same time as setting the $\bf{6}+\overline{\bf{6}}$ of 2-form potentials to zero.
The bosons that remain, namely gravity, the $\bf{15}$ Yang-Mills fields,
and the $\bf{20}^\prime$ of scalars, come just from the metric plus
the self-dual 5-form sector of the original type IIB theory.  We shall obtain
complete results for the consistent embedding of this sub-sector of the
gauged $\cN=8$ theory, with all fifteen gauge fields
$A_\1^{ij}$ of the $SO(6)$, and the twenty scalars that parameterize the full
$SL(6,\R)/SO(6)$ sub-manifold of the complete scalar coset.  These can
be parameterized by a unimodular symmetric tensor $T_{ij}$.

    Another way of expressing the truncation that we shall consider in
this section is as follows.  The type IIB theory itself can be
consistently truncated in $D=10$ so that just gravity and the
self-dual 5-form remain.  The fifteen Yang-Mills gauge fields and
twenty scalars that we retain in our ansatz are the full set of
massless fields associated with Kaluza-Klein reduction from this
ten-dimensional starting point.  (The counting of massless fields is
the same as the one that arises from a toroidal reduction from the
same ten-dimensional starting point.)  We shall see below that the
self-duality condition on the 5-form plays an essential r\^ole in the
consistency of the $S^5$ reduction.
This subsector of type IIB supergravity is particularly relevant
for the AdS/CFT correspondence, because it is the metric and the self-dual 
5-form that couple to the D3-brane as we will see in chapter IV.  

   We parameterize the fields for this truncated theory as follows.
The twenty scalars, which are in the $\bf{20}^\prime$ representation
of $SO(6)$,
are represented by the symmetric unimodular tensor $T_{ij}$, where $i$
is a 6 of $SO(6)$.  The fifteen $SO(6)$ Yang-Mills gauge fields will
be represented by the 1-form potentials $A_\1^{ij}$, antisymmetric in
$i$ and $j$.  The inverse of the scalar matrix $T_{ij}$ is denoted by
$T^{-1}_{ij}$.  In terms of these quantities, we find that the
Kaluza-Klein reduction ansatz is given by
\bea
d\hat s_{10}^2 &=& \Delta^{1/2}\, ds_5^2 + g^{-2}\, 
\Delta^{-1/2}\, T^{-1}_{ij}\, 
\cD\mu^i\, \cD\mu^j\;,\label{iib-metric-ans-s5}\\
\hat H_\5 &=& \hat G_\5 + {{\hat *}\hat G_\5}\;,\label{iib-h5-ans-s5}\\
\hat G_\5 &=& -g\, U\, \ep_\5 + g^{-1}\, (T^{-1}_{ij}\, {*\cD}\, T_{jk})\wedge 
(\mu^k\, \cD\mu^i)\nn\\
&& -\fr12 g^{-2}\, 
T^{-1}_{ik}\, T^{-1}_{j\ell}\, {*F_\2}^{ij}\wedge
\cD\mu^k\wedge \cD\mu^\ell\;,\label{iib-gauge-ans-s5}\\
{{\hat *}\hat G_\5} &=& \fr1{5!}\, \ep_{i_1\cdots i_6}\, \Big[
g^{-4}\, U\, \Delta^{-2}\, \cD\mu^{i_1}\wedge \cdots \wedge \cD\mu^{i_5}\,
\mu^{i_6}\nn\\
&& -5 g^{-4}\, \Delta^{-2}\, \cD\mu^{i_1}\wedge \cdots \wedge \cD\mu^{i_4}
\wedge \cD T_{i_5 j}\, T_{i_6 k}\, \mu^j\, \mu^k \nn\\
&&- 10 g^{-3}\, \Delta^{-1}\, 
 F_\2^{i_1 i_2}\wedge \cD\mu^{i_3}\wedge \cD\mu^{i_4}\wedge \cD\mu^{i_5}\, 
T_{i_6 j}\, \mu^j \Big]\;,\label{iib-gdualans-s5}
\eea
where
\bea
&&U \equiv 2 T_{ij}\, T_{jk}\, \mu^i\, \mu^k -\Delta\, T_{ii}\;, \qquad 
\Delta \equiv T_{ij}\, \mu^i\, \mu^j\;,\nn\\
&&F_\2^{ij} = dA_\1^{ij} + g\, A_\1^{ik}\wedge A_\1^{kj}\;,
\qquad \cD T_{ij} \equiv dT_{ij} + g\, A_\1^{ik}\, T_{kj} + g\, A_\1^{jk}\, 
T_{ik}\;,\nn\\
&& \mu^i\, \mu^i = 1\;,\qquad 
\cD\mu^i \equiv d\mu^i +g\,  A_\1^{ij}\, \mu^j\;,
\eea
and $\ep_\5$ is the volume form on the five-dimensional space-time.
Note that ${{\hat *}\hat G_\5}$ is derivable from the given
expressions (\ref{iib-metric-ans-s5}) and (\ref{iib-gauge-ans-s5}); 
we have presented it here because it is quite an involved computation.  
The coordinates $\mu^i$,
subject to the constraint $\mu^i\, \mu^i=1$, parameterize points in
the internal 5-sphere.  In obtaining the above ansatz we have been
guided by previous results in the literature, including the $S^4$
reduction ansatz from $D=11$ that was constructed in 
\cite{Nastase:1999cb,Nastase:1999kf}.

   It is consistent to truncate the fields of type IIB supergravity to
the metric and self-dual 5-form $\hat H_\5$.  The ten-dimensional
equations for motion for these fields can be read off from
Eqs.~(\ref{iib-einstein-eq})-(\ref{iib-5form-eq}). The ansatz
presented above satisfies these equations of motion if and
only if the five-dimensional fields satisfy the equations
\bea
\cD(T^{-1}_{ik}\, {*\cD} T_{kj}) &=& -2g^2\, (2T_{ik}\, T_{jk} - T_{ij}\,
T_{kk})\,\ep_\5 + T^{-1}_{ik}\, T^{-1}_{\ell m}\, {*F_\2^{\ell k}}
\wedge F_\2^{mj}\nn\\
&&-\fr16 \delta_{ij}\, \Big[ -2g^2\, (2T_{k\ell}\, T_{k\ell} -
(T_{kk})^2)\, \ep_\5 +
T^{-1}_{pk}\, T^{-1}_{\ell m}\, {*F_\2^{\ell k}} \wedge
F_\2^{mp} \Big]\;,\nn\\
\cD(T^{-1}_{ik}\, T^{-1}_{j\ell}\, {*F_\2^{k\ell})} &=&
-2g\, T^{-1}_{k[i}\, {*\cD}T_{j]k} - \fr18 \ep_{ij k_1\cdots k_4}\, 
F_\2^{k_1 k_2}\wedge F_\2^{k_3 k_4}\;,\label{d5eom}
\eea
together with the five-dimensional Einstein equation.  These equations
of motion can all be derived from the five-dimensional Lagrangian
\bea
{\cal L}_5 &=& R\, {*\oneone} - \fr14 T^{-1}_{ij}\, {*\cD T_{jk}}\wedge
T^{-1}_{k\ell}\,\cD T_{\ell i} - \fr14 T^{-1}_{ik}\,
T^{-1}_{j\ell}\, {* F_\2^{ij}}\wedge F_\2^{k\ell} 
-V\, {*\oneone}\nn\\
&&- \fr1{48}\, \ep_{i_1\cdots i_6}\, 
\Big(F_\2^{i_1 i_2}\, F_\2^{i_3 i_4}\, A_\1^{i_5 i_6} -
 g\, F_\2^{i_1 i_2}\, A_\1^{i_3 i_4}\, A_\1^{i_5 j}\, A_\1^{j i_6} \nn\\
&&
+\fr25 g^2\, A_\1^{i_1 i_2} \, A_\1^{i_3 j}\, A_\1^{j i_4}\, A_\1^{i_5
k}\, A_\1^{k i_6} \Big)\;,\label{d5lag}
\eea
where the potential $V$ is given by
\be
V = \fr12 g^2\, \Big(2 T_{ij}\, T_{ij} - (T_{ii})^2 \Big)\;.
\ee
In (\ref{d5lag}) we have omitted the wedge symbols in the final
topological term due to its cumbersome.  The Lagrangian is in agreement 
with the one for five-dimensional gauged $SO(6)$ supergravity in
\cite{Gunaydin:1986cu}.

   To see this, we look first at the ten-dimensional equation $d{\hat
H_\5}=0$, with the ansatz given above.  The terms involving structures
of the form $X^{ij}_\4 \wedge \cD\mu^i\wedge \cD\mu^j$, where $X^{ij}_\4$
represents a 4-form in the five-dimensional space-time, give rise to
the Yang-Mills equations above.  Contributions of this kind come from
$d{\hat G_\5}$ and also from the final term in $d{\hat *\hat G_\5}$.
The terms involving structures of the form $X^{ij}_\5\wedge (\mu^i\,
\cD\mu^j)$, where $X_\5^{ij}$ represents a 5-form in the five-dimensional
space-time, give rise to the scalar equations of motion in
(\ref{d5eom}).  Since $\mu^i\, \cD\mu^i = \mu^i\, d\mu^i = \fr12
d(\mu^i\, \mu^i)=0$, there is a trace subtraction in the scalar
equation, and we read off $X_\5^{ij} - \fr16 \delta_{ij} X_\5^{kk}=0$
as the five-dimensional equation of motion.  Contributions of this
structure come only from $d\hat G_\5$.  Finally, all other structures
arising from calculating $d{\hat H_5}=0$ vanish identically, without
the use of any five-dimensional equations of motion.  In deriving
these results one needs to make extensive use of the Schoutens
over-antisymmetrization identity, $\ep_{[i_1\cdots i_6}\, V_{i_7]}=0$.

   It is worth remarking that it is essential for the consistency of
the reduction ansatz that the 5-form $\hat H_\5$ should be self-dual.
One cannot simply consider a reduction ansatz for a ten-dimensional
theory consisting of gravity plus a non-self-dual 5-form, whose ansatz
is given by $\hat G_\5$ in (\ref{iib-gauge-ans-s5}).  Although the Bianchi
identity $d\hat G_\5=0$ would give perfectly acceptable equations of
motion for $F_\2^{ij}$ and $T_{ij}$, the field equation $d{\hat *\hat
G_\5}=0$ would produce the (unacceptable) constraint
\be
\ep_{ijk_1\cdots k_4}\, F_\2^{k_1 k_2}\wedge F_\2^{k_3 k_4}=0\;.
\label{ffterm}
\ee
It is only by combining $\hat G_\5$ and ${\hat *\hat G_\5}$ together
into the self-dual field $\hat H_\5$ that a consistent
five-dimensional result is obtained, with (\ref{ffterm}) now combining
with terms from $d\hat G_\5$ to form part of the five-dimensional
Yang-Mills equations given in (\ref{d5eom}). Notice that if one were to
consider the reduction of a ten-dimensional theory with a
non-self-dual 5-form there would be additional fields present in a
complete massless truncation.  These would comprise
$\bf{10}=\bf{4}+\bf{6}$ vector potentials and $\bf{5}=\bf{1}+\bf{4}$ 
scalars.  However the inclusion of these fields
would still not achieve a consistent reduction ansatz, since the current
(\ref{ffterm}), which is in the $\bf{15}$ of $SO(6)$, could still not
acquire an interpretation as a source term for the additional fields.
Thus the additional requirement of self-duality seems to be essential
for consistency. Of course, anti-self-duality would be equally good.
It is interesting, therefore, that self-duality of the 5-form is
apparently forced on us by the requirements of Kaluza-Klein
consistency, if we try to ``invent'' a ten-dimensional theory that can
be reduced on $S^5$.  Thus once again we see that supersymmetry and
Kaluza-Klein consistency for sphere reductions seem to go hand in
hand. Of course, the initial truncation of the type IIB theory
to its gravity plus self-dual 5-form sector is itself a
non-supersymmetric one, but the crucial point is that the consistency
of the Kaluza-Klein $S^5$ reduction is singling out a starting point
that is itself a subsector of a supersymmetric theory.

   The Kaluza-Klein $S^5$ reduction that we have obtained here retains
the full set of massless fields that can result from the reduction of
gravity plus a self-dual 5-form in $D=10$.  In other words, after the
initial truncation of the type IIB theory in $D=10$, no further
truncation of massless fields has been performed.

   It should also be emphasized that it would be inconsistent to omit
the fifteen $SO(6)$ gauge fields when considering the embedding of the
twenty scalars $T_{ij}$.  This can be seen from the Yang-Mills
equations in (\ref{d5eom}), which have a source term $g\,
T^{-1}_{k[i}\, {*\cD}T_{j]k}$ appearing on the right-hand side.  This is
a quite different situation from a toroidal reduction, where it is
always consistent to truncate to the scalar sector, setting the gauge
fields to zero.  The new feature here in the sphere reduction is that
the scalar fields are charged under the gauge group.  This is a
general feature of all the sphere reductions, to $D=4$, $D=5$ and
$D=7$, and thus in all cases it is inconsistent to include the full
set of scalar fields without including the gauge fields as well (see also
\cite{Nastase:1999cb,Nastase:1999kf,Nastase:2000tu}).  (One can 
consistently truncate to the
diagonal scalars in $T_{ij}$, setting all the gauge fields to zero, as
in \cite{Cvetic:1999xx,Cvetic:2000eb}, since then 
$T^{-1}_{k[i}\, {*\cD}T_{j]k}=0$.)

    Our testing of the consistency of the reduction ansatz
(\ref{iib-metric-ans-s5})-(\ref{iib-gdualans-s5}) has so far been restricted to
checking the equations of motion for $\hat H_\5$.  A full testing of
the consistency of the ten-dimensional Einstein equations would be
quite involved, and will be addressed in future work. The
experience in all previous work on consistent reductions is that
although the actual checking of the higher-dimensional Einstein
equations is the most difficult part from a computational point of
view, the consistency seems to be assured once it has been achieved
for the equations of motion for the antisymmetric tensor fields, which
itself is an extremely stringent requirement.

The $S^5$ reduction ansatz that we
have constructed here is also
consistent if we include the dilaton $\hat\phi$ and axion $\hat\chi$ 
of the type IIB theory.  These simply reduce according to the ans\"atze
\be
\hat\phi = \phi\;,\qquad \hat \chi= \chi\;,
\ee
where the unhatted quantities denote the fields in five dimensions.  In this
reduction they do not appear in the previous ans\"atze for the metric and
self-dual 5-form, and in $D=5$ they just give rise to the additional
$SL(2,\R)$-invariant Lagrangian
\be
{\cal L}_{(\phi,\chi)} = -\fr12 {*d\phi}\wedge d\phi - \fr12 e^{2\phi}\, 
{*d\chi}\wedge d\chi\;,
\ee
which is added to (\ref{d5lag}).  Note in particular that $\phi$ and $\chi$
do not appear in the five-dimensional scalar potential $V$.

%% file: ch4.tex
\chapter{Some applications}
\vspace{1cm}

\section{Super Yang-Mills operators via AdS/CFT correspondence}

There are two known ways to
determine the SYM operators that correspond to given supergravity
modes via AdS/CFT~
\cite{Maldacena:1998re,Gubser:1998bc,Witten:1998qj}. 
In the first approach \cite{Witten:1998qj,Ferrara:1998bp}, one 
considers the
representations of the super Lie algebra, $SU(2,2|4)$, of the
fields in SYM theory and IIB supergravity  respectively, and
matches the supergravity modes with the SYM operators by comparing
the various quantum numbers. The other approach was proposed by Das and 
Trivedi \cite{Das:1998ei}. 
They considered the lowest Kaluza-Klein mode of the NS-NS 2-form
fields polarized along the D3-brane worldvolume. They worked out
the corresponding SYM operators by expanding the Dirac-Born-Infeld
(DBI) action plus Wess-Zumino (WZ) 
terms around an $AdS_5 \times S^5$ background. They noted that
the expansion around this background, as opposed to a flat
background, is crucial in order to obtain the correct SYM
operators. A similar method was used in~\cite{Das:1996wn,Callan:1997tv}

The relevance of curved backgrounds in AdS/CFT was already noticed
in~\cite{Ferrara:1998ej,Liu:1998bu} and was further motivated 
in~\cite{Park:1999xz}. Following~\cite{Park:2000du}, we consider another
supergravity mode, and work out the SYM operators using the method
presented in~\cite{Das:1998ei}.

The ansatz
given in Eqs.~(\ref{iib-metric-ans-s5})-(\ref{iib-gdualans-s5}) is for 
a general $S^5$ reduction, but for our discussion, we choose the
space-time $ds_5^2$ to be a five-dimensional anti-de Sitter space, $AdS_5$.
Furthermore, to simplify our discussion, we consider
configurations with $A^{ij}_\1=0$, which in turn enables us to
put $T_{ij}$ into diagonal form 
\be 
T_{ij} = \mbox{diag}(X_1,X_2,X_3,X_4,X_5,X_6)\;;
\;\;\;\; \prod_{i=1}^6 X_i = 1\;. 
\ee
We now adopt the parameterization used in~\cite{Cvetic:1999xx}
\be 
X_i = \exp\left(-\fr12 \vec{b}_i\cdot\vec{\vp}\right)\;, 
\ee
where $\vec{b}_i$ are the weight vectors of the fundamental
representation of $SL(6, \R)$ which satisfy the following relations
\be
\vec{b}_i\cdot\vec{b}_j =
8\,\d_{ij} - \fr43\;,\; \sum_{i=1}^6\,\vec{b}_i = 0\;, \; \mbox{and}
\;\sum_{i=1}^6 (\vec{u}\cdot\vec{b}_i)\,\vec{b}_i = 8\,\vec{u}\;,
\ee
where $\vec{u}$ is an arbitrary vector and $\vec{\vp} = (\vp_1, \vp_2,
\vp_3, \vp_4, \vp_5)$ are five independent scalars. The explicit
expressions for $\vec{b}_i$ can be taken as follows
\bea 
\vec{b}_1 &=& \left(2, \fr2{\sqrt{3}}, \fr2{\sqrt{6}},
\fr2{\sqrt{10}}, \fr2{\sqrt{15}}\right),\hspace{0.3cm}  \vec{b}_2
= \left(-2, \fr2{\sqrt{3}},
                             \fr2{\sqrt{6}}, \fr2{\sqrt{10}},
\fr2{\sqrt{15}}\right)\;, \nn\\
\vec{b}_3 &=& \left(0, -\fr4{\sqrt{3}}, \fr2{\sqrt{6}},
\fr2{\sqrt{10}}, \fr2{\sqrt{15}}\right), \vec{b}_4 = \left(0, 0,
-\sqrt{6}, \fr2{\sqrt{10}},
\fr2{\sqrt{15}}\right)\;, \nn\\
\vec{b}_5 &=& \left(0, 0, 0, -\fr8{\sqrt{10}},
\fr2{\sqrt{15}}\right), \hspace{0.9cm} \vec{b}_6 = \left(0, 0, 0,
0, -\fr{10}{\sqrt{15}}\right)\;. 
\eea
With the diagonal form of $T_{ij}$, the Lagrangian~(\ref{d5lag}) and
equations~(\ref{d5eom}) of motion up to fourth order in $\vec{\vp}$ 
are
\bea 
\rme^{-1} {\cal L} &=&  - \fr12 (\pa\vec{\vp})\cdot(\pa\vec{\vp}) +
12 g^2 + \fr14 g^2 \sum_{i=1}^6 (\vec{b}_i\cdot\vec{\vp})^2 +
\fr1{24} g^2 \sum_{i=1}^6 (\vec{b}_i\cdot\vec{\vp})^3\nn\\
& & - \fr5{192} g^2 \sum_{i=1}^6 (\vec{b}_i\cdot\vec{\vp})^4 +
\fr1{128} g^2 \left[\sum_{i=1}^6 (\vec{b}_i\cdot\vec{\vp})^2\right]^2\nn\\
&=&  - \fr12  \pa_\m\vec{\vp}\cdot\pa^\m\vec{\vp} + 12 g^2 + 2
g^2  \vec{\vp}\cdot\vec{\vp} + V_3+ V_4\;, \nn\\
\label{5sugra} 
\Box\vec{\vp} &=& - 4 g^2 \vec{\vp} - \fr18 g^2
\sum_{i=1}^6 \vec{b}_i\,(\vec{b}_i\cdot\vec{\vp})^2 + \fr5{48} g^2
\sum_{i=1}^6
\vec{b}_i\,(\vec{b}_i\cdot\vec{\vp})^3\nn\\
& & - \fr14 g^2 \vec{\vp} \sum_{i=1}^6 (\vec{b}_i\cdot\vec{\vp})^2\;,
\eea
where $V_3$ and $V_4$ are third-order and fourth-order polynomials
in $\vec{\varphi}$ respectively. We do not present here the explicit forms of
$V_3$ and $V_4$ due to their cumbersome. However, their explicit forms
can be found in~\cite{Park:2000du}.

In \cite{Lee:1998bx}, two- and three-point correlators for various
chiral primary operators were computed using a linear ansatz in
\cite{Kim:1985ez}. As a consequence of using the linear ansatz,
non-linear field redefinitions were required.  The advantage of
having a non-linear ansatz is obvious from (\ref{5sugra}): it
renders such field redefinitions unnecessary. One can easily read
off two- and three-point functions using the formulae in
\cite{Freedman:1998tz,Muck:1998rr}.

Keeping only the terms linear in $\varphi$, the metric
ansatz~(\ref{iib-metric-ans-s5}) can be written as
\bea 
ds_{10}^2 &\simeq& \left(1-\fr{1}{4}\sum_i(\m^i)^2\vec{b}_i
                      \cdot\vec{\varphi}\right)ds_5^2 \nn\\
          & &    +\fr{1}{g^2}\left(1+\fr{1}{4}\sum_i\vec{b}_i\cdot
              \vec{\varphi}(\m^i)^2\right)\sum_i\left(1+\fr{1}{2}\vec{b}_i
                      \cdot\vec{\varphi} \right)(d\m^i)^2\;.
\label{metlinear} 
\eea
As previously mentioned, the ansatz quoted in Eq.~(\ref{iib-metric-ans-s5})
is for general $S^5$ reductions. However, we choose the
five-dimensional space-time to be $AdS_5$ for our discussion. The
structure of the D3-brane solution of type IIB supergravity in the
near-horizon region is such that the ``radius'' of the $AdS_5$ is
the same as that of the internal sphere. Therefore we choose
$ds_5^2$ to be an $AdS_5$ of radius $1/g$. Then we have
\be 
ds_5^2+\fr{1}{g^2}\sum_i (d\m^i)^2=g^2r^2 \sum_a(dx^a)^2
           +\fr{1}{g^2r^2}\left( dr^2+r^2\sum_i(d\m^i)^2\right)\;.
\ee
The metric ansatz (\ref{metlinear}) in its written form, i.e. in
the $\m$-coordinate system, does not make manifest the $SO(6)$
covariance of the conformal field theory. A more suitable
coordinate system is one that reveals the brane structure more
transparently; not surprisingly, such a coordinate system is of
Cartesian type,
\be \m^i=\fr{\F^i}{r} \hspace{.5in}{\rm and}\hspace{.5in}
r^2=\sum_{i=1}^6 (\F^i)^2\;. 
\label{nc} 
\ee
Using the $\F$-coordinate system, one can rewrite
(\ref{metlinear}) in the form
\be ds_{10}^2=g^2r^2f \, \sum_a(dx^a)^2 +
\fr{1}{g^2r^2}\,\sum_{i,j=1}^6\, g_{ij}\, d\F^id\F^j\;,
\label{metric} 
\ee
where
\bea f       &\equiv&
1-\fr{1}{4r^2}\sum_i\vec{b}_i\cdot\vec{\varphi}
                                   \,(\F^i)^2\;, \nn\\
g_{ij}  &\equiv&\fr{1}{g^2r^2}\left[
                \d_{ij}+\fr{1}{2}\vec{b}_i\cdot\vec{\varphi}\,\d_{ij}
                -\fr{1}{r^2}\vec{b}_i\cdot\vec{\varphi}\,\F^i\F^j
                +\fr{1}{4r^2}\sum_k\vec{b}_k\cdot\vec{\varphi}\,
                                   (\F^k)^2\,\d_{ij}
                 \right]\;.
\label{fg} 
\eea
The action for D3 branes in a general type IIB background was
studied in \cite{Cederwall:1997pv,Bergshoeff:1997tu,Aganagic:1997zk,Metsaev:1998hf}. For our discussion, it is
enough to keep the metric and the 4-form field  since the scalars
come only from these. The relevant part of the action is
\bea 
I = -\int d^4\xi\, \sqrt{-{\rm det}(G_{ab}+{F}_{ab})}
    +\int \;\hat{C}_\4\;,
         \label{dbi}
\eea
where
\bea G_{ab}        &=&\frac{\partial
Z^M}{\partial\xi^a}\frac{\partial
                  Z^N}{\partial\xi^b}\,E^{\;\;A}_{M}\,E^{\;\;B}_N\,\eta_{AB}\;, \nn \\
C_{abcd}      &=& C_{MNPQ} \, \pa_aZ^M\pa_bZ^N\pa_cZ^P\pa_dZ^Q\;.
\label{defs} \eea
It is the 4-form potential $C_{(4)}$ that appears in (\ref{dbi}),
whereas the ansatz in Eqs.~(\ref{iib-gauge-ans-s5}) and 
(\ref{iib-gdualans-s5}) was obtained in terms of the 5-form 
field strength, $\hat{H}$. It is possible to obtain $C_\4$ from 
Eqs.~(\ref{iib-gauge-ans-s5}) and (\ref{iib-gdualans-s5}). Since we
eventually consider only the leading order in the derivative
expansion, it turns out that the contribution from $C_\4$ can be neglected.
Therefore, we can concentrate solely on the contributions coming from the DBI
action.

To obtain $G_{ij}$, we substitute (\ref{metric}), (\ref{fg}) and
the super vielbeins obtained in \cite{Metsaev:1998it,Kallosh:1998nx}
into the first equation of (\ref{defs}). The result is 
\bea 
G_{ab}=g^2r^2f\,\eta_{ab} + g_{ij}\,\pa_a \Phi^i \pa_b \Phi^j
  + F_{ab}\;.
\eea
There are no fermionic terms in the above formula. The reason is that
fermionic fields do not have any contributions with conformal dimension 2,
which is a relevant dimension to our discussion.

After some algebra, the relevant part
of the action is given by
\bea 
I[\f]
  &=&-\int
    -\frac{g^4 r^2}{2}\sum_i\,\vec{b}_i\cdot\vec{\varphi}\,(\Phi^i)^2+
      \fr{1}{2}\sum_{ij}\left(\fr{1}{2}\vec{b}_i\cdot\vec{\varphi}\,\d_{ij}
          -\fr{1}{r^2}\vec{b}_i\cdot\vec{\varphi}\,\F^i \F^j\right)
             \partial_a \Phi^i\partial^a \Phi^j \nn\\
&& + \cdots\;,
\label{Odiag} 
\eea
where the ellipses refer to the terms with higher numbers of
derivatives.

The discussion of the full 20 scalars $T_{ij}$, without imposing
$A^{ij}_\1=0$, goes very similarly, since eventually we will be
interested only in terms that are coupled to $T_{ij}$, but do not
have any factors of $A^{ij}_\1$.   In fact the computation is
almost identical, except that one needs the new parameterization
\be T_{ij}\equiv (\rme^{S})_{ij}\;, \ee
where $S_{ij}$ is symmetric and traceless. Using this
parameterization, one gets
\bea I[S_{ij}]  && =-\int\;
    {g^4 r^2}\Phi^i \F^j S_{ij}+
       \fr{1}{2}\left(-\pa_a \F^i \pa^a \F^j
          +\fr{1}{r^2}\left[\F^i \F^k
             \partial_a \Phi^j\partial^a \Phi^k
                     +i\leftrightarrow j  \right] \right)S_{ij} \nn\\
          &&       \hspace{3in}       - \;(\mbox{trace})\;.
\label{O} 
\eea
To leading order in the derivative expansion, we have
\be \hspace{1.5in}I[S_{ij}] = -\int g^4r^2\left(
\F^i\F^j-\fr{1}{6}\d^{ij}\F^k\F_k
                      \right)S_{ij}+\hspace{.1in}\cdots\;.
\label{OforS} \ee
As one can easily show, $S_{ij}\sim \fr{1}{r^2}$ in the boundary
region, i.e. $r\rightarrow \infty$. Therefore we impose the
following boundary condition,
\be 
S_{ij} \equiv \fr{1}{r^2}S^o_{ij}\;, 
\ee
which leads to the correct CFT operator.
\be 
{\cal O}^{ij}\equiv \left(
\F^i\F^j-\fr{1}{6}\d^{ij}\F^k\F_k
                      \right)+\hspace{.1in}\cdots\;.
\label{final} 
\ee
The result is given in Eq.~(\ref{final}) is in agreement with the CFT
operators obtained based on the conformal symmetry
argument~\cite{Witten:1998qj}.


\section{Supergravity duals of $\cN=2$ SCFTs in $D=3$ and $D=5$}


We start this section by reviewing the general idea
of supergravity description of field theories on a curved
manifold presented in~\cite{Maldacena:2000mw}. Generally, if we put a
supersymmetric field theory on a curved manifold ${\cal M}$, we will
break supersymmetry because we will not have a covariantly constant
spinor which obeys $(\pa_\m + \omega_\m)\,\ep =0$. If the
supersymmetric field theory has a global R-symmetry, we then can
couple the theory to an external gauge field that coupled to
R-symmetry current and the spinor obeys the equation 
$(\pa_\m + \omega_\m - A_\m)\,\ep =0$. Obviously, if
we choose the external gauge field to be equal to the spin connection
$\omega_\m = A_\m$, then we can find a covariantly constant spinor. 
The resulting theory is the 
so-called ``twisted'' theory, since we can view the coupling to the external 
gauge field as effectively changing the spins of all fields. The
supersymmetry parameter becomes a scalar. This is precisely the way that
branes wrapping on non-trivial cycles in M-theory or string 
compactifications manage to preserve some
supersymmetries~\cite{Bershadsky:1996qy}. A supersymmetric cycle 
is defined by the condition that a worldvolume theory is
supersymmetric~\cite{Hughes:1986fa,Becker:1995kb}. The condition to
have a supersymmetric (p+1)-cycle is 
\be
\left(1-\frac{\im}{(p+1)!}\,h^{-1/2}\,\vep^{\a_1\cdots\a_{p+1}}\pa_{\a_1}X^{m_1}\cdots
\pa_{\a_{p+1}}X^{m_{p+1}}\Gamma_{m_1\cdots m_{p+1}}\right)\Psi = 0\;,
\label{super-cycle}
\ee
where $\Psi$ is a covariantly constant ten-dimensional spinor,
$h_{\a\b}$ is an induced metric on p-brane and $\Gamma_{m_1\cdots
m_{p+1}}$ are ten-dimensional $\Gamma$-matrices. The cycles satisfying
Eq.~(\ref{super-cycle}) minimize the p-brane
worldvolume~\cite{Becker:1995kb}. In the case of branes wrapping on
cycles, ${\cal M}$ is the worldvolume geometry of the cycle and the
external gauge field takes into account the fact that the normal to the
cycle forms a non-trivial bundle, the normal bundle, and $A_\m$ is the
connection on this normal bundle. The condition that the cycle
preserves some supersymmetries then boils down to the condition that the
spin connection is equal to the gauge connection, which is exactly the
equation mentioned earlier. So if we are
interested in understanding field theories arising on the branes
wrapping on some non-trivial cycles, we will then have to study these
twisted field theories.

In this section we construct various supergravity solutions, some of
which are duals to $\cN=2$ SCFTs in three and five dimensions. These
cases provide new examples of AdS/CFT duality involving twisted field
theories.

\subsection{Supergravity duals of three-dimensional $\cN=2$ SCFT}

In this subsection, we focus on the D4-D8 system, which was studied
from the gauge theory side in~\cite{Intriligator:1997pq} and from the
gravity/CFT viewpoint in~\cite{Ferrara:1998gv,Brandhuber:1999np}. We
will wrap D4-brane of the D4-D8 system on 2- and
3-cycles. These systems are the
gravity duals of superconformal field theories in five dimensions 
with 8 supercharges~\cite{Intriligator:1997pq}, 
which in the infrared (IR) flow to superconformal
field theories in three dimensions with 4 supercharges, which is
the ${\cal {N}}=2$ theory in three 
dimensions~\cite{Aharony:1997bx,deBoer:1997mp,deBoer:1997ck,
deBoer:1997kr,deBoer:1997ka}.

We start with a brief review of D4-D8 system~\cite{Brandhuber:1999np}.
Consider type I theory on $R^9 \times S^1$ with $N$ coinciding
D5-branes wrapping on the circle. The six-dimensional field theory on the
worldvolume of D5-branes is $\cN=1$ with $Sp(N)$ gauge
group~\cite{Witten:1996gx}. This theory has one hypermultiplet in the
antisymmetric representation of $Sp(N)$ from Dirichlet-Dirichlet
sector and 16 hypermultiplets in the fundamental representation from
the Neuman-Dirichlet sector. T-dualizing D5-branes on the circle
results in type I$^\prime$ theory compactified on the interval
$S^1/\Z_2$ with two orientifolds (O8-planes) located at fixed
points~\cite{Polchinski:1996df}. The D5-branes become D4-branes and
there are 16 D8-branes located at points on the interval. The
existence of 16 D8-branes is necessary in order for R-R charge to be
conserved and furthermore the theory is 
anomaly-free~\cite{Polchinski:1996fm,Polchinski:1996na}. 
The relative positions of the
D8-branes with respect to the D4-brane correspond to masses for 
the hypermultiplet in the fundamental representation. The
hypermultiplet in the antisymmetric representation is massless. 

The field theory on the worldvolume of D4-brane is a five-dimensional 
supersymmetric field theory~\cite{Seiberg:1996bd,Morrison:1997xf}.
The Lorentz group of D4-brane worldvolume is $SO(4,1)$. 
The spinor representation of $SO(1,4)$ is pseudoreal and is
four-dimensional. This theory in five dimensions is related by 
dimensional reduction to the 
$\cN=1$ theory in six dimensions described above. All these theories 
have an $SU(2)_R$ symmetry of automorphism algebra which can be
associated with the gauged symmetry in the gravitational set-up of
\cite{Romans:1986tw}. The vector multiplet in five dimensions consists of
a real scalar component, a vector field and a spinor,
while the hypermultiplet comprises four real scalars and a spinor. This
theory has a Coulomb branch when the real scalar has a vacuum
expectation value (VEV),
and a Higgs branch when the scalar in the hypermultiplet is
excited. The theory on the D4-brane has $Sp(N)$ gauge group with one
vector multiplet (whose scalar component describes the Coulomb
branch $\R^+$) and hypermultiplets whose first components describe
the Higgs branch. The global symmetries of the theory are $SU(2)_R
\times SU(2)\times SO(2N_f)\times U(1)$. The $SU(2)_R$ factor is 
the R-symmetry (the supercharges and the scalars in the
hypermultiplets are doublets under this group). The other $SU(2)$ is
associated with the hypermultiplet in the antisymmetric
representation which is only present if there are more than one
D4-brane (e.g. $N>1$). The $SO(2N_f)$ is the symmetry group of the
$N_f$ massless hypermultiplets in the fundamental representation, and
the $U(1)$ is associated with the instanton number. When the
D4-brane is in the origin of the Coulomb branch we have a fixed
point and the global symmetry is enhanced to $ SU(2)\times E_{N_f
+1}$. The gravitational system is given by $N_f$ D8-branes
situated on a O8-plane, $(16 - N_f)$ D8-branes in another fixed
plane, and a D4-brane which can move between them. The position of
the D4-brane is parameterized by the scalar in the vector
multiplet. If $<\phi>$ is not zero, we have a theory with a $U(1)$
symmetry with $N_f$ ``electrons'', on the fixed planes; the theory
recovers its $SU(2)$ R-symmetry, and it will have $N_f$ quarks.

The gravity theory is a fibration of $AdS_6$ over $S^4$, with
isometries $SO(2,5)\times SU(2)\times SU(2)$. The six-dimensional
$SU(2)$-gauged supergravity has  an $AdS_6$  vacuum solution. It
can be up-lifted to massive IIA theory~\cite{Cvetic:1999un}
leading to the configuration described above~\cite{Nieder:2000kc}.
Other solutions in massive type IIA supergravity were obtained in~\cite{Janssen:1999sa,Massar:1999sb}.
Here we will consider solutions of Romans' theory which,
when uplifted, will give ten-dimensional configurations with $N_f
=0$. These theories were studied in depth in~\cite{Cachazo:2000ey} 
in the context of algebraic geometry. There,
the authors studied the moduli spaces of the type I$^\prime$ string and
the heterotic string. It would be interesting to understand the case of
$N_f \neq 0$ and see whether there is a gravity solution dual to
the theory with enhanced symmetry $E_1^\prime$. It would be also useful
to clarify the role of D0-branes in our gravity solutions. In
principle they should correspond to excitations of the
six-dimensional Abelian gauge field. 

In order to analyze the flow of the theory from five dimensions to
a superconformal theory in three dimensions with 4 supercharges, 
we consider a configuration where the
geometry ``loses'' two dimensions at low energies. The
idea is to start with a six-dimensional gauged
supergravity theory which has an $AdS_6$ vacuum with 8 preserved
supercharges~\cite{Nunez:2001pt}. This vacuum solution is dual to the 
five-dimensional
SCFT mentioned above and its ten-dimensional interpretation is
given by a D4-D8 system. When one moves to the IR of the gauge
theory (flowing in the radial coordinate on the gravity dual) two
of the dimensions of the theory become very small and no low-energy
massless modes are excited on this two-dimensional space. Therefore,
the gauge theory is effectively three-dimensional. Further, since
the D4-brane is wrapped on a curved surface, we need to twist the
theory. The effect of this twisting is the breaking of some
supersymmetries, as can be seen from the projections of the
Killing vectors.

Under the group $SO(1,4) \times SO(3)_R$ the fields in the vector
multiplet on the D4-brane worldvolume transform as 
$(\bf{1},\bf{1})$ for the scalar field, $(\bf{3},\bf{1})$ for the
gauge field and
$(\bf{4},\bf{2})$ for the fermion. In the IR, the metric is taken to 
have $SO(1,2)\times SO(2)_D$ symmetry. The twisting involves ``mixing'' the
$SO(2)$ group of the metric with an $SO(2)$ subgroup of $SO(3)_R$.
This mixing is responsible for the breaking of some
supersymmetries because the only spinors that can be defined on
the curved manifold are those that have scalar properties on the
curved part. In fact, it breaks $1/2$ of the supersymmetries of
the five-dimensional SCFT.

Below we construct a gravity dual to the five-dimensional SCFT
theory by finding a solution of Romans' theory with excited
$F^I_{\m\n}$ fields. Then, we find a fixed-point solution for the
massive field configuration. We postpone the study of the massless
case to the next subsection.

Let us consider the metric ans\"{a}tze of the form
\beq 
ds^2 = \rme^{2\,f} \, (dt^2 - dr^2 - dz^2 - dv^2 ) -
\frac{\rme^{2\,h}}{y^2} \, (dx^2 + dy^2) \;, 
\eeq
for $AdS_4 \times H_2$  and
\beq 
ds^2 = \rme^{2\,f} \, (dt^2 - dr^2 - dz^2 - dv^2 ) -
\rme^{2\,h} \, (d\theta^2 + \sin^2 \theta\, d\varphi^2) \;,
\eeq
for $AdS_4 \times S^2$. The only non-vanishing non-Abelian gauge 
potential components are taken to be
\bea
A^{(3)}_x &=& \frac{a}{y}\;\;\;\;\;{\rm for}\;\;AdS_4 \times H_2\;\;{\rm case}\;,\\
A^{(3)}_\varphi &=& - a\cos\theta\;\;\;\;\;{\rm for}\;\;AdS_4 \times
S^2\;\;{\rm case}\;. 
\eea
One can treat
both cases together by introducing a parameter $\lambda$ that can
take two values 
\be
\lambda=+1 \to H_2\;, \;\;\;\;\;\;
\lambda=-1 \to S^2 \;.
\ee 
We impose the following projections
\be
\Gamma_{45}(T^{(3)})_i^{\;\;j}\,\epsilon_j =
\frac{\lambda}{2}\,\epsilon_i\;, \;\;\;\;\;\;
\Gamma_2\Gamma_7\,\epsilon_i= \epsilon_i \;. 
\ee
A solution must
satisfy the equations obtained by setting to zero the
supersymmetry transformations for gauginos and gravitinos. After
some algebra, they lead to 
\bea 
\varphi^\prime &=&
\frac{1}{4\sqrt{2}}\,\rme^f\,[g\, \rme^\varphi - 3\,m\,
\rme^{-3\varphi} + 2\, a\,\lambda \rme^{-2 h -\varphi}] \;,
\label{martinn1}\\
h^\prime &=&
\frac{1}{4\sqrt{2}}\,\rme^f\,[-g\, \rme^\varphi - m\, \rme^{-3\varphi}
+ 6\,a\,\lambda\,\rme^{-2 h -\varphi}] \;, 
\label{martinn2}\\
f^\prime &=& -\frac{1}{4\sqrt{2}}\,\rme^f\,[g\,
\rme^\varphi + m\, \rme^{-3\varphi} + 2\, a\, \lambda\,\rme^{-2 h
-\varphi}] \;, 
\label{martinn3} 
\eea
where $\varphi \equiv \phi/\sqrt{2}$.

A fixed-point solution of these equations is given by 
\bea
\rme^{4\varphi} & = & \frac{2 m}{g} \;, \nn \\
\rme^{-2h} & = & \sqrt{\frac{g\,m}{8}} \frac{1}{a\lambda} \;, \nn \\
\rme^{- f(r)} & = & \frac{g\,\rme^\varphi}{2\sqrt{2}}\,r \;.
\label{solfixed} 
\ea 
In fact, it only happens if $\lambda \, a >
0$. This solution satisfies the second order equations derived
from the six-dimensional Lagrangian.

We can analyze the structure of the solutions of the system above
near the UV of the theory ($r=0$). Indeed, we can see that an
expansion leads to the following behavior for the different fields
\be
f\approx - \log(r) + c_1 r^2 +\cdots, \;\;\;\; g\approx -
\log(r) + c_2 r^2 +\cdots,\;\;\;\; \phi\approx  c_3 r^2 +\cdots\;, 
\ee
where $c_i$s are constants. The interpretation of these
expansions is that the scalar field
$\phi$ describes the coupling of the five-dimensional field theory
to an operator of conformal weight $\Delta=3$ and mass squared
$m^2=-6$~\cite{Balasubramanian:1998sn}. 
The operator is not the highest component of its
supermultiplet, thus it is a deformation that breaks some
supersymmetry.

The ten-dimensional metric corresponding to the fixed-point
solution described above is easily obtained by using the results
of \cite{Cvetic:1999un}, 
\bea 
ds^2_{10}&=&(3\,m\,g^2)^{-1/8}\,(\sin\chi)^{1/12}
X^{1/8}\left\{\frac{\Delta^{3/8}\,\rme^{2f_0}}{2\,r^2} (-dt^2 +
dr^2 +
dv^2 + dz^2) \right.\nonumber\\
& &\left. + \frac{\sqrt{2}\Delta^{3/8}}{\sqrt{g^3\,m}\,y^2}(dx^2 +
dy^2 )
+ \Delta^{3/8}\,\sqrt{2\,m\,g^3}\,d\chi^2\right.\nonumber\\
& &\left. +\frac{\Delta^{-5/8}}{g\,X}\,\cos^2\chi
\left[\left(\sigma_1 -\frac{dx}{y}\right)^2 +\sigma_2^2 + \sigma_3
^2\right]\right\} \;. 
\eea
The metric has symmetries of the
form $SO(2) \times SO(3)$. The corresponding expression for the R-R
and NS-NS fields and the ten-dimensional dilaton can be read off from
(\ref{massive-iia-forms-ans-s4}). 

\begin{figure}
\centering
\includegraphics[width=9cm]{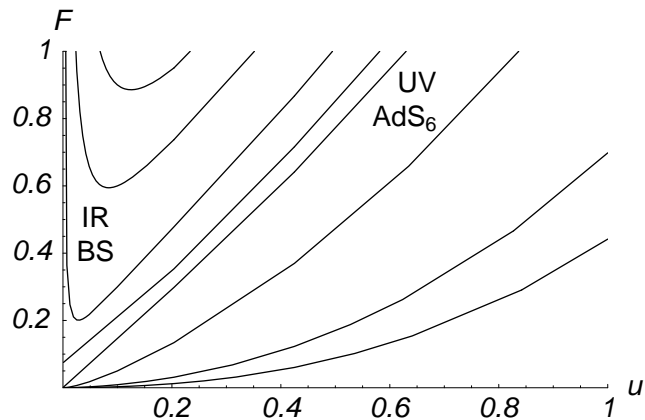}
\caption{We show the
behavior of the orbits for the case $S^2$. Similar comments as in
the previous case are also valid here. For small $u$-values we can
see the region corresponding to $F \approx \frac{1}{u}$.}
\label{b0-sphere}
\end{figure}

Although we could not find an exact solution, we can carry out an 
analysis similar to \cite{Acharya:2000mu}. Actually, the two cases
$\lambda=\pm 1$ can be gathered again. Thus defining $F\equiv u^2
\rme^{-2 \varphi}\;,\;u\equiv \rme^{2 h(r)}$ and using
Eqs.~(\ref{martinn1})-(\ref{martinn2}) we obtain the following
differential equation 
\beq 
\frac{d F}{d u} = \left(\frac{ 3 g u^4
- m F^2 - 10 a \lambda u F}{ g u^4 + m F^2- 6 a \lambda u F}
\right) \frac{F}{u} \;. 
\label{ORBITS} 
\eeq 
For the case
$\lambda=+1$ (hyperbolic plane), we can solve this equation in the
approximations when $u\to \infty$ and $u\to 0$. In fact, in the UV
one has
\be
F\approx  u^2 + \cdots\;, \,\,\,\,\;\;
\rme^{-2\varphi}\approx 1+\cdots \;,   
\ee 
and the metric results to
be the $AdS_6$ limit of our original metric. In the IR we have
\be
F\approx F_0 u^{5/3} + \cdots\;, \;\;\,\,\,\,
e^{-2\varphi}\approx F_0 u^{-1/3}+\cdots\;, 
\ee
and the metric is
\be
ds_6^2= \frac{1}{u^{1/3}}(dt^2 - dz^2 - dv^2) -
\frac{u}{y^2}(dx^2 + dy^2) - \frac{2}{9 \, F_0 \, a \,
u^{1/3}}du^2 \;.   
\ee
\begin{figure}
\centering
\includegraphics[width=9cm]{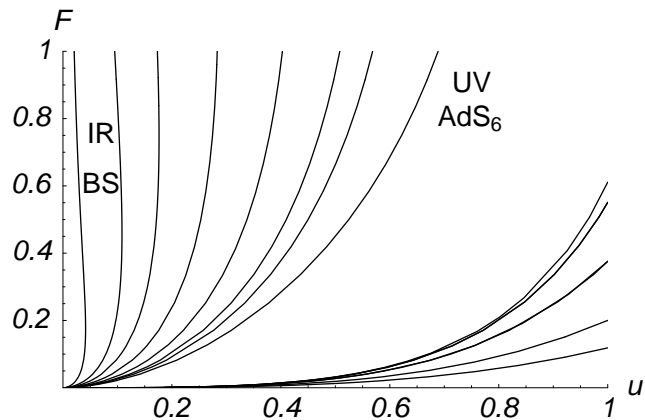}
\caption{Orbits for the
case $AdS_4 \times H_2$. In the UV limit $F$ behaves as $u^2$
leading to an $AdS_6$-type region. On the other hand, it flows to
``bad'' singularities (BS) in the IR. We have used $g = 3 m$ with
$m=\sqrt{2}$, and $a=1$.}
\label{b0-hyper}
\end{figure}
When up-lifted, the singularities turn out to be of the
``bad''type. They correspond to the curves flowing to the origin
of the plots in Fig.~\ref{b0-hyper}. On the other hand, in both cases of the
hyperbolic plane and sphere there are other kinds of ``bad''
singularities that correspond to the orbits going like $F \approx
\frac{1}{u}$ as $u$ approaches to zero. Particularly, one can see
this kind of behavior in Fig.~\ref{b0-sphere}. For the two-sphere we have the
set of orbits depicted in Fig.~\ref{b0-sphere}.

This also constitutes a part of the IR behavior of the flow. As we
can see in Fig.~\ref{b0-hyper} and Fig.~\ref{b0-sphere}, there is a
smooth solution 
interpolating between the UV and the IR limits of the system, thus
realizing our initial set up. The fixed-point solution is out of
the range of this plot.

In summary, we have found a fixed-point solution for the
hyperbolic plane. In the UV, both geometries flow to their
corresponding asymptotic $AdS_6$. In the IR limit all the
singularities in both geometries are of the ``bad'' type according to the
criterion of~\cite{Maldacena:2000mw}. Similar behavior is
expected for the sphere. Note that the solutions studied here only
represent a special case of solutions that, in general, have the
form of a Laurent series. The fact that these solutions have
non-acceptable singularities can be easily seen from the
construction of an effective potential corresponding to this
effective four-dimensional system, and following the analysis
outlined by Gubser in \cite{Gubser:2000nd} one can see that
these kinds of solutions will not be interpretable as the IR of a
gauge theory. However, this does not exclude the possibility of
finding a different solution that could be intepreted as
the transcendental solution found in \cite{Acharya:2000mu}.

\subsection{A non-Abelian solution in massive type IIA theory}

In this subsection we present a non-Abelian solution in the massive
IIA theory. We begin with a gravity dual of a
five-dimensional SCFT with 8 supercharges that flows through the
dimensions to a two-dimensional (1,1) CFT with 2 supercharges at
the IR. The field content consists of a real scalar, a gauge field
and a corresponding fermionic partner. The idea is to find a
compactification from six-dimensional supergravity on $AdS_3
\times H_3$. Under $SO(1,4)\times SO(3)_R$ the charges transform
as $(\bf{4},\bf{2})$ and $(\bf{\bar{4}},\bf{2})$. After the
twisting $ SO(1,4)\times SO(3)_R \to SO(1,1)\times SO(3) \times
SO(3)_R \to SO(1,1) \times SO(3)_D$ we are left with 2
supercharges which are scalars under the diagonal subgroup.
$SO(3)_D$ results from the identification  of the normal bundle of
the D4-brane worldvolume with the spin bundle of $\Sigma_3$. They
are the two ``twisted'' supercharges that survive the twisting
process. This solution is similar to
the solution obtained in \cite{Acharya:2000mu} for massive
seven-dimensional gauged supergravity. In that case the solution
up-lifts to M-theory. This suggests some connection
between theories in the line of~\cite{Lavrinenko:1999xi} that
merits further exploration.

Our configuration is 
\be
ds^2 = \rme^{2f}(dt^2 - dr^2 - du^2 )-
\frac{\rme^{2 h}}{y^2}(dx^2 + dz^2 + dy^2)\;,\;\; A_x^{(1)}=
\frac{a}{y}\;,~ A_z^{(3)}= \frac{b}{y}\;. 
\ee
Similarly, an $AdS_3 \times S^3$ for the case $\lambda = -1$ can also
be considered. The projections (all the indices are flat indices) are given by 
\be
\Gamma_{65}
(T^{(3)})^{\;\;j}_i\,\epsilon_j = \frac{1}{2}\,\epsilon_i\;,\;\; \Gamma_{64}
(T^{(2)})^{\;\;j}_i\,\epsilon_j = \frac{1}{2}\,\epsilon_i\;,\;\; \Gamma_{45}
(T^{(1)})^{\;\;j}_i\,\epsilon_j = \frac{1}{2}\,\epsilon_i\;,
\ee
and
\be
\Gamma_7\Gamma_2\,\epsilon_i = \epsilon_i \;, 
\ee
we use $\xi=
\rme^{\phi/\sqrt{2}}$. The BPS equations are 
\bea
h^\prime &=& - \frac{1}{4
\sqrt{2}}\,\rme^f\,[ g\,\xi + m\,\xi^{-3} - 10\,\lambda\,a\,\xi^{-1}\,\rme^{-2
h}] \;,\\ 
\phi^\prime &=& \frac{1}{4\sqrt{2}}\,\rme^f\, [ g\,
\xi  - 3\,m\,\xi^{-3} + 6\,\lambda\,a\,\xi^{-1}\,\rme^{-2 h}]
\;,\\ 
f^\prime &=& - \frac{1}{4 \sqrt{2}}\,\rme^f\,[g\,\xi +
m\,\xi^{-3} + 6\,\lambda\,a\,\xi^{-1}\,\rme^{-2 h}] \;, 
\eea
and $a\,b = 1/g$. A fixed-point solution is 
\be
\rme^{-2 h}= \frac{\sqrt{m\,g^3}}{2\sqrt{6}}\;, \,\,\,\;\;
\xi = \left(\frac{3m}{2g}\right)^{1/4} \;.
\ee
This configuration can be uplifted to massive type IIA in ten
dimensions using the results in \cite{Cvetic:1999un} 
\bea 
ds^2_{10}&=&\left(\frac{1}{3\,m\,g^2}\right)^{1/8}\,(\sin\chi)^{1/12}
X^{1/8}\left\{\frac{\Delta^{3/8}}{2\,r^2} (-dt^2 + dr^2 +
dv^2)\right.\nonumber\\
& & \left. + \frac{\rme^{2h}}{2\,y^2}\,\Delta^{3/8}\,(dx^2 +dz^2+
dy^2) + g^2\,\xi\,\Delta^{3/8}\,d\chi^2 \right.\nonumber\\
& & \left. + \frac{\Delta^{-5/8}}{g\,X}\,\cos^2\chi \left[
\left(\sigma_1 - \frac{dx}{y}\right)^2 +\sigma_2^2 +
\left(\sigma_3 -\frac{dz}{y}\right)^2\right]\right\} \;, 
\eea
where $\sigma_i$ are the three left-invariant forms in the
3-sphere whose explicit forms are given in 
Eq.~(\ref{cartan-maurice-euler}). We can see that the metric 
has an $SO(2) \times SO(3)$ invariance.

\begin{figure}
\centering
\includegraphics[width=9cm]{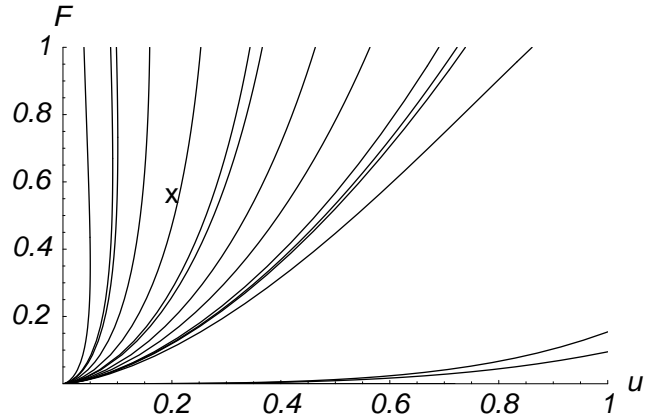}
\caption{Behavior of
the orbits for the non-Abelian massive case. The $AdS_6$-type
region is at the UV, i.e. for large $u$ and $F$-values. In
the IR it flows to ``bad'' singularities (BS). The cross indicates
the fixed point-solution aforementioned. We again set $g = 3 m$,
$m= \sqrt{2}$, and $a=b=\lambda=1$.}
\label{fig-massive}
\end{figure}

Since we are not able to integrate the above BPS system, we will
repeat the analysis of the singularities as we did in the previous
section. Again, both cases (hyperbolic plane and sphere) are
treated together by using the parameter $\lambda = \pm 1$.
Proceeding as above we get the following first order differential
equation 
\be
\frac{d F}{d u} = \left( \frac{3 g u^4 - m F^2
-14 a \lambda u F}{g u^4 + m F^2 - 10 a \lambda u F} \right)
\frac{F}{u}\;. 
\ee
 When $\lambda = +1$ we solve this equation
in two approximations, $u \to \infty$ and $u \to 0$. The large-$u$
approximation leads to the corresponding asymptotic $AdS_6$
whereas the other case gives
\be
F\approx  F_0 \frac{1}{u} + \cdots\;, \,\,\,\,\;\;
\rme^{-2\varphi}\approx \frac{F_0}{u^3}+\cdots\;,\;\; 
\ee
\be
F\approx - 2 \frac{a}{m}u + \cdots\;, \;\;\,\,\,\,
e^{-2\varphi}\approx -\frac{2 a}{m u}+\cdots 
\ee 
\be
F\approx F_0
u^{7/5} + \cdots\;, \;\;\,\,\,\, e^{-2\varphi}\approx F_0 u^{3/5}+\cdots
\;. 
\ee
 Note that all the singularities are of the ``bad'' type
in the IR. They correspond to the curves flowing to the origin of the plots in Fig.~\ref{fig-massive}.
Here we should make a comment similar to that in the previous
section: Gubser's criteria tells us to construct the effective
potential of the three-dimensional gravity theory. With this
potential we can see that the solutions in the form of a Laurent
series shown above will not have an interpretation as the IR
limit  of a gauge theory.

We can study a problem similar to the one in the previous section.
One can write down a similar ansatz for the metric, i.e. a
wrapped product of $AdS_4 \times \Sigma_2$, where $\Sigma_2$ is a
hyperbolic plane or sphere, as before. The system of
Eqs.~(\ref{martinn1})-(\ref{martinn3}) can be studied in the
massless case, i.e. $m=0$, while $g$ and $a$ are
non-vanishing quantities. We have the solution 
\be
f =-
\varphi\;,  \,\,\,\,\,\,\, \varphi = \frac{g}{4\sqrt{2}}\,r +
\frac{1}{8}\log(r)\;,\;\;\,\,\,\,\, h = -\frac{g}{4\sqrt{2}}\,r +
\frac{3}{8}\,\log(r) \;. 
\ee
Therefore, the six-dimensional metric is given by
\be
ds^2 = \frac{1}{r^{1/4}}\rme^{-\frac{g\,r}{2\sqrt{2}}}\,
[dt^2 - dr^2 - dz^2 - dv^2 -  r\,d\Omega^2_{\lambda}]\;.
\label{badmetric}
\ee
This metric has a singularity. One way to resolve it is to look for
a solution that has more degrees of freedom excited such as a
non-Abelian solution. Another way to have an acceptable gravity
solution is to construct a black hole solution such that the
singularity is hidden by the horizon~
\cite{Buchel:2001gw,Gubser:2001ri,Buchel:2001qi}. In the next
section we will discuss a resolution of the singularity in the
$S^2$ case, i.e. $\lambda =-1$ in Eqs.~(\ref{martinn1})-(\ref{martinn3}).

\subsection{A non-Abelian solution}

We would like to present a non-Abelian solution that resolves the
``bad'' singularity of the solution in Eq.~(\ref{badmetric}). This
solution is useful for the study of the gravity dual for the
three-dimensional ${\cal{N}}=2$ super Yang-Mills theory. This
solution, which relates to the solution in~\cite{Maldacena:2000yy}, 
represents a smeared NS5-brane on $S^2$
after a T-duality. In the IR, the theory living on the brane will
be an ${\cal{N}}=2$ super Yang-Mills theory in three dimensions plus
Kaluza-Klein  modes that do not decouple. The solution in the
string frame is
\bea 
ds^2 &=& 2\, [dt^2 - dr^2 - d\vec{x}_2^2 -
\rme^{2\,h}(d\theta^2 +
\sin\theta^2 d\varf^2) ]\;,\nn\\
F &=& -w^\prime \s^2 dr\wedge d\theta + w^\prime \sin\theta \s^1
dr\wedge d\varphi + (w^2 -
1)\,\s^3\,\sin\theta d\theta \wedge d\varphi\;,\nn\\
\phi_S &=& \frac{1}{2} \log\left[\frac{\sinh(r)}{R(r)}\right]\;,
\;\;\; w(r)=\frac{r}{\sinh(r)}, w^\prime = \frac{dw}{dr}\;,\nn\\
\rme^{2\,h} &=& R(r)^2, R(r)=\sqrt{2r \coth(r) - \left[\frac{r}{\sinh(r)}\right]^2 -
1} \;. 
\eea
This is the same solution obtained by Chamseddine and Volkov in
\cite{Chamseddine:1997nm,Chamseddine:1998mc}.

The relations between the scalars and metric in the Einstein and
string frames are
\[\varphi = -\fr12 \phi_S\;,\;\;\;\;\; g_{\mu\nu}(E) =
{\rm e}^{\phi_S}g_{\mu\nu}(S)\;.\]
In Einstein frame the solution is
\begin{equation}
ds^2_E = 2 {\rm e}^{\phi_S}(-dt^2 + dr^2 + d\vec{x}_2^2) + {\rm
e}^{2\,G(r)}(d\theta^2 + \sin^2\theta d\varphi^2)\;,
\end{equation}
where the dilaton is given by
\begin{eqnarray}
\phi_S &=&
\frac{1}{2}\log\left[\frac{\sinh(r)}{R(r)}\right]\;,\nonumber\\
R(r) &=& \sqrt{2 \, r \, \coth(r) - w(r)^2 -1}\;,\;\;\; {\rm
e}^{2\,G} = R(r)^2\;,\;\;\; w(r) = \frac{r}{\sinh(r)} \;.
\end{eqnarray}
The field strength components are
\begin{eqnarray}
F^1_{(2)} &=& w^{\prime} \sin\theta dr\wedge d\varphi\;,\nonumber\\
F^2_{(2)} &=& -w^{\prime} dr\wedge d\theta\;,\nonumber\\
F^3_{(2)} &=& (w^2-1)\sin\theta d\theta\wedge d\varphi \;.
\end{eqnarray}
Collecting all, we have the following ten-dimensional solution:
\begin{eqnarray}
d\hat{s}_{10}^2 &=& {\rm e}^{-\frac{3\sqrt{2}}{8}\phi}\left[-dt^2
+ dr^2 + d\vec{x}_2^2 + {\rm e}^{2G}(d\theta^2 + \sin^2\theta
d\varphi^2) + \frac{1}{g^2}\,
\sum_{i=1}^3\,(\sigma^i - \frac{g}{\sqrt{2}}\,A^i_{(1)})^2\right]\nonumber\\
& & + {\rm e}^{\frac{5\sqrt{2}}{8}\phi} dZ^2\;,\nonumber\\
\hat{F}_{(4)} &=& (-\frac{1}{g^2}h^1\wedge h^2\wedge h^3 +
\frac{1}{\sqrt{2}\,g}\,F^i_{(2)}\wedge h^i)\wedge dZ\;,\nonumber\\
\hat{\phi} &=& \frac{1}{\sqrt{2}}\phi = -
\frac{1}{4}\log\left[\frac{\sinh(r)}{R(r)}\right]\;,
\end{eqnarray}
where $h^i = \sigma^i - \fr{g}{\sqrt{2}}\,A^i_\2$ and $\sigma^i$ are
left-invariant 1-forms in the $S^3$ as given in
Eqs.~(\ref{cartan-maurice})-(\ref{cartan-maurice-euler}). 

\subsection{A black hole solution in the massless theory}

The solutions of the massless six-dimensional Romans' theory, the
string frame metric, and 2-form ans\"atze are
\bea ds^2 &=& \frac{f(r)}{r}dt^2 - \frac{1}{r\,f(r)}\,dr^2 -
R^2\,(d\theta_1^2 +
\sin^2\theta_1\,d\varphi_1^2) - dx^2 - dy^2,\nn\\
F_2 &=& h(r)\,dr\wedge dt + \frac{R\,Q_m}{\sqrt{2}}\sin\theta_1\,
d\theta_1\wedge d\varphi_1\;, 
\eea
where $f$ and $h$ are functions of $r$ while $\g$ and $Q_m$ are
constants.

The solution in the string frame is
\be 
h(r) = \frac{Q_e}{r^2}\;,\;\; f(r) = - M + \frac{2\,Q_e^2}{r} +
\left(\frac{g^2}{2} + \frac{Q_m^2}{R^2}\right)r\;,\;\;\f(r) =
\frac{1}{2}\,\log(r)\;. 
\ee
Similar solutions were discussed previously 
in~\cite{Cvetic:1999pu}. Our solution has a different geometry
and carries both electric and magnetic charges.

One can up-lift the above solution to type IIA theory using the
ans\"atze~(\ref{6d-dual-iia-s1xs3}), and the
results are
\bea
d\hat{s}_{10}^2 &=& -\frac{1}{2}\,r^{-1/8}\,
\left[ \frac{f(r)}{\sqrt{r}}\,dt^2 -
\frac{1}{\sqrt{r}\,f(r)}\,dr^2 - R^2\,\sqrt{r}\,(d\theta_1^2 +
\sin^2\theta_1\,d\varphi^2_1)\right]\nn\\
& & +\fr12\,r^{3/8}\,(dx^2 + dy^2) + r^{-5/8}\,dZ^2\nn\\
& & + \frac{r^{3/8}}{g^2}\,\left[(\s^1)^2 + (\s^2)^2 + \left(\s^3
+ \frac{g\,Q_e}{\sqrt{2}\,r}\,dt +
\frac{g\,R\,Q_m}{2}\,\cos\theta_1\,d\varphi_1\right)^2\right]\;,\nn\\
\hat{\f} &=& -\frac{1}{4}\,\log(r)\;,\nn\\
\hat{F}_4 &=& -\fr1{g^2}\,\s^1\wedge \s^2\wedge h^3\wedge dZ +
\frac{1}{\sqrt{2}\,g}\,F_2\wedge h^3\wedge dZ\;,\nn\\
h^3 &=& \s^3 + \frac{g\,Q_e}{\sqrt{2}\,r}\,dt + \frac{g
R\,Q_m}{2}\,\cos\theta_1\,d\varphi_1 \;. 
\eea 
For different
values of the constants $Q_e, Q_m, R, g$, we will have either a
horizon or a naked singularity. It should be instructive to
compute the entropy  and the Hawking temperature of this black
hole and comparing them with those of the M-theory black holes calculated in
\cite{Klemm:1998in}. Since the system analyzed in that reference
is the same in string variables as the one we analyze here, it is
expected that those results will be repeated. Indeed, the
ten-dimensional interpretation suggested in \cite{Klemm:1998in}
agrees with the one we have described above.

\subsection{A solution with excited B-fields}

Here we consider the six-dimensional Romans' theory with the mass
parameter and all fields except the scalar and the 3-form
field set to zero.  In addition, we take 3-form $G_3$ to be a constant.
The geometry of the $AdS_3 \times R^3$ space-time is given by
\beq 
ds^2= \rme^{2\,f}\,(dt^2 - dr^2 - dz^2) - \rme^{2\,h}\,(dx^2
+ dy^2 + dv^2) \;. 
\eeq
Let us consider a spinor satisfying the following constraints
\beq
\Gamma_2 \Gamma_7\,\epsilon_i= \epsilon_i\;,\;\;\;\; \Gamma_{4 5
6}\,\epsilon_i=\epsilon_i \;, 
\eeq
and 
\beq 
G_{xyv}= G = {\rm constant}\;. 
\eeq 
From the vanishing of the
supersymmetric variation of gravitinos and gauginos we obtain two
independent equations
\bea 
h^\prime &=& -  \rme^{f}\, \left( \frac{g}{4 \sqrt{2}}\,
\rme^\varphi +  \frac{G}{2}\,  \rme^{-3 h + 2\varphi} \right)\;,\\
f^\prime &=& -  \rme^{f}\, \left( \frac{g}{4 \sqrt{2}}\, 
\rme^\varphi -  \frac{G}{2} \,  \rme^{-3 h + 2\varphi} \right)\;,\\
\varphi^\prime &=&   - h^\prime\;. 
\label{edels} 
\eea
A fixed-point solution can be easily obtained
\beq
 \rme^{\frac{\phi}{\sqrt{2}}}=  \frac{g\,\rme^{3h}}{2\sqrt{2} G}\;.
\eeq
The values of $h(r)$ remain undetermined since the equations for
$\phi^\prime$ and $h^\prime$ are proportional to each other. In
this case, $f(r)$ is given by
\beq 
f(r) = - \log \left( \frac{ g^2 \, \rme^{3h} \,r}{8 \, G} \right) \;,
\eeq
therefore the metric is given by
\beq
 ds^2 = \left(\frac{8\,G}{g^2\,\rme^{3h}\,r}\right)^2
 (dt^2 - dr^2 - dz^2)- \rme^{2h} (dx^2 + dy^2 + dv^2) \;.
\eeq
Eq.~(\ref{edels}) implies that $\varphi = -h$, where, for
simplicity, we omit integration constants. The system of equations
above can be solved leading to
\beq 
f(h)=  - h + \frac{1}{2} \, \log\Xi \;, 
\eeq
where
\be
\Xi = \frac{g}{4 \sqrt{2}}\, \rme^{4h} +\frac{G}{2}\;.
\ee
Next, if we make the change of the integration variable $r
\rightarrow h$, such that
\beq 
\rme^f\, dr = - \frac{\rme^{5h }}{\Xi}\, dh \;, 
\eeq
in terms of $h$ the metric reads
\beq
 ds^2 =- \frac{\rme^{10h}}{\Xi^2}
\,dh^2 + \rme^{-2h}\,\Xi\,(dt^2 - dz^2)- \rme^{2h}\,(dx^2 + dy^2 + dv^2)\;. 
\eeq
Using Eq.~(\ref{6d-dualization}) we can write down the field strength
of the dual theory as follows
\be 
F_{trz} = -G\,\rme^{2\sqrt{2}\f + 3\,f-3\,h} \;. 
\ee
The ten-dimensional solution using Eq.~(\ref{6d-dual-iia-s1xs3}) is
\bea ds_{10}^2 &=& -\frac{1}{2}\rme^{\sqrt{2}\f/4}ds_6^2 
+ \rme^{5\sqrt{2}\f/4}\,dz_1^2\nn\\
& & +
\fr1{g^2}\,\rme^{-3\sqrt{2}\f/4}\,[d\theta^2 + \sin\theta^2
d\varphi^2
+ (d\psi + \cos\theta d\varphi)^2]\;,\nn\\
F_4 &=& -G\,\rme^{2\sqrt{2}\f + 3\,f-3\,h}\,dt\wedge dr\wedge
dz\wedge dz_1 - \fr1{g^2}\,\sin\theta\,d\psi\wedge
d\theta\wedge d\varphi\wedge dz_1\;,\nn\\
\hat{\f} &=& \fr1{\sqrt{2}}\f \;. 
\eea
The metric has the form of a warped product of a six-dimensional
space, a 3-sphere and a single coordinate. We also note that
the $G_3$ field does not play a r\^ole similar to the previous
cases. In fact, there is no twisting in this case, since the space
is not curved. One can view this solution as a NS5-brane of
type IIB after wrapping one of the directions on a circle. It
would be interesting to find an example in which the twisting on
the curved manifold is performed by a $B_2$ field.

\section{Supergravity duals of three-dimensional $\cN=1$ SYM theory on
a torus}

In this section, we concentrate on a system which, when up-lifted to
ten dimensions, can be interpreted as type IIB NS5-branes
wrapped on $S^3 \times T^2$. In particular, this $S^3$ is embedded
in a seven-dimensional manifold with $G_2$ holonomy. We consider the
decoupling limit of $N$ NS5-branes wrapped on $S^3 \times T^2$
\cite{Itzhaki:1998dd}, keeping the radii fixed. Since the brane
worldvolume is curved, in order to define covariantly constant
Killing spinors, the resulting field theory on the brane
worldvolume will be twisted. In order to describe the flows
between the six-dimensional field theory (defined in the
NS5-brane worldvolume in the UV) and the three-dimensional field
theory in the IR, we will start with the $SO(4)$-symmetric
solution, obtained by \cite{Chamseddine:2001hk} of the
five-dimensional $SU(2)$-gauged ${\cal{N}}=4$ supergravity
constructed by Romans \cite{Romans:1986ps}. That solution only
contains magnetic non-Abelian and electric Abelian fields. 
The dual twisted field theory is
defined on the NS5-brane worldvolume wrapped on $S^3$, whereas
the other two spatial directions are wrapped on a torus. In the IR
limit it corresponds to a three-dimensional twisted gauge field
theory on $\R^1 \times T^2$, with 2 supercharges. It is worth
noting that this theory do not come from an $AdS_4$-like
manifold since its spatial directions does not live in the spatial
sector of five-dimensional supergravity, but on the torus in the
ten-dimensional theory. 
Several aspects of this theory as seen from
the gauge field theory point of view, including some dual twisted
gauge field theories, have been analyzed in \cite{Nunez:2001pt}. 
We then up-lift the previously mentioned
solution to massless type IIA supergravity on $S^1 \times S^3$.

If we turn off the electric Abelian fields, it is possible to
find a solution for the five-dimensional gauged supergravity
\cite{Chamseddine:2001hk}, which is singular. Using the
criterion given in \cite{Maldacena:2000mw}, one can see
that the singularity of that solution is a ``bad'' type, so that in the
IR, this solution does not represent a gauge field theory.
Therefore, one may say that the electric Abelian fields 
remove the singularity. It would be interesting to know whether
the non-singular solution with non-vanishing Abelian 2-form is
related to the rotation of the NS5-brane. If it were the case,
it would probably be related to the mechanism studied in 
\cite{Maldacena:2000dr}, leading to a de-singularization by rotation.

\subsection{Supergravity duals of ${\cal{N}}=1$ SYM
theory in $D=3$ on a torus}

In this subsection, we study the supergravity dual of a three-dimensional
${\cal{N}}=1$ SYM theory on a torus~\cite{Schvellinger:2001ib}. 
The gravitational system we
are dealing with can be understood as follows. Let us consider $N$
type IIB NS5-branes. If the 5-branes were flat, the
isometries of this system would be $SO(1,5) \times SO(4)$. The
first corresponds to the Lorentz group on the flat 5-brane
worldvolumes, while the second one is the corresponding rotation
group of the $S^3$ transverse to the 5-brane directions. Since
the NS5-branes are not flat but wrapped on a second $S^3$ (in
the five-dimensional Romans' theory), we have the following chain
of breaking of the isometries $SO(1,9) \rightarrow SO(1,5) \times
SO(4) \rightarrow SO(4) \times SO(4)$. There is also an additional
isometry group corresponding to the torus, where the two
additional spatial directions of the 5-brane are wrapped. On the
other hand, the supergravity solution that we consider here has a
global $SO(4)$ symmetry, and its corresponding ansatz for the
five-dimensional metric has the $\R^1 \times S^3 \times \R^1$
geometry. The $\R^1$s correspond to the time and the radial
coordinate, respectively. In ten dimensions, the solution has the
geometry of the form $(\R^1_0 \times S^3_{1,2,3} \times \R^1_4)
\times T^2_{5,6} \times S^3_{7,8,9}$, where the lower indices
label the coordinates. Recall from the previous chapters that the
seven-dimensional supergravity is related to the five-dimensional
one through a $T^2$ reduction, whereas the up-lifting to
10-dimensional theory is obtained through an $S^3$. In
Table~\ref{tab1}, 
we schematically show the global structure of the
ten-dimensional metric. The first five coordinates are arbitrarily
chosen to represent the five-dimensional metric for the Romans'
theory.
\begin{table}
\caption{Structure of ten-dimensional metric}
\begin{center}
\begin{tabular}{|c|c|c|c|c|c|}
\hline
 0 & 1 \, 2 \, 3 & 4 & 5 \, 6 & 7 \, 8 \, 9   \\
\hline
 $\R^1_0$ & $S^3_{1,2,3}$ & $\R^1_4$ & $T^2_{5,6}$ & $S^3_{7,8,9}$  \\
\hline
\end{tabular}
\end{center}
\label{tab1}
\end{table}
From Table~\ref{tab1}, one can see that the NS5-brane is wrapped
on the $S^3$ (which belongs to the five-dimensional $SU(2)$-gauged
supergravity metric ansatz), while its other two spatial
directions are wrapped on $T^2_{5,6}$, i.e. the fifth and sixth 
directions.

Now, we focus on the twisting preserving 2 supercharges. As
already mentioned above, there are three spatial directions of the
NS5-branes wrapped on $S^3$. Therefore, the supersymmetry will
be realized through a twisting. Also notice that the NS5-branes
have two directions on a torus, so that these are not involved in
a twisting. The brane worldvolume is on $\R^1 \times S^3 \times
T^2$. The non-trivial part of the spin connection on this
worldvolume is the $SU(2)$ connection on the spin bundle of $S^3$.
On the other hand, the normal bundle to the NS5-brane in the
$G_2$ manifold is given by $SU(2) \times SU(2)$, one of them being
the spin bundle of $S^3$. In this case, the twisting consists in
the identification of the $SU(2)$ group of the spin bundle with
one of the factors in the R-symmetry group of the 5-brane, 
i.e. $SO(4)_R \rightarrow SU(2)_L \times SU(2)_R$. It leads to a
diagonal group $SU(2)_D$, so that it gives a twisted gauge theory.
The resulting symmetry group is $SO(1,2) \times SU(2)_D \times
SU(2)_R$. In the UV limit, the global symmetry is $SO(1,5) \times
SU(2)_L \times SU(2)_R$, so that the four scalars transform as the
representation $({\bf 1}, {\bf 2}, {\bf 2})$ and there are also 16
supercharges. After the twisting we get 2 fermions (which are the
2 supercharges of the remaining unbroken supersymmetry)
transforming in the $({\bf 2}, {\bf 1}, {\bf 1})$ representation
of $SO(1,2) \times SU(2)_D \times SU(2)_R$. There are no scalars
after twisting, while we get one vector field as it is before the
twisting. Therefore, $1/8$ of the supersymmetries are preserved,
which is related to the fact that the two $S^3$s, together with
the radial coordinate are embedded in a $G_2$ manifold. In this
way, since our IR limit corresponds to setting the radial coordinate
to be zero, it implies (as we will see) that the  $S^3$ part of
the 5-brane will reduce to a point. This is in contrast with the
fact that the transverse $S^3$ and the torus get fixed radii. It
shows that when one moves to the IR of the gauge theory (flowing
in the radial coordinate on the gravity dual) three of the
dimensions become very small  and no low energy massless modes are
excited on this two-space. Therefore, far in the IR
the gauge theory is effectively three-dimensional.

In order to show explicitly how the theory flows to a
three-dimensional SYM theory on a torus, we briefly describe the
$SO(4)$-symmetric solution of the five-dimensional Romans'
supergravity presented in
\cite{Chamseddine:2001hk}. Following~\cite{Chamseddine:2001hk}, 
we consider a static field configuration,
invariant under the $SO(4)$ global symmetry group of spatial
rotations. As already mentioned, the metric ansatz has the
structure $\R \times S^3 \times \R$ and it can be written as 
\be 
ds^2_5 = \rme^{2 \nu(r)} \, dt^2 - \frac{1}{M(r)} \, dr^2 - r^2 \, 
d\Omega^2_3 \;, 
\label{metricfive} 
\ee 
where $d \Omega^2_3$ is the metric on $S^3$. In terms of
left-invariant
1-forms~(\ref{cartan-maurice})-(\ref{cartan-maurice-euler}), $d
\Omega^2_3$ can be written as $\sum_{i=1}^3\,\s^i\,\s^i$. We
consider the non-Abelian gauge potential components written in
terms of the left-invariant 1-forms 
\be 
A^i = A^i _\m \, d x^\m =
[w(r)+1] \, \s^i \;, 
\ee 
so that they are invariant under the
combined action of the $SO(4)$ rotations  and the $SU(2)$ gauge
transformations. The corresponding field strength is purely
magnetic and is given by 
\be 
F^i = dw \wedge \s^i - [w(r)^2-1]\,d\s^i \;, 
\ee 
while for the Abelian gauge potential we
consider a purely electric ansatz 
\be 
f(r) = Q(r) \, dt \, \wedge
\, dr \;. 
\ee 
All other functions, i.e. $\n$,
$M$, $w$, $Q$ and the dilaton $\phi$ are functions of the
radial coordinate $r$. From the equation of motion of the dilaton, one gets
\be 
\n(r) = \sqrt{\frac{2}{3}} \, (\phi(r) - \phi_0)\;,
\ee 
where $\phi_0$ is an integration constant. On
the other hand, from the equation of motion of the Abelian field,
we obtain
\be 
Q(r) = \frac{e^{5 \n(r)}}{r^3\,\sqrt{M(r)}} \, 
[2 \, w(r)^3 - 6 \, w(r) + H]\;, 
\ee 
where $H$ is
an integration constant. Other equations of motion with the
field configuration and the metric ansatz described above can be
found in \cite{Chamseddine:2001hk}. In order for the
solution to preserve some supersymmetries, it must
satisfy the equations obtained by setting to zero the
supersymmetry transformations for gauginos and gravitinos
Eqs.~(\ref{5d-romans-susy-gravitinos-transf})-(\ref{5d-romans-susy-gauginos-transf}).
After some algebra, we obtain the following first-order differential 
equations 
\ba 
M(r)
& = & \left( \frac{1}{3} \, \zeta^2 \, V - w  \right)^2 +
2 \, \zeta^2\, (w^2-1)^2 - \frac{2}{3} \, (w^2 -1)+ \frac{1}{18
\zeta^2}\;, \nn \\
\frac{d w(r)}{d \log{r}} & = & \frac{1}{6 \, \zeta^2 \, M} \left\{
- 2 \, V \,  (w^2 -1) \, \zeta^4 + (H - 4 \, w^3) \,  \zeta^2 - w
\right\} \;, \nn \\
\frac{d \zeta(r)}{d \log{r}} & = & -\frac{\zeta}{3\,M}[V^2 \,\zeta^4 + 12 \, \zeta^2 \, (w^2-1)^2 - 4 \, V
\, w \, \zeta^2 + w^2 +2] \;, 
\label{BOGO} 
\ea 
where
we have defined $\zeta(r) = \rme^\n/r$ and $V(r) = 2 \, w(r)^3 -
6 \, w(r) +H$. These equations are compatible with the equations
of motion derived from the Romans' five-dimensional Lagrangian
given in appendix F, and any solution of these first-order
differential equations preserves two supersymmetries.

Since we are interested in the IR limit, i.e. when $r
\rightarrow 0$, we obtain the expansions of the functions defining
the metric, the magnetic non-Abelian and the electric Abelian
fields for the five-dimensional ansatz. They are 
\ba
w(r) & = & 1 - \frac{1}{24} \, r^2 + \cdots\;, \nn \\
\zeta(r) & = & \frac{1}{r} + \frac{7}{288} \, r + \cdots\;, \nn \\
M(r) & = & 1 + \frac{5}{144} \, r^2 + \cdots\;, 
\ea 
and straightforwardly 
\be 
\n(r)  =  \frac{7}{288} \,
r^2 + \cdots\;,
\ee 
while for the dilaton
we obtain 
\be 
\phi(r)  =  \phi_0 +\frac{7}{288} \,
\sqrt{\frac{3}{2}} \, \, r^2 + \cdots\;. 
\ee 
In this case we have taken $H$ to be 4. Also, we get 
\be
Q(r)=\frac{1}{96} \, r+ \frac{13}{13824} \, r^3+ \cdots\;. 
\ee 
In this way, one can see that in the IR limit the
non-Abelian gauge potential has a core, while both field
strengths, i.e. the Abelian and the non-Abelian one are of
the order $r$ around $r=0$. Furthermore, the above solution can be
up-lifted, following the ans\"atze presented in chapter II and in chapter
III, to either type IIA or type IIB theory.
As we will see later, the IR limits in these two cases turn out to be
the same. 

\subsection{Up-lifting to type IIB theory}

First, we consider the case when the solution is up-lifted to
type IIB supergravity. From Eq.(\ref{5d-no-g1-iib-s3xt2}) the
ten-dimensional metric is
\ba 
d\hat{s}_{10}^2 &=& - \rme^{\frac{13}{5\sqrt{6}}\f}\, \left(
\rme^{2 \nu(r)} \, dt^2 - \frac{1}{M(r)} \, dr^2 - r^2
\, d \Omega^2_3 \right) \nn \\
& & + \rme^{\frac{3}{5\sqrt{6}}\f}\,(dY^2 + dZ^2) +
\frac{1}{4g^2}\,\rme^{-\frac{3}{\sqrt{6}}\f}\,\sum_{i=1}^3(\s^i -
g\,A_1^i)^2\;, \nn\\
\hat{\phi} &=& \sqrt{6}\,\phi\;. 
\label{metricfinalIIB} 
\ea
Using the previously calculated IR, expansion we can
obtain the radii of the different manifolds. Thus, for the $S^3$
involving the coordinates 1, 2 and 3, the radius is given by
\be 
R^2_{1,2,3} = \rme^{13 \phi_0/5 \sqrt{6}} \, r^2 + {\cal
{O}}(r^4)\;.
\label{r123iib} 
\ee
On the other hand, the radii of $T^2_{5,6}$, $S^3_{7,8,9}$ are given by
\ba 
R^2_{T} &=&\rme^{\frac{\sqrt{3}}{5\sqrt{2}}\f_0}\left(
1 - \frac{7}{960}\, r^2 + {\cal {O}}(r^3)\right) \;, \nn \\
R^2_{7,8,9} &=&
\frac{1}{4g^2}\rme^{-\frac{3}{\sqrt{6}}\phi_0}\,\left( 1 -
\frac{21}{576}\, r^2 + {\cal {O}}(r^3)\right) \;.
\label{rtiib} 
\ea
Without loss of generality, we can set $\phi_0$ to zero. It is
obvious from Eqs.~(\ref{r123iib}) and (\ref{rtiib}) that in the
limit $r \rightarrow 0$, $R_T$ and $R_{7,8,9}$ remain finite,
while $R_{1,2,3} \rightarrow 0$. Since the type IIB NS5-brane
is wrapped on $S^3_{1,2,3}$, $T^2$, and in the IR limit
$S^3_{1,2,3}$ effectively reduces to a point,
in this limit we obtain a twisted gauge field theory defined on the torus.

\subsection{Up-lifting to type IIA theory}

Using the ansatz~(\ref{5d-no-g1-iia-s1xs3xs1}) after up-lifting to
type IIA theory, the ten-dimensional solution is
\begin{eqnarray}
d\hat{s}_{10}^2 &=& - {\rm e}^{\frac{7}{8\sqrt{6}}\phi}\,
\left(\rme^{2 \nu(r)} \, dt^2 - \frac{1}{M(r)} \, dr^2 - r^2
\, d \Omega^2_3 \right)  \nn \\
& & + \frac{1}{4g^2}\,{\rm e}^{-\frac{9}{8\sqrt{6}}\phi}\,
\sum_{i=1}^3\,\left(\sigma^i - g\,A^i_1\right)^2 + {\rm
e}^{-\frac{9}{8\sqrt{6}}\phi}\,dZ^2 +
\rme^{\frac{15}{8\sqrt{6}}\phi}\,dY^2 \;, \nonumber\\
\hat{\phi\baselineskip=20pt plus 1pt minus 1pt } &=&
\frac{3}{4\sqrt{6}}\,\phi \;. 
\label{metricfinal} 
\ea
As we did for the type IIB case, we can obtain the
radius
\be 
R^2_{1,2,3} = {\rm e}^{\frac{7}{8\sqrt{6}}\f_0} \, r^2 + {\cal
{O}}(r^4) \;, 
\ee
which shrinks to zero in the IR limit. In addition, the radii of
$S^1_{5}$, $S^3_{6,7,8}$ and $S^1_9$ are finite
\ba 
R^2_{5} &=&{\rm e}^{\frac{15}{8\sqrt{6}}\f_0}\,\left(
1 + \frac{105}{4608}\, r^2 + {\cal {O}}(r^3)\right) \;, \nn \\
R^2_{6,7,8} &=& \frac{1}{4g^2}\,{\rm
e}^{-\frac{9}{8\sqrt{6}}\phi_0}
\left( 1 - \frac{27}{4608}\, r^2 + {\cal {O}}(r^3)\right) \;, \nn \\
R^2_{9} &=& {\rm e}^{-\frac{9}{8\sqrt{6}}\f_0}\,\left(1 -
 \frac{27}{4608}\, r^2 + {\cal {O}}(r^3)\right) \;.
\ea
Again, by considering $\phi_0=0$, in the IR, the radii
$R_{5}=R_{9}$ and $R_{6,7,8}=1/(2 g)$, while $R_{1,2,3}
\rightarrow 0$. Since the type IIA NS5-brane is wrapped on
$S^3_{1,2,3}$, $S^1_5$ and $S^1_9$, and in the IR limit
$S^3_{1,2,3}$ effectively shrinks to a point as in the type IIB
case, we get the same geometric reduction as in the previous case.
Note that this can be obtained when $\phi_0=0$, so that the radii
of the torus (in type IIB case) and the two $S^1$s (in type IIA
case) are exactly the same.

In addition, in both cases, one can use the criterion for
confinement given in 
\cite{Kinar:1998vq,Sonnenschein:1999if}, in order to show that the
corresponding static potential is confining.

\subsection{The singular $SO(4)$-symmetric solution}

A solution with no electric Abelian fields can be obtained by
setting $H$ to zero. It implies that $w$, $V$ and also $Q$ are
zero, as we expected since no electric field is excited. In this
way, the first-order differential equations (\ref{BOGO}) can be
easily integrated, yielding the relation 
\be 
r=r_0 \, \frac{{\rm
e}^{1/24 \zeta^2}}{\sqrt{\zeta}} \;, 
\ee 
where $r_0$ is an
integration constant. The metric is given by 
\be 
d s^2_5 = r_0^2
\,\, {\rm e}^{1/(12 \zeta^2)} \, \left( \zeta \, dt^2 - \frac{1}{8
\, \zeta^5} \, d \zeta^2 - \frac{1}{\zeta} \, d\Omega^2_3 \right)
\;. 
\ee 
In addition, for the dilaton we have the following
relation 
\be {\rm e}^{\sqrt{2} \phi/\sqrt{3}} = r_0 \,\, {\rm
e}^{1/24 \zeta^2} \,\, \sqrt{\zeta} \;. 
\ee 
Using the
criterion of \cite{Maldacena:2000mw}, it is
obvious that the IR singularity is not acceptable,
in both type IIA and type IIB theories.

%% file: ch5.tex
\chapter{Conclusions}
\vspace{1cm}

We have presented a number of reduction ans\"atze of
type IIA, type IIB supergravities in ten dimensions, and
eleven-dimensional supergravity on various spaces of the form
$S^m\times T^n$. Cases that were presented are reductions of type IIA
supergravity on $S^3$, $S^4$, $S^1\times S^3$ and $S^1\times S^3\times
S^1$ which give rise to seven-dimensional, $SO(4)$-gauged supergravity,
six-dimensional, $SO(5)$-gauged supergravity, the dual $SU(2)$-gauged
supergravity in $D=6$ and the Romans' theory without the $U(1)$ 
gauge-coupling in $D=5$, respectively; reductions of type IIB supergravity
on $S^5$ and $S^3\times T^2$. Reduction of type IIB supergravity on
one deformation of $S^5$ gives rise to the $\cN=4$, 
$SU(2)\times U(1)$-gauged supergravity in $D=5$ and on another 
deformation of $S^5$ 
produces the subset of the fields of five-dimensional $\cN=8$ 
$SO(6)$-gauged supergravity, consisting of gravity, the fifteen $SO(6)$ gauge 
fields, and the twenty scalars of the $SL(6,\R)/SO(6)$ submanifold of 
the full scalar manifold. This embedding can equivalently be viewed as
a complete reduction ansatz (with no truncation of massless fields)
for the ten-dimensional theory, comprising just gravity plus a
self-dual 5-form, which itself is a consistent truncation of type IIB
supergravity. In a full $S^5$ reduction of type IIB supergravity there
will be additional fields coming from the reduction of the NS-NS and
R-R 2-form potentials, and from the dilaton and axion. A compete
analysis of the $S^5$ reduction can therefore be expected to be
extremely complicated. In particular, for example, the {\bf 10} and
$\overline{\mbox{\bf 10}}$ of pseudo-scalars lead to a considerably more
complicated metric reduction ansatz. The ansatz for a subset of the
fields that included one scalar and one pseudo-scalar was derived
in~\cite{Cvetic:2000tb}, and in~\cite{Pilch:2000ej}.

These cases presented here represent other elements
in the accumulating body of examples where ``remarkable'' Kaluza-Klein
sphere reductions exist, even though there is no known group-theoretic
explanation for their consistency. What is still lacking is a deeper
understanding of why they should work. One might be tempted to think
that supersymmetry could provide the key, but this evidently cannot in
general be the answer, since there are some examples such as the
consistent $S^3$ and $S^{D-3}$ reductions of the $D$-dimensional
low-energy limit of the bosonic string in arbitrary dimension $D$ that
are obviously unrelated to supersymmetry~\cite{Cvetic:2000dm}.

Albeit we have no theoretical explanation of why the spherical
reductions should work, knowing the non-linear ans\"atze has 
proven to be very useful in a number of contexts. We showed that using
the non-linear Kaluza-Klein ansatz considerably simplifies the
calculation of $n$-point correlation functions in AdS/CFT
correspondence. Applying the method in~\cite{Das:1998ei} and using
the non-linear ansatz, we derived the CFT operators which are in
agreement with those obtained based on the conformal
symmetry~\cite{Witten:1998qj}. One implication of our result is to
find the operators that deform CFT theory from supergravity
solutions. This procedure is an exact opposite to what has been
presented in the literature, where one first deforms the CFT by adding
some operators, and then tries to find the corresponding supergravity
solution via the AdS/CFT duality.

We also showed that non-linear Kaluza-Klein reduction ans\"atze are
very useful in studying the supergravity duals of certain field theories.
These investigations provide more examples to support the well-known
AdS/CFT correspondence. In particular, we found certain supergravity 
solutions in the Romans' gauged supergravities in $D=5$ and $D=6$. 
Using the non-linear Kaluza-Klein
ans\"atze, these solutions, after uplifting to type IIA, type IIB and
massive type IIA supergravities, can be interpreted as a D4-brane and 
a NS5-brane wrapped on $S^2$, $S^3$ and $S^3\times T^2$. These solutions 
are dual to certain twisted field theories. D4-brane wrapped on 
$S^2$ is supergravity dual to five-dimensional $\cN=2$ SCFT, which 
in the IR flows to $\cN=2$ SCFTs in $D=3$. On the other hand,
D4-brane wrapped on $S^3$ is supergravity dual to $\cN=2$ SCFT in
$D=5$, which flows to a two-dimensional $(1, 1)$ CFT. We also found a
solution that has ``good'' singularity in the sense
of~\cite{Maldacena:2000mw,Gubser:2000nd}. After uplifting to type IIB,
this solution is interpreted as NS5-brane wrapped
on $S^2$ which is dual to three-dimensional $\cN=2$ super Yang-Mills
theory. We also studied the NS5-brane of type IIB wrapped on
$S^3$. This solution is dual to three-dimensional $\cN=1$ SYM theory on
a torus.

%% file: appendA.tex
\appendix{Eleven-dimensional supergravity}

Eleven is a maximum space-time dimension in which a consistent
supersymmetric theory can be constructed~\cite{Nahm:1978tg}. The
${\cal N}=1$ supergravity in eleven dimensions was constructed by
Cremmer, Julia and Scherk~\cite{Cremmer:1978km}. The field content of
eleven-dimensional supergravity consists of a graviton $g_{MN}$, a
3-form potential $A_\3$ whose field strength is $F_\4 = d\,A_\3$ (or
in terms of indexes $F_{MNPQ} = 4\,\pa_{[M}\,A_{NPQ]}$), and a
gravitino $\Psi_M$ with 44, 84 and 128 physical degrees of
freedom. For a counting number degrees of freedom in various
dimensions, see Table.~\ref{Degrees}. The Lagrangian of the bosonic
sector is
\bea
{\cal L} &=& \sqrt{-g}(R - \frac1{2\cdot4!}\,F_{MNPQ}\,F^{MNPQ})\nn\\
& &- \fr1{3!^2\cdot 4!^2}\e^{M_1\cdots M_{11}}F_{M_1\cdots
M_4}F_{M_5\cdots M_8}A_{M_9\cdots M_{11}}\;,
\label{11d-lag-component}
\eea
where $g = {\rm det}(g_{MN})$. Eq.~(\ref{11d-lag-component}) can also be
written in form notation
\be
{\cal L}_{11} = R\,{*\oneone} - \fr12\,{*F}_\4\wedge F_\4 - \fr16\,F_\4\wedge
F_\4\wedge A_\3\;.
\label{11d-lag-form}
\ee
The supersymmetry transformation of gravitino in the absence of
fermion is
\be
\d\Psi_M = \left[\pa_M\Psi_N +
\fr14(\w_{AB})_N\G^{AB}\Psi_N\right]\ep -
\fr1{288}(\G_M^{\;\;PQRS} - 8\d_M^{\;\;P}\G^{QRS})F_{PQRS}\,\ep\;.
\label{11d-susy-transf}
\ee
The equations of motion of the eleven-dimensional supergravity are
\be
R_{MN} = \fr1{12}\left(F_{MPQR}\,F_N^{\;\;PQR} -
\fr1{12}\,g_{MN}\,F_{PQRS}\,F^{PQRS}\right)\;,
\label{11d-einstein-eq}
\ee
and
\be 
d{*F}_\4 = \fr12\,F_\4\wedge F_\4\;.
\label{11d-4form-eq}
\ee
\newpage
\vspace*{-1cm}
\begin{table}
\caption{On-shell degrees of freedom in $D$ dimensions. $\alpha=\fr12\,D$ if 
$D$ is even,
$\alpha=\fr12\,(D-1)$ if $D$ is odd.}
\begin{center}
\begin{tabular}{|l|l|l|}\hline
Field & Notation & \# degrees of freedom\\
\hline
$D$-bein&$e_{M}{}^{A}$&$\fr{D(D-3)}{2}$\\
\hline
Majorana fermions&$(\Psi_M)_{Maj}$&$2^{(\alpha-1)}(D-3)$\\
\hline
Majorana-Weyl fermions& $(\Psi_M)_{Maj-Weyl}$& $2^{(\a - 2)}(D-3)$\\
\hline
$p$-form&$A_{M_{1}M_{2}\cdots M_{p}}$& $\fr{(D-2)!}{p!\,(D-2-p)!}$\\
\hline
self-dual $p$-form &$A^+_{M_{1}M_{2}\cdots M_{p}}$& $\fr{(D-2)!}{2\,p!\,(D-2-p)!}$\\
\hline
Majorana spinor&$\chi_{Maj}$&$2^{(\alpha-1)}$\\
\hline
Majorana-Weyl spinor&$\chi_{Maj-Weyl}$&$2^{(\alpha-2)}$\\
\hline
\end{tabular}
\end{center}
\la{Degrees}
\end{table}

%% file: appendB.tex
\appendix{Ten-dimensional type IIA supergravity}

Ten-dimensional type IIA supergravity can be obtained by
dimensional reducing eleven-dimensional supergravity on a
circle~\cite{Huq:1985im}. The
field content of type IIA supergravity comprises a vielbein
$e_M^{\;\;A}$, a scalar $\f$, an R-R 1-form potential $\cA_\1$ whose
field strength is $\cF_\2 = d\,\cA_\1$, an NS-NS 2-form potential $B_\2$
whose field strength is $F_\3 = d\,B_\2$ and an R-R 3-form potential
$A_\3$ whose field strength is $F_\4 = d\,A_\3 + \cA_\1\wedge
F_\3$, two gravitini, which have opposite chirality, $\Psi_M^1$ and 
$\Psi_M^2$, and two dilatini $\l_1$ and $\l_2$. The bosonic Lagrangian 
of the theory is
\bea
\cL_{\rm IIA} &=& R\,{*\oneone} - \fr12\,{*d}\f\wedge d\f -
\fr12\,\rme^{\fr32\f}{*\cF}_\2\wedge\cF_\2-\fr12\,\rme^{\fr12\f}{*F}_\4\wedge
F_\4\nn\\
& & -\fr12\,\rme^{-\f}{*F}_\3\wedge F_\3 + \fr12\,dA_\3\wedge
dA_\3\wedge A_\2\;, 
\la{iia-lag-form}
\eea
where
\be
F_\4 = dA_\3 - dA_\2\wedge \cA_\1\;,\;\;\;F_\3 = dA_\2\;,\;\;\;\cF_\2
= d\cA_\1\;.
\ee
In the absence of the fermionic background, the gravitino
transformation rule is
\bea
\d\,\Psi_M &=& \left(\pa_M + \fr14\,(\w_{AB})_M\G^{AB}\right)\ep
+ \fr1{64}\,\rme^{-\fr32\f}(\G_M^{\;\;NP} -
14\,\d_M^{\;\;N}\G^P)\G^{11}\,F_{NP}\,\ep\nn\\
& &+\fr1{96}\,\rme^{\fr12\f}(\G_M^{\;\;NPQ} -
9\,\d_M^{\;\;N}\G^{PQ})\G^{11}\,F_{NPQ}\,\ep\nn\\
& &+\fr{\im}{256}\,\rme^{-\fr14\f}(\G_M^{\;\;NPQR} -
\fr{20}{3}\,\d_M^{\;\;N}\,\G^{PQR})F_{NPQR}\,\ep\;,
\label{iia-gravitino-transf}
\eea
and the supersymmetry transformation of the dilatino is
\bea
\d\,\l &=&
-\fr{\sqrt{2}}{4}\left(\pa_M+\fr14\,(\w_{AB})_M\G^{AB}\right)\f\G^M\G^{11}\ep
+ \fr3{16\sqrt{2}}\,\rme^{-\fr34\f}\,\G^{MN}F_{MN}\,\ep\nn\\
& & +\fr{\im}{24\sqrt{2}}\,\rme^{\fr12\f}\,\G^{MNP}\,F_{MNP}\,\ep -
\fr{\im}{192\sqrt{2}}\,\rme^{\fr14\f}\G^{MNPQ}F_{MNPQ}\,\ep\;,
\label{iia-dilatino-transf}
\eea
where $\G^M$ are the ten-dimensional Dirac matrices and $\G^{11} =
\im\,\G^0\G^1\cdots\G^9$.

The equations of motion of the Lagrangian~(\ref{iia-lag-form}) are
\bea
d{*d}\f &=& \fr12\,\rme^{-{\f}}\,{*F}_\3\wedge F_\3 - \fr34\,
\rme^{\fft32\f}\,{*\cF}_\2\wedge \cF_\2 - \fr14\, {\rm
e}^{\fft12\f}\,{*F}_\4\wedge F_\4 \;,\nn\\
d(\rme^{\fr12{\f}}\,{*F}_\4) &=& F_\4\wedge F_\3 \;,\nn\\
d(\rme^{\fr32{\f}}\,{*\cF}_\2) &=& - \rme^{\fr12{\f}}\,{*F}_\4\wedge
F_\3 \;,\nn\\
d({\rme}^{-{\f}}\,{*F}_\3) &=& \fr12\, {F}_\4\wedge {F}_\4 -
{\rme}^{\fr12{\f}}\,{*F}_\4\wedge\cF_\2\;.
\label{iia-eqs}
\eea

%% file: appendC.tex
\appendix{Ten-dimensional type IIB supergravity}

The field content of ten dimensional type IIB supergravity
consists of the metric, a self-dual 5-form field strength, a scalar, an
axion, an R-R 3-form, an NS-NS 3-form field strength, two gravitini, which have the same chirality, $\Psi_M^1$, $\Psi_M^2$ and two dilatini $\l_1$, $\l_2$.  
There is no simple covariant Lagrangian for type IIB supergravity, 
on account of
the self-duality constraint for the 5-form.  However, one can write a
Lagrangian in which the 5-form is unconstrained, which must then be
accompanied by a self-duality condition which is imposed by hand at
the level of the equations of motion \cite{Bergshoeff:1995as}.  The bosonic 
Lagrangian of this type IIB is
\bea
{\cal L}_{\rm IIB} &=& R\,{*\oneone} - \fr12
{*d}\phi\wedge d\phi - \fr12
\rme^{2\phi}\,{*d}\chi\wedge d\chi - \fr14 {*H}_\5
\wedge H_\5 \nn\\
& & -\fr12 \rme^{-\phi} \, {*F}^2_\3\wedge F^2_\3 -
\fr12 \rme^{\phi}\, {*F}^{1}_\3\wedge F^{1}_\3 -
\fr12 B_\4\wedge dA_\2^{1}\wedge dA_\2^{2}\;,
\label{iib-lag-can}
\eea
where 
\bea
F^2_\3 &=& dA^2_\2\;,\;\;\; F^1_\3=d A^1_\2 - \chi\, d A^2_\2\;,\nn\\ 
H_\5 &=& dB_\4 - \ft12 A_\2^1\wedge dA_\2^2 + \ft12 A_\2^2\wedge dA_\2^1\;.
\label{iib-notation-can}
\eea
The equations of motion following from the Lagrangian, together
with the self-duality condition, are
\bea
 R_{MN} &=& \ft12\pa_M\phi\, \pa_N\phi + \ft12 \rme^{2
\phi}\, \pa_M\chi\, \pa_N\chi + \ft1{96}  H^2_{MN}\nn\\
& &+ \ft14 \rme^{\phi}\, \Big((F^{1}_\3)^2_{MN} -
\ft1{12}(F^1_\3)^2 {g}_{MN}\Big)\nn\\
& & + \ft14 \rme^{-\phi}\,
\Big((F^{2}_\3)^2_{MN} - \ft1{12}
(F^2_\3)^2{g}_{MN}\Big)\;,
\label{iib-einstein-eq}
\eea
\be
d{*d\phi} = - \rme^{2\phi}\, {*d\chi}\wedge
d\chi - \ft12 \rme^{\phi}\, {*F_\3^1}\wedge
F_\3^1 + \ft12 \rme^{-\phi}\, {*F_\3^2}\wedge
F_\3^2\;,
\label{iib-dilaton-eq}
\ee
\be
d\Big(\rme^{2\phi}\, {*d\chi}\Big) =
\rme^{\phi}\, {*F_\3^1}\wedge F_\3^2\;,
\label{iib-axion-eq}
\ee
\be
d\Big(\rme^{\phi}\, {*F_\3^1}\Big) = {H}_\5\wedge
F_\3^2\,,\qquad
d\Big(\rme^{-\phi}\, {*F_\3^2} -
\chi\, \rme^{\phi}\, {*F_\3^1}\Big) = -
{H}_\5\wedge ( F_\3^1 + \chi\, F_\3^2)\;,
\label{iib-3form-eq}
\ee
\be
d(*H_\5) = - F_\3^1\wedge
F_\3^2\;, \qquad  H_\5 = *H_\5\;.
\label{iib-5form-eq}
\ee
The equations (\ref{iib-einstein-eq})-(\ref{iib-5form-eq}) are precisely those
which were found in \cite{Schwarz:1983qr}.

 In the absence of fermionic backgrounds and using complex notations the supersymmetry transformation for dilatino and gravitino are
\bea
\d\Psi_M &=& {\cal D}_M\,\epsilon + 
\fr{\im}{480}\G^{N_1\cdots N_5}\G_M \epsilon\,F_{M_1\cdots M_5}
+\fr1{96}\left(\G_M^{\;\;NPQ}G_{NPQ} - 9\G^{NP}G_{MNP}\right)\epsilon^*\;,\nn\\
\d\l &=& \im\,\G^M\epsilon^*\, P_M - \fr{\im}{24}\G^{MNP}\epsilon\, G_{MNP}\;,
\label{iib-susy-transf}
\eea
where 
\bea
G_{MNP} &=& f\,(F_{MNP} - B\,F_{MNP}^*)\;,\;\;\; F_{MNP} =
3\pa_{[M}A_{NP]}\;,\nn\\
F_{MNPQR} &=& 5 \pa_{[M}A_{NPQR]} - \fr54\, {\rm
Im}\,(A_{[MN}F^*_{PQR]})\;,\nn\\
Q_M &=& f^2\,{\rm Im}\,(BP_M^*)\;,\;\;\; P_M = f^2\,\pa_MB\;,\;\;\; f =
\fr1{\sqrt{1-|B|^2}}\;.
\label{iib-notation-complex}
\eea
In the above Eq.~(\ref{iib-notation-complex}), $B$ is a complex
scalar. Using a simple field redefinition~\cite{Bergshoeff:1995as}, 
$B$ can be written in terms of dilaton $\f$ and axion $\chi$ as 
\be
B = \frac{1-\rme^{-2\f}-\chi^2+2\im\chi}{(1+\rme^{-\f})^2+\chi^2}\;.
\label{iib-scalar-can-complex}
\ee
The relation between 4-form potentials of two notations is $B_{MNPQ} =
4 A_{MNPQ}$. The complex 3-form field strength $F_\3$ can be written in
terms of NS-NS 2-form potential $A^2_\2$ and R-R 2-form potential
$A^1_\2$ as $F_{MNP}=3\pa_{[M}(A^2_{NP]} - \im A^1_{NP]})$.

%% file: appendD.tex
\appendix{Ten-dimensional massive type IIA supergravity}

The bosonic sector of massive type IIA
supergravity~\cite{Romans:1986tw} consists of a graviton, a scalar, a
2-form field strength, 3-form and 4-form field strengths. In the
language of differential forms, the bosonic Lagrangian is
given by \cite{Lavrinenko:1999xi}
\bea
{\cal L}_{{\rm massive}} &=& R\,{*\oneone} -
\ft12 {*d}\phi\wedge d\phi
- \ft12 e^{\fft32\phi}\, {*F_\2}\wedge F_\2 - 
\ft12 e^{-\phi}\, {*F_\3}\wedge F_\3\nn\\
& & - 
\ft12 e^{\fft12\phi}\, {*F_\4}\wedge F_\4
-\ft12 dA_\3\wedge dA_\3 \wedge A_\2 - \ft16 m\, 
dA_\3 \wedge (A_\2)^3\nn\\
& & -\ft1{40} m^2\, (A_\2)^5 -\ft12 m^2\, e^{\fft52\phi}\, 
{*\oneone}\;,
\label{massive-iia-lag}
\eea
where the field strengths are given in terms of potentials by
\bea
F_\2 &=& dA_\1 + m\, A_\2\;,\qquad F_\3 = 
dA_\2\;,\nn\\
F_\4 &=& dA_\3 + A_\1\wedge dA_\2 + \ft12 m\, 
A_\2\wedge A_\2\;.
\label{massive-iia-fields}
\eea
It follows that the equations of motion and Bianchi identities are
\bea
&&d(e^{\fft12\phi}\, {*F_\4}) = -F_\3\wedge 
F_\4\;,\qquad d(e^{\fft32\phi}\, {*F_\2}) =-
e^{\fft12\phi}\, {*F_\4}\wedge F_\3\;,\nn\\
&&d(e^{-\phi}\, {*F_\3}) = -\ft12 F_\4\wedge 
F_\4 -m\, e^{\fft32\phi}\, {*F_\2}-
e^{\fft12\phi}\, {*F_\4}\wedge F_\2\;,\nn\\
&&d{*d\phi}= -\ft54m^2\, e^{\fft52\phi}\, 
{ *\oneone}-\ft34 e^{\fft32\phi}\, {*F_\2}\wedge
F_\2 + \ft12 e^{-\phi}\, {*F_\3}\wedge F_\3
-\ft14 e^{\fft12\phi}\, {*F_\4}\wedge F_\4\;,\nn\\
&&dF_\4=F_\2\wedge F_\3\,,\qquad
dF_\3=0\,,\qquad dF_\2=m\,F_\3\;,
\label{massive-iia-eqs}
\eea
for the form fields and dilaton, together with the Einstein equation
(in vielbein components)
\bea
 R_{AB} &=& \ft12 \del_A\phi\, \del_B\phi + 
\ft1{16}m^2\, e^{\fft52\phi}\, \eta_{AB} + \ft1{12}
e^{\fft12\phi}\, \Big( F^2_{\4 AB} -\ft3{32} F_\4^2\,
\eta_{AB} \Big)\label{massive-iia-einstein-eq}\\
&& + \ft14 e^{-\phi}\, \Big( F^2_{\3 AB} -\ft1{12} 
F_\3^2\, \eta_{AB}\Big) + \ft12 e^{\fft32\phi}\, \Big( 
F^2_{\2 AB} -\ft1{16} F_\2^2\, \eta_{AB}\Big)\;.\nn
\eea

%% file: appendE.tex
\appendix{Six-dimensional gauged supergravity}

The existence of six-dimensional gauged supergravity possessing
a ground state with unbroken $F_4$ supersymmetry was predicted by
Nahm~\cite{Nahm:1978tg} and by DeWitt and van Nieuwenhuizen~\cite{DeWitt:1982wm}.
The six-dimensional gauged ${\cal{N}}=4$ 
supergravity constructed by Romans~\cite{Romans:1986tw}. Further
developments of $F_4$ gauged supergravity in six dimensions coupled to
matter were recently done by D'Auria et
al.~\cite{D'Auria:2000ad,Andrianopoli:2001rs}. Here we will review the
theory constructed by Romans~\cite{Romans:1986tw}, whose conventions
we follow. The theory consists of a graviton
$e_\m^\a$, three $SU(2)$ gauge potentials $A_\m^I$, an Abelian
potential  ${\cal A}_\m$, a two-index tensor gauge field
$B_{\m\n}$, a scalar $\phi$, four gravitinos $\psi_{\m\, i}$ and
four gauginos $\c_\m$. The bosonic Lagrangian is
\begin{eqnarray} 
e^{-1}\,{\cal L} &=& -\frac{1}{4} \, R +\frac{1}{2} \,
(\partial^\mu\phi) \, (\partial_\mu\phi) - \frac{1}{4} \, {\rm
e}^{-\sqrt{2}\f}\, ({\cal H}^{\mu\nu} \, {\cal H}_{\mu\nu} +
F^{I \, \mu\nu} \, F^I_{\mu\nu})\nn\\
& &  + \frac{1}{12}\,{\rm e}^{2\sqrt{2}\f}\,G_{\m\n\r}\,G^{\m\n\r}
+ \fr18\,(g^2\,{\rm e}^{\sqrt{2}\f} + 4 g\,m\,{\rm
e}^{-\sqrt{2}\f} -
m^2\,{\rm e}^{-3\sqrt{2}\f})\nn\\
& & - \frac{1}{8} \,e^{-1}\,\varepsilon^{\m\n\r\s\t\k}\, B_{\m\n} \,
\left({\cal F}_{\r\s}\,{\cal F}_{\t\k} + m \, B_{\r\s} \, {\cal
F}_{\t\k}\right.\nn\\
& & \left. + \frac{1}{3} \, m^2 \, B_{\r\s} \, B_{\t\k} +
F^I_{\r\s} \, F^I_{\t\k}\right)\;, 
\label{6d-romans-lag} 
\end{eqnarray}
where $e$ is the determinant of the vielbein, $g$ is the $SU(2)$
coupling constant, $m$ is the mass parameter associated with the
two-index tensor field $B_{\m\n}$ and $\varepsilon_{\m\n\r\s\t\k}$
is a Levi-Civita tensor density. The Abelian field strength ${\cal
F}_{\m\n}$, the non-Abelian one $F_{\m\n}^I$, the three-form
$G_{\m\n\r}$ and the field ${\cal H}_{\m\n}$ are given by
\bea
{\cal F}_{\m\n}&\equiv& \pa_\m{\cal A}_\n - \pa_\n{\cal A}_\m \,\,\, ,\;\;\;
F_{\m\n}^I \equiv \pa_\m A_\n^I - \pa_\n A_\m^I + g \,
\e^{IJK}\,A^J_\m\,A^K_\n \;, \nn\\
G_{\m\n\r} &\equiv& 3\,\pa_{[\m}B_{\n\r]} \;,\;\;\;
{\cal H}_{\m\n}\equiv {\cal F}_{\m\n} + m\,B_{\m\n} \;, 
\eea
respectively. The supersymmetry transformations for the gauginos
is
\bea 
\d\c_i = \left( \fr1{\sqrt{2}}\, \G^\m \, \pa_\m\f + A \,
\G_7 - \fr1{12} \,{\rm e}^{\sqrt{2}\f}\G_7\,\G^{\m\n\r}\,
G_{\m\n\r}\right) \e_i + \fr1{2\sqrt{2}}\,\G^{\m\n}\,
({\hat{H}}_{\m\n})_i^{\,\,\, j} \, \e_j \;,
\label{6d-romans-gauginos-transf}
\eea
and that for the gravitinos is
\bea
\d\psi_{\m\, i} & = & \left( \nabla_\m  + T \, \G_\m \, \G_7
 - \frac{1}{24}\,{\rm
e}^{\sqrt{2}\f}\,\G_7\,\G^{\n\r\s}\,G_{\n\r\s}\,\G_\m
\right) \, \e_i  \nonumber \\
&  & + \left[ g \, A_\mu^I \, (T^I)_i^{\,\,\, j}
-\frac{1}{4\sqrt{2}} \, (\Gamma_\mu^{\,\,\,\, \nu\rho} - 6 \,
\delta_\mu^{\,\,\,\, \nu} \, \Gamma^\rho)
({\hat{H}}_{\nu\rho})_i^{\,\,\, j} \right] \epsilon_j \;,
\label{6d-romans-gravitino-transf} 
\eea
where $A$, $T$, and $\hat{H}$ are defined as follows
\bea 
A &\equiv& \frac{1}{4\sqrt{2}} \, ( g \,
\rme^{\frac{1}{\sqrt{2}}\f} - 3 \, m \,
\rme^{\frac{-3}{\sqrt{2}}\f})\;,\;\;\;\; T \equiv -
\frac{1}{8\sqrt{2}} \, (g \, e^{\frac{1}{\sqrt{2}}\f} + m \,
\rme^{\frac{-3}{\sqrt{2}}\f}) \;,\label{6d-romans-at-def}\\
({\hat{H}}_{\mu\nu})_i^{\,\,\, j} &\equiv&
\rme^{-\frac{1}{\sqrt{2}}\f} \, \left[ \fr12 {\cal H}_{\mu\nu} \,
\delta_i^{\,\,\, j} + \Gamma_7 \, F_{\mu\nu}^I \, (T^I)_i^{\,\,\,
j} \right] \;. \label{6d-romans-h-def} 
\ea
The gauge-covariant derivative ${\cal D}_\m$ acting on the Killing
spinor is
\beq 
{\cal D}_\m\,\e_i = \nabla_\m\,\e_i +
g\,A^I_\m\,(T^I)_i^{\;\;j}\,\e_j \;, 
\eeq
with
\beq 
\nabla_\mu \e_i \equiv (\partial_\mu+\frac{1}{4} \,
\omega^{\, \, \, \, \alpha \beta}_\mu  \,
      \Gamma_{\alpha \beta} ) \, \e_i \;,
\eeq
where $\omega^{\;\;\alpha \beta}_\mu$ is the spin connection.
Indices $\a$ and $\b$ are tangent space (or flat) indices, while $\m$
and $\n$ are space-time (or curved) indices. The
$\Gamma_{\alpha\beta\cdots}$ are the six-dimensional Dirac
matrices,
\[ \Gamma_{\a_1 \cdots\a_n}=\frac{1}{n\,!} \,
\Gamma_{[\a_1}\,\cdots\,\Gamma_{\a_n]}\;,\;\;\;\;\;\; n = 1, \cdots,
6\;.\]
The Einstein equation of the Lagrangian~(\ref{6d-romans-lag}) is
\bea 
R_{\m\n} &=& 2\,\pa_\m\f\,\pa_\n\f +
\fr18\,g_{\m\n}\,(g^2\,{\rm e}^{\sqrt{2}\f} + 4\,g\,m\,{\rm
e}^{-\sqrt{2}\f} - m^2\,{\rm
e}^{-3\sqrt{2}\f})\nn\\ 
& & + {\rm e}^{2\sqrt{2}\f}\left(G_\m^{\;\;\r\s}\,G_{\n\r\s} -
\fr16\,g_{\m\n}\, G^{\r\s\t}\,G_{\r\s\t}\right)\nn\\
& & - 
2\,{\rm e}^{-\sqrt{2}\f}\left({\cal H}_\m^{\;\;\r}\,{\cal
H}_{\n\r} - \fr18\,g_{\m\n}\,{\cal H}_{\r\s}\,{\cal H}^{\r\s}\right)\nn\\
& & - 2\,{\rm e}^{-\sqrt{2}\f}\left(F_\m^{I\;\r}\,F^I_{\n\r} -
\fr18\,g_{\m\n}\,F_{\r\s}^I\,F^{I\;\r\s}\right)\;.
\label{6d-romans-einstein-eq}
\eea
The dilaton equation is
\bea
\Box\f &=& \fr1{4\sqrt{2}}\,(g^2\,{\rm e}^{\sqrt{2}\f} -
4\,m\,g\,{\rm e}^{-\sqrt{2}\f} + 3\,m^2\,{\rm e}^{-3\sqrt{2}\f}) 
+ \fr1{3\sqrt{2}}\,{\rm e}^{2\sqrt{2}\f}
G^{\m\n\r}\,G_{\m\n\r}\nn\\
& & + \fr1{2\sqrt{2}}\,{\rm
e}^{-\sqrt{2}\f}\,({\cal H}^{\m\n}\,{\cal H}_{\m\n} +
F^{I\;\m\n}\,F_{\m\n}^I)\;,
\label{6d-romans-scalar-eq}
\eea
and the equations of motion for gauge fields are
\bea
{\cal D}_\n\,({\rm e}^{-\sqrt{2}\f}\,{\cal H}^{\n\m}) &=&
\fr16\,e\,\varepsilon^{\m\n\r\s\t\k}\,{\cal H}_{\n\r}\,G_{\s\t\k}\;,
\label{6d-romans-abelian-eq}\\
{\cal D}_\n\,({\rm e}^{-\sqrt{2}\f}\,F^{I\;\n\m}) &=&
\fr16\,e\,\varepsilon^{\m\n\r\s\t\k}\,F^I_{\n\r}\,G_{\s\t\k}\;,
\label{6d-romans-nonabelian-eq}\\
{\cal D}_\r\,({\rm e}^{2\sqrt{2}\f}\,G^{\r\m\n}) &=& - m\,{\rm
e}^{-\sqrt{2}\f}\,{\cal H}^{\m\n} -
\fr14\,e\,\varepsilon^{\m\n\r\s\t\k}\,({\cal H}_{\r\s}\,{\cal
H}_{\t\k} + F^I_{\r\s}\,F^I_{\t\k})\;.
\label{6d-romans-2form-eq}
\ea
Depending upon the values of the gauge coupling and mass
parameter, there are five distinct theories: ${\cal N} = 4^+
$ (for $g > 0, m>0$), ${\cal N} = 4^-$ (for $g
< 0, m > 0$), ${\cal N} = 4^g$ (for $g > 0, m =
0$), ${\cal N} = 4^m$ (for $g = 0, m > 0$),
and ${\cal N} = 4^0\,\,\,$ (for $g = 0, m = 0$). The ${\cal N} =
4^g$ theory coincides with a theory~\cite{Giani:1984dw}
obtained by dimensional reduction of gauged ${\cal N} = 2$
supergravity in seven dimensions, followed by truncation. It was
pointed out in~\cite{Romans:1986tw} that there is a dual version
of the ${\cal N} = 4^g$ theory, which has a similar field content
but cannot be obtained from the ${\cal N} = 4^g$ theory by a field
redefinition. The $B_{\m\n}$ field is replaced by a new tensor
field $A_{\m\n}$, whose field strength is $\tilde{F}_{\m\n}^I$. In
absence of the Abelian field strength, the bosonic Lagrangian of
the dual ${\cal N} = \tilde{4}^g$ theory is
\bea
\tilde{e}^{-1}\,\tilde{{\cal L}} &=& -\fr14\,\tilde{R} +
\fr12\,\del_\m\tilf\del^\m\tilf - \fr14\,{\rm
e}^{-\sqrt{2}\tilf}\,\tilde{F}_{\m\n}^I\tilde{F}^{I\;\m\n}\nn\\
& & +
\fr1{12}\,{\rm e}^{-2\sqrt{2}\tilf}\,
\tilde{F}_{\m\n\r}\tilde{F}^{\m\n\r} + \fr18\,\tilde{g}^2\,{\rm
e}^{\sqrt{2}\tilf}\;, 
\label{6d-dual-lag} 
\eea
where
\be 
\tilde{F}_{\m\n\r}\equiv 3\,\left(\del_{[\m}\,A_{\n\r]} -
F^I_{[\m\n}\,A^I_{\r]} -
\fr13\,g\,\epsilon^{IJK}\,A^I_\m\,A^J_\n\,A^K_\r\right)\;. 
\ee
The Lagrangian~(\ref{6d-dual-lag}) can be obtained from the
Lagrangian~(\ref{6d-romans-lag}) by formally writing
$\tilde{F}_{\m\n\r}$ as follows
\be 
\tilde{F}_{\m\n\r} = \fr16\,{\rm
e}^{2\sqrt{2}\,\tilf}\,e\,\varepsilon_{\m\n\r\s\t\k}\,G^{\s\t\k}
\;. 
\label{6d-dualization} 
\ee
Since the Lagrangian~(\ref{6d-dual-lag}) differs from ${\cal N} =
4^g$ theory only by a sign of dilaton coupling to three-form
field, in the absence of the three-form field ${\cal N} = 4^g$
and ${\cal N} = \tilde{4}^g$ are identical. Furthermore, the
Eq.~(\ref{6d-dualization}) tells us that any solution of ${\cal
N}=4^g$ theory with the two-form fields not being excited,
which is also a solution of ${\cal N} = \tilde{4}^g$ theory,
can be up-lifted into higher dimensional theory. 

After re-scaling, the Lagrangian~(\ref{6d-romans-lag}) can be written 
in terms of form notation
\bea
\cL &=& R{*\oneone} - \fr12 {*d}\f\wedge d\f -
\fr12\,\rme^{\fr1{\sqrt{2}}\f} ({*{\cal H}}_2\wedge {\cal H}_\2 +
{*F}^I_\2\wedge F^i_\2)\nn\\
& & - \fr12\,\rme^{-\sqrt{2}\f}\,{*G}_\3\wedge G_\3 + 
\fr1{16}(4g^2\,\rme^{-\fr1{\sqrt{2}}\f}+2mg\,\rme^{\fr1{\sqrt{2}}\f} -
m^2\,\rme^{-\fr{3}{\sqrt{2}}\f}){*\oneone}\nn\\
& & - B_\2\wedge \left(\fr12
\cF_\2\wedge \cF_\2 \right. \nn\\
& & \left. + \fr12 m B_\2\wedge \cF_\2
+ \fr13 m^2 B_\2\wedge B_\2 + \fr12 F_\2^I\wedge F^I_\2\right)\;.
\la{6d-romans-lag-form}
\eea
The dilaton equation in terms of form notation is
\bea
d{*d}\f &=& \fr1{\sqrt{2}}\rme^{-\sqrt{2}\f}\,{*G}_\3\wedge G_\3 - 
\fr1{2\sqrt{2}}\,\rme^{\fr1{\sqrt{2}}\f}({*{\cal H}}_\2\wedge {\cal H}_\2 
+ {*F}_\2^I\wedge F_\2^I) \nn\\
& & - \fr1{16}\left(\fr3{\sqrt{2}}\,m^2\,\rme^{-\fr3{\sqrt{2}}\f} + 
\sqrt{2}\,m\,g\,\rme^{\fr1{\sqrt{2}}\f} -
2\sqrt{2}\,g^2\,\rme^{-\fr1{\sqrt{2}}\f}\right){*\oneone}\;.  
\label{6d-scalar-eq-form}
\eea
the equations of motion for gauge fields are
\bea
d(\rme^{-\sqrt{2}\f}\,{*G}_\3) &=& -\fr12\,{\cal H}_\2\wedge {\cal H}_\2
- \fr12\,F^I_\2\wedge F^I_\2 - m\,\rme^{\fr1{\sqrt{2}}\f}\,{*{\cal
H}}_\2\;,\nn\\
d(\rme^{\fr1{\sqrt{2}}\f}\,{*{\cal H}}_\2) &=& - G_\3\wedge {\cal
H}_\2\;,\nn\\
{\cal D}(\rme^{-\fr1{\sqrt{2}}\f}\,{*F}^I_\2) &=& - F^I_\2\wedge G_\3\;,
\label{6d-gauge-eq-form}
\eea
and the Einstein equation is
\bea
R_{\m\n} &=& -\fr12\,\pa_\m\f\pa_\n\f +
\left(\fr18\,m^2\,\rme^{\fr3{\sqrt{2}}\f}-m\,g\,\rme^{\fr1{\sqrt{2}}\f}
- \fr12\,g^2\,\rme^{-\fr1{\sqrt{2}}\f}\right)\,g_{\m\n}\nn\\
& & + \fr14\,\rme^{-\sqrt{2}\f}\left(G_{\m\r\s}G^{\;\;\r\s}_\n -
\fr16\,g_{\m\n}\, G_{\r\s\t}G^{\r\s\t}\right)\nn\\
& & +
\fr12\,\rme^{\fr1{\sqrt{2}}\f}\left({\cal H}_{\m\r}{\cal
H}^{\;\;\r}_\n - \fr18\,g_{\m\n}\,{\cal H}_{\r\s}{\cal
H}^{\r\s}\right)\nn\\
& & + \fr12\,\rme^{\fr1{\sqrt{2}}\f}\left(F^I_{\m\r}F^{I\;\;r}_\n -
\fr18\,g_{\m\n}\,F^I_{\r\s}F^{I\;\r\s}\right)\;.
\label{6d-einstein-eq-can}
\eea
As shown in~\cite{Cvetic:1999un}, the six-dimensional Romans'
theory~(\ref{6d-romans-lag-form}) can be obtained by reducing massive 
type IIA on a local $S^4$. We present the results which are written in
terms of two parameters $m$ and $g$. The metric ansatz is
\bea
d\hat{s}_{10}^2 &=&
\left(\fr1{3mg^2}\right)^{1/8}\,(\sin\xi)^{1/12}\,X^{1/8}\,\left[
-\fr12\,\Delta^{3/8}\,ds_6^2 + g^2\,\Delta^{3/8}\,X^2\,d\xi^2\right.\nn\\
& &\left. + \fr1{g}\,\Delta^{-5/8}\,X^{-1}\,\cos^2\xi\,\sum_{i=1}^3(\sigma^i
- g\,A_\1^i)^2\right]\;.
\label{massive-iia-metric-ans-s4}
\eea
The ans\"atze for 4-form, 3-form and 2-form are
\bea
\hat{F}_\4 &=&
\left(\fr1{3mg^3}\right)^{3/4}\,\left[
-\fr23\,s^{1/3}\,c^3\,\Delta^{-2}\,U\,d\xi\wedge\ep_\3 -
12mg\,s^{4/3}\,c^4\,\Delta^{-2}\,X^{-3}\,dX\wedge\ep_\3
\right.
\nn\\
& & + 2g^2\,c\,s^{1/3}\,X^4 {*G}_\3\wedge d\xi -
\fr{3mg^2}{2}\,s^{4/3}\,X^{-2} {*{\cal H}}_\2\nn\\
& & \left. + 2g\,c\,s^{1/3}\,F^i_\2\wedge h^i\wedge d\xi -
\fr{3mg}{2}\,c^2\,s^{4/3}\,\Delta^{-1}\,X^{-3}\,F_\2^i\wedge
h^j\wedge h^k\,\ep_{ijk}\right]\;,\nn\\
\hat{F}_\2 &=&
\left(\fr{2g}{3m}\right)^{1/4}\,s^{2/3}\,{\cal
H}_\2\;,\nn\\
\hat{F}_\3 &=& \left(\fr1{3mg}\right)^{1/2}\,[3m\,s^{2/3}\,F_\3 +
2\,c\,s^{-1/3}\,{\cal H}_\2\wedge d\xi]\;,\nn\\
\rme^{\,\hat{\f}} &=&
\left(\fr{3m}{g^2}\right)^{1/4}\,s^{-5/6}\,\Delta^{1/4}\,X^{-5/6}\;, 
\label{massive-iia-forms-ans-s4}
\eea
where 
\bea
s &\equiv& \sin\xi\;,\;\; c\equiv\cos\xi\;,\;\; X =
\rme^{-\frac{1}{2\sqrt{2}}\f}\;,\nn\\
\Delta &\equiv& 2g\,X\,c^2 +
3m\,X^{-3}\,s^2\;,\;\;\; h^i \equiv \sigma^1 - g\,A_\1^i\;,\nn\\
U &\equiv& 9m^2\,s^2\,X^{-6} - 3g^2\,c^2\,X^2 + 12mg\,X^{-2}\,c^2 - 
18mg\,X^{-2}\;.
\eea

%% file: appendF.tex
\appendix{Five-dimensional $\cN =1, SU(2)\times U(1)$-gauged supergravity}

The bosonic sector of the $\cN=4$ gauged supergravity in 
$D=5$~\cite{Romans:1986ps} consists of the metric, a
 scalar,  the $SU(2)$ Yang-Mills  potentials $A^i_\m$, a $U(1)$ gauge
 potential $\cA_\m$, and two 2-form potentials
 $B^\a_{\m\n}$ which transform as a charged doublet under the $U(1)$.
The Lagrangian is given by
\bea 
e^{-1}{\cal L} &=& -\fr14\,R + \fr12\,(\pa_\m\f)(\pa^\m\f) -
\fr14\,\rme^{-\fr8{\sqrt{6}}\f}\,\cF_{\m\n}\cF^{\m\n} -
\fr14\,\rme^{\fr4{\sqrt{6}}\f}\,(F^I_{\m\n}F^{I\;\m\n} +
B^\a_{\m\n}B^{\a\;\m\n})\nn\\
& & + \fr18\,g_2(g_2\,\rme^{-\fr4{\sqrt{6}}\f} +
2\sqrt{2}\,g_1\,\rme^{\fr2{\sqrt{6}}\f})\nn\\
& & +
\fr14\,e^{-1}\,\vep^{\m\n\r\s\t}\left(\fr1{g_1}\,\vep_{\a\b}B^\a_{\m\n}{\cal
D}_\r B^\b_{\s\t} - F^I_{\m\n}\,F^I_{\r\s}\,\cA^\t\right)\;,
\label{5d-romans-lag}
\eea
where $e$ is the determinant of the vielbein, $g_1$ and $g_2$ are the
gauged couplings of $U(1)$ and $SU(2)$ respectively, and
$\vep_{\m\n\r\s\t}$ is a Levi-Civita tensor density. The Abelian field
strength $\cF_{\m\n}$ and non-Abelian field strength $F_{\m\n}^I$ are
given by
\be
\cF_{\m\n}\equiv \pa_\m\,\cA_\n - \pa_\n\,\cA_\m\;,\;\;\; F_{\m\n}^I \equiv
\pa_\m\,A^I_\n - \pa_\n\,A^I_\m + g\,\vep^{IJK}\,A_\m^J\,A_\n^k\;.
\ee
The supersymmetry transformation for the gravitinos is
\be
\d\psi_{\m\,a} = {\cal D}_\m\,\ep_a + \G_\m\,T_{ab}\,\ep^b -
\fr1{6\sqrt{2}}(\G_\m^{\;\;\n\r} -
4\,\d_\m^{\;\;\n}\,\G^\r)\left(H_{\n\r\;ab} +
\fr1{\sqrt{2}}\,h_{\n\r\;ab}\right) \ep^b\;,
\label{5d-romans-susy-gravitinos-transf}
\ee
and that for the gauginos is
\be
\d\chi_a = \fr1{\sqrt{2}}\G^\m\pa_\m\f\,\ep_a + A_{ab}\,\ep^b -
\fr1{2\sqrt{6}}\,\G^{\m\n}(H_{\m\n\;ab} -
\sqrt{2}\,h_{\m\n\;ab})\,\ep^b\;,
\label{5d-romans-susy-gauginos-transf}
\ee
where $T_{ab}$, $A_{ab}$, $H_{\m\n\;ab}$ and $h_{\m\n\;ab}$ are
defined as follows
\bea
T^{ab} &\equiv& \left(\fr1{6\sqrt{2}}\,g_2\,\rme^{-\fr2{\sqrt{6}}\f} +
\fr1{12}\,g_1\,\rme^{\fr4{\sqrt{6}}\f}\right)(\G_{45})^{ab}\;,\nn\\
A^{ab} &\equiv& \left(\fr1{2\sqrt{6}}\,g_2\,\rme^{-\fr2{\sqrt{6}}\f} -
\fr1{2\sqrt{3}}\,g_1\,\rme^{\fr4{\sqrt{6}}\f}\right)(\G_{45})^{ab}\;,\nn\\
H_{\m\n}^{ab} &\equiv& \rme^{\fr2{\sqrt{6}}\f}\,[F^I_{\m\n}(\G_I)^{ab}
+ B^\a_{\m\n}(\G_\a)^{ab}]\;,\;\; h_{\m\n}^{ab} \equiv
\rme^{-\fr4{\sqrt{6}}\f}\,\W^{ab}\,\cF_{\m\n}\;,
\label{5d-romans-tahH-def}
\eea
and $\W^{ab}$ is a charge conjugation matrix. The gauge covariant
derivative ${\cal D}_\m$ acting on Killing spinor is defined by
\be
{\cal D}_\m\,\ep_a = \nabla_\m\,\ep_a +
\fr12\,g_1\,\cA_\m\,(\G_{45})_a^{\;\;b}\,\ep_b +
\fr12\,g_2\,A^I_\m\,(\G_{45})_a^{\;\;b}\,\ep_b\;,
\ee
with $\nabla_\m$ is the usual space-time derivative
\[ \nabla_\m\,\ep_a \equiv \left(\pa_\m +
\fr14\,\w_\m^{\;\a\b}\,\G_{\a\b}\right)\,\ep_a\;, \]
where $\w_\m^{\;\;\a\b}$ is the spin connection. Indices $\a$ and $\b$ are
tangent indices while $\m$ and $\n$ are space-time indices. The
$\G_{\a\b\cdots}$ are the five-dimensional Dirac matrices,
\be
\G_{\a_1\cdots\a_n} \equiv
\fr1{n!}\,\G_{[\a_1}\cdots\G_{\a_n]}\;,\;\;\; n = 1, \ldots, 5\;.
\ee
The Einstein equation of the Lagrangian~(\ref{5d-romans-lag}) is
\bea
R_{\m\n} &=& 2\,\pa_\m\f\,\pa_\n\f +
\fr16\,g_{\m\n}\,(g_2^2\,\rme^{-\fr4{\sqrt{6}}\f} +
2\sqrt{2}\,g_1g_2\,\rme^{\fr2{\sqrt{6}}\f})\nn\\
& & + \rme^{\fr4{\sqrt{6}}\f}\left[2\,F^I_{\m\r}F_\n^{I\;\r} +
2\,B^\a_{\m\r}B^{\a\;\r}_\n -
\fr13\,g_{\m\n}(F^I_{\r\s}F^{I\;\r\s}+B^\a_{\r\s}B^{\a\;\r\s})\right]\nn\\
& & +
\rme^{-\fr8{\sqrt{6}}\f}\left(2\,\cF_{\m\r}\cF_\n^{\;\;\r} -
\fr13\,g_{\m\n}\,\cF_{\r\s} \cF^{\r\s}\right)\;,
\label{5d-romans-einstein-eq}
\eea
the dilaton equation is
\bea
\Box\f &=& \left(\fr1{2\sqrt{3}}\,g_1g_2\,\rme^{\fr2{\sqrt{6}}\f} -
\fr1{2\sqrt{6}}\,g_2^2\,\rme^{-\fr4{\sqrt{6}}\f}\right) +
\fr2{\sqrt{6}}\,\rme^{-\fr8{\sqrt{6}}\f}\,\cF_{\m\n}\cF^{\m\n}\nn\\
& & -
\fr1{\sqrt{6}}\,\rme^{\fr4{\sqrt{6}}\f}(F^I_{\m\n}F^{I\;\m\n} + 
B^\a_{\m\n}B^{\a\;\m\n})\;,
\label{5d-romans-scalar-eq}
\eea
the equations for the gauge fields are
\bea
{\cal D}_\n(\rme^{-\fr8{\sqrt{6}}\f}\,\cF^{\n\m}) &=&
\fr14\,e^{-1}\,\ep^{\m\n\r\s\t} (F^I_{\n\r}\,F^I_{\s\t} +
B^\a_{\n\r}\,B^\a_{\s\t})\;,\nn\\
{\cal D}_\n(\rme^{\fr4{\sqrt{6}}\f}\,F^{I\;\n\m}) &=&
\fr12\,e^{-1}\,\vep^{\m\n\r\s\t}\,F^I_{\n\r}\,\cF_{\s\t}\;,\nn\\
\rme^{\fr4{\sqrt{6}}\f}\,(B^\a)^{\m\n} &=&
\fr1{g_1}\,e^{-1}\,\vep^{\m\n\r\s\t}\,\vep^{\a\b}\,{\cal
D}_\r\,B^\b_{\s\t}\;.
\label{5d-romans-gauge-eq}
\eea
Note that the fields $B^\a_\2$ satisfy a first-order equation of
motion, of the kind referred to as ``odd-dimensional self-duality''
in \cite{Townsend:1984xs}.

   As discussed in \cite{Romans:1986ps}, there are three in-equivalent
theories, corresponding to different choices for the gauge couplings
in (\ref{5d-romans-lag}): $\cN=4^0$ in which $g_2=0$; $\cN=4^+$ in which
$g_2=\sqrt{2}g_1$ and $\cN=4^-$ in which $g_2=-\sqrt{2}g_1$ (see also
\cite{Gunaydin:1986cu}).  The $\cN=4^+$ theory is obtained by truncating 
the gauged $SO(6),\ \cN=8$ supergravity theory in five dimensions, while the
$\cN=4^0$ and $\cN=4^-$ theories arise as truncations of non-compact $\cN=8$
supergravities. From the Lagrangian of the
Romans' theory one might conclude that it is not possible to set the
$U(1)$ coupling constant $g_1$ to be zero. However, as was pointed out
in \cite{Cowdall:1998rs}, after appropriate rescalings the limit can be
taken.

In the language of differential form, the
Lagrangian~(\ref{5d-romans-lag}) can be written as
\bea
{\cal L} &=& R\, {*\oneone} - 3 X^{-2} {*dX}\wedge dX
-\ft12 X^4\,  {*\cF_\2}\wedge \cF_\2
-\ft12 X^{-2}\, ({*F^i_\2}\wedge F^i_\2 + {*B^\a_\2}\wedge B^\a_\2)\nn\\
& & +\fr1{2g_1} \epsilon_{\a\b}\, B^\a_\2\wedge dB^\b_\2 - \ft12
B^\a_\2\wedge B_\2^\a\wedge \cA_\1 - \ft12 F^i_\2\wedge F^i_\2\wedge 
\cA_\1\nn\\
& &  + 2 g_2( g_2\, X^2 + 2\sqrt{2} g_1\, X^{-1})\, {*\oneone}\;,
\label{5d-romans-lag-form}
\eea
where $X$ parameterizes the scalar degree of freedom, and can be
written in terms of a canonically-normalized dilaton $\phi$ as
$X=\rme^{-\ft1{\sqrt6}\,\phi}$.  The 2-form field strengths are given by
\be
F_\2^i = dA_\1^i + \ft12 g_2\, \ep^{ijk}\, A_\1^j\wedge
A_\1^k\;,\qquad
\cF_\2= d\cA_\1\;.
\ee
    Without loss of generality we may set
$g_1=g$ and $g_2=\sqrt{2}g$, since the two independent gauge coupling
constants may be recovered by appropriate rescalings.  We also find it
advantageous to adopt a complex notation for the two 2-form
potentials, which form a charged doublet with respect to the
gauge field $\cA_\1$.  Thus we define
\be
B_\2 \equiv B_\2^1 + \im\, B_\2^2\;.
\label{5d-romans-complexa}
\ee
The equations of motion in differential form notations are
\bea
d(X^{-1}\, {*dX}) &=& \ft13 X^4\,  {*\cF_\2}\wedge \cF_\2 -
\ft16 X^{-2} \, ({*F^i_\2}\wedge F^i_\2 + {*{\bar B}_\2}\wedge B_\2)\nn\\
& & - \ft{4}{3}g^2\, (X^2 - X^{-1})\, {*\oneone}\;,\nn\\
d(X^4\, {*\cF_\2}) &=& - \ft12 F^i_\2\wedge F^i_\2 -
\ft12 {\bar B}_\2\wedge B_\2\;,\nn\\
d(X^{-2}\, {*F^i_\2}) &=& \sqrt{2}\, g \,
\epsilon^{ijk}\, X^{-2}\, {*F^j_\2}\wedge A^k_\1
- F^i_\2\wedge \cF_\2\;,\nn\\
X^{2}\, {*F_\3} &=& - \im \, g\, B_\2 \;,\nn\\
R_{\m\n} &=& 3 X^{-2}\,  \pa_\m X\, \pa_\n X - \ft{4}{3}g^2\,(X^2 +
2  X^{-1})\, g_{\m\n}\nn\\
& & + \ft12 X^4 \, (\cF_\m{}^\r \cF_{\n \r} -\ft16 g_{\m\n} \,
\cF_\2^2) + \ft12 X^{-2}\,
(F^{i\ \r}_\m F^{i}_{\n\r}  - \ft16 g_{\m\n}\, (F^i_\2)^2)\nn\\
& & + \ft12 X^{-2}\,  ({\bar B}_{(\m}{}^\r\,  B_{\n)\r} - \ft16
g_{\m\n}\, |B_\2|^2)\;,
\label{5d-romans-eqs-form}
\eea
where
\be
F_\3 ={\cal D}B_\2 \equiv dB_\2  -\im\, g\, \cA_\1\wedge B_\2\;.
\label{5d-romans-gauge-def}
\ee
The operator ${\cal D}$ defined in this equation is the $U(1)$ gauge-covariant
exterior derivative.

%% file: vita.tex
\vita{Tuan Anh Tran received a Bachelor of Science in Physics
from Hanoi University, Vietnam, in
July 1992. From 1993 to 1994, he was a graduate student at the 
Institute of Physics, National Center for Science and Technology, 
Hanoi, Vietnam.
He joined the Diploma Program at the Abdus Salam International
Center for Theoretical Physics (ICTP), Trieste, Italy in September 1994 
and completed the program in August 1995. After spending most of his time 
from September 1995 to August 1996 as a visiting researcher at ICTP, he 
joined the graduate program at Texas A\&M University in September 1996.
His address is the Institute for Fundamental Theory, 
University of Florida, P.O. Box 118440, Gainesville, FL 32611-8440. 

}

\typist{Tuan A. Tran}

%% file: thesis.bbl
\begin{thebibliography}{100}

\bibitem{Green:1987sp}
M.B. Green, J.H. Schwarz,  E.~Witten, Superstring Theory, vol.~1-2.
Cambridge University Press, Cambridge, 1987.

\bibitem{Polchinski:1998rq}
J.~Polchinski, String Theory, vol.~1-2.
Cambridge University Press, Cambridge, 1998.

\bibitem{Cremmer:1978km}
E.~Cremmer, B.~Julia,  J.~Scherk,  Supergravity theory in 11 dimensions,  Phys.
  Lett.  B 76 (1978) 409--412.

\bibitem{Schwarz:1983qr}
J.H. Schwarz,  Covariant field equations of chiral {${\cal N}=2$} {$D = 10$}
  supergravity,  Nucl. Phys.  B 226 (1983) 269--288.

\bibitem{Duff:1986hr}
M.J. Duff, B.E.W. Nilsson,  C.N. Pope,  Kaluza-{K}lein supergravity,  Phys.
  Rept.  130 (1986) 1--142.

\bibitem{Salam:1989fm}
A.~Salam,  E.~Sezgin, Supergravities in Diverse Dimensions, vol.~1-2.
North-Holland/World Scientific, Singapore, 1989.

\bibitem{Cremmer:1998ct}
E.~Cremmer, B.~Julia, H.~L{\"u},  C.N. Pope,  Dualisation of dualities. {I},
  Nucl. Phys.  B 523 (1998) 73--144, hep-th/9710119.

\bibitem{Cremmer:1998em}
E.~Cremmer, I.V. Lavrinenko, H.~L{\"u}, C.N. Pope, K.S. Stelle,  T.A. Tran,
  Euclidean-signature supergravities, dualities and instantons,  Nucl. Phys.  B
  534 (1998) 40--82, hep-th/9803259.

\bibitem{Hull:1998br}
C.M. Hull,  B.~Julia,  Duality and moduli spaces for timelike reductions,
  Nucl. Phys.  B 534 (1998) 250--260, hep-th/9803239.

\bibitem{Duff:1983gq}
M.J. Duff,  C.N. Pope,  Kaluza-{K}lein supergravity and the seven sphere,  in
  Supersymmetry and {S}upergravity (S.~Ferrara, J.G. Taylor,  P.~van
  Nieuwenhuizen, eds.), World Scientific, Singapore, 1983.

\bibitem{Pilch:1984xy}
K.~Pilch, P.~van Nieuwenhuizen,  P.K. Townsend,  Compactification of {$D = 11$}
  supergravity on {$S^4$} (or {$11 = 7 + 4$}, too),  Nucl. Phys.  B 242 (1984)
  377--392.

\bibitem{Gunaydin:1985fk}
M.~G{\"u}naydin,  N.~Marcus,  The spectrum of the {$S^5$} compactification of
  the chiral {$\cN = 2$}, {$D = 10$} supergravity and the unitary
  supermultiplets of {$U(2, 2/4)$},  Class. Quantum Grav.  2 (1985) L11--L17.

\bibitem{Kim:1985ez}
H.J. Kim, L.J. Romans,  P.~van Nieuwenhuizen,  The mass spectrum of chiral
  {${\cal N} = 2$ $D = 10$} supergravity on {$S^5$},  Phys. Rev.  D 32 (1985)
  389--399.

\bibitem{Pope:1987ad}
C.N. Pope,  K.S. Stelle,  Zilch currents, supersymmetry and {K}aluza-{K}lein
  consistency,  Phys. Lett.  B 198 (1987) 151--155.

\bibitem{deWit:1987iy}
B.~de~Wit,  H.~Nicolai,  The consistency of the {$S^7$} truncation in {$D =
  11$} supergravity,  Nucl. Phys.  B 281 (1987) 211--240.

\bibitem{Maldacena:1998re}
J.M. Maldacena,  The large {$N$} limit of superconformal field theories and
  supergravity,  Adv. Theor. Math. Phys.  2 (1998) 231--252, hep-th/9711200.

\bibitem{Gubser:1998bc}
S.S. Gubser, I.R. Klebanov,  A.M. Polyakov,  Gauge theory correlators from
  non-critical string theory,  Phys. Lett.  B 428 (1998) 105--114,
  hep-th/9802109.

\bibitem{Witten:1998qj}
E.~Witten,  Anti-de {S}itter space and holography,  Adv. Theor. Math. Phys.  2
  (1998) 253--291, hep-th/9802150.

\bibitem{Cvetic:1999xp}
M.~Cveti\v{c}, M.J. Duff, P.~Hoxha, J.T. Liu, H.~L{\"u}, J.X. Lu,
  R.~Martinez-Acosta, C.N. Pope, H.~Sati,  T.A. Tran,  Embedding {$AdS$} black
  holes in ten and eleven dimensions,  Nucl. Phys.  B 558 (1999) 96--126,
  hep-th/9903214.

\bibitem{Nastase:1999cb}
H.~Nastase, D.~Vaman,  P.~van Nieuwenhuizen,  Consistent nonlinear {KK}
  reduction of {11D} supergravity on {$AdS_7\times S^4$} and self-duality in
  odd dimensions,  Phys. Lett.  B 469 (1999) 96--102, hep-th/9905075.

\bibitem{Nastase:1999kf}
H.~Nastase, D.~Vaman,  P.~van Nieuwenhuizen,  Consistency of the {$AdS_7\times
  S^4$} reduction and the origin of self-duality in odd dimensions,  Nucl.
  Phys.  B 581 (2000) 179--239, hep-th/9911238.

\bibitem{Cvetic:1999un}
M.~Cveti\v{c}, H.~L{\"u},  C.N. Pope,  Gauged six-dimensional supergravity from
  massive type {IIA},  Phys. Rev. Lett.  83 (1999) 5226--5229, hep-th/9906221.

\bibitem{Cvetic:2000cj}
M.~Cveti{\v{c}}, H.~L{\"u}, C.N. Pope,  J.F. Vazquez-Poritz,  {AdS} in warped
  spacetimes,  Phys. Rev.  D 62 (2000) 122003, hep-th/0005246.

\bibitem{Cvetic:2000yp}
M.~Cveti{\v{c}}, H.~L{\"u},  C.N. Pope,  Consistent warped-space
  {K}aluza-{K}lein reductions, half-maximal gauged supergravities and {$CP^n$}
  constructions,  Nucl. Phys.  B 597 (2001) 172--196, hep-th/0007109.

\bibitem{Cvetic:2000ah}
M.~Cveti\v{c}, H.~L{\"u}, C.N. Pope, A.~Sadrzadeh,  T.A. Tran,  {$S^3$} and
  {$S^4$} reductions of type {IIA} supergravity,  Nucl. Phys.  B 590 (2000)
  233--251, hep-th/0005137.

\bibitem{Romans:1986tw}
L.J. Romans,  The {$F(4)$} gauged supergravity in six dimensions,  Nucl. Phys.
  B 269 (1986) 691--711.

\bibitem{Lu:1999bc}
H.~L{\"u},  C.N. Pope,  Exact embedding of {${\cal N} = 1, D = 7$} gauged
  supergravity in {$D = 11$},  Phys. Lett.  B 467 (1999) 67--72,
  hep-th/9906168.

\bibitem{Lu:1999bw}
H.~L{\"u}, C.N. Pope,  T.A. Tran,  Five-dimensional {${\cal N} = 4$,
  $SU(2)\times U(1)$} gauged supergravity from type {IIB},  Phys. Lett.  B 475
  (2000) 261--268, hep-th/9909203.

\bibitem{Cvetic:1999au}
M.~Cveti\v{c}, H.~L{\"u},  {C.N.} Pope,  Four-dimensional {${\cal N} = 4,
  SO(4)$} gauged supergravity from {$D = 11$},  Nucl. Phys.  B 574 (2000)
  761--781, hep-th/9910252.

\bibitem{Lu:2000xc}
H.~Lu,  C.N. Pope,  Branes on the brane,  Nucl. Phys.  B 598 (2001) 492--508,
  hep-th/0008050.

\bibitem{Cvetic:2000gj}
M.~Cveti{\v{c}}, H.~L{\"u},  C.N. Pope,  Brane-world {K}aluza-{K}lein
  reductions and branes on the brane, hep-th/0009183.

\bibitem{Cvetic:1999xx}
M.~Cveti\v{c}, S.S. Gubser, H.~L{\"u},  C.N. Pope,  Symmetric potentials of
  gauged supergravities in diverse dimensions and coulomb branch of gauge
  theories,  Phys. Rev.  D 62 (2000) 086003, hep-th/9909121.

\bibitem{Cvetic:2000eb}
M.~Cveti\v{c}, H.~L{\"u}, C.N. Pope,  A.~Sadrzadeh,  Consistency of
  {K}aluza-{K}lein sphere reductions of symmetric potentials,  Phys. Rev.  D 62
  (2000) 046005, hep-th/0002056.

\bibitem{Cvetic:2000tb}
M.~Cveti\v{c}, H.~L{\"u},  C.N. Pope,  Geometry of the embedding of
  supergravity scalar manifolds in {$D = 11$} and {$D = 10$},  Nucl. Phys.  B
  584 (2000) 149--170, hep-th/0002099.

\bibitem{Cvetic:2000nc}
M.~Cveti\v{c}, H.~L{\"u}, {C.N.} Pope, A.~Sadrzadeh,  T.A. Tran,  Consistent
  {$SO(6)$} reduction of type {IIB} supergravity on {$S^5$},  Nucl. Phys.  B
  586 (2000) 275--286, hep-th/0003103.

\bibitem{Park:2000du}
I.Y. Park, A.~Sadrzadeh,  T.A. Tran,  Super {Y}ang-{M}ills operators from the
  {D3}-brane action in a curved background,  Phys. Lett.  B 497 (2001)
  303--308, hep-th/0010116.

\bibitem{Lee:1998bx}
S.~Lee, S.~Minwalla, M.~Rangamani,  N.~Seiberg,  Three-point functions of
  chiral operators in {$D = 4$}, {$\cN = 4$} {SYM} at large {$N$},  Adv. Theor.
  Math. Phys.  2 (1998) 697--718, hep-th/9806074.

\bibitem{Bershadsky:1996qy}
M.~Bershadsky, C.~Vafa,  V.~Sadov,  D-branes and topological field theories,
  Nucl. Phys.  B 463 (1996) 420--434, hep-th/9511222.

\bibitem{Maldacena:2000mw}
J.M. Maldacena,  C.~N{\'u\~ne}z,  Supergravity description of field theories on
  curved manifolds and a no go theorem,  Int. J. Mod. Phys.  A 16 (2001)
  822--855, hep-th/0007018.

\bibitem{Maldacena:2000yy}
J.M. Maldacena,  C.~N{\'u\~n}ez,  Towards the large {$N$} limit of pure {${\cal
  N} = 1$} super {Y}ang {M}ills,  Phys. Rev. Lett.  86 (2001) 588--591,
  hep-th/0008001.

\bibitem{Alishahiha:1999ds}
M.~Alishahiha,  Y.~Oz,  {AdS/CFT} and {BPS} strings in four dimensions,  Phys.
  Lett.  B 465 (1999) 136--141, hep-th/9907206.

\bibitem{Fayyazuddin:1999zu}
A.~Fayyazuddin,  D.J. Smith,  Localized intersections of {M}-fivebranes and
  four-dimensional superconformal field theories,  JHEP  04 (1999) 030,
  hep-th/9902210.

\bibitem{Fayyazuddin:2000em}
A.~Fayyazuddin,  D.J. Smith,  Warped {AdS} near-horizon geometry of completely
  localized intersections of {M5}-branes,  JHEP  10 (2000) 023, hep-th/0006060.

\bibitem{Brinne:2000fh}
B.~Brinne, A.~Fayyazuddin, S.~Mukhopadhyay,  D.J. Smith,  Supergravity
  {M5}-branes wrapped on {R}iemann surfaces and their {QFT} duals,  JHEP  12
  (2000) 013, hep-th/0009047.

\bibitem{Klebanov:2000nc}
I.R. Klebanov,  A.A. Tseytlin,  Gravity duals of supersymmetric {$SU(N)\times
  SU(N+M)$} gauge theories,  Nucl. Phys.  B 578 (2000) 123--138,
  hep-th/0002159.

\bibitem{Klebanov:2000hb}
I.R. Klebanov,  M.J. Strassler,  Supergravity and a confining gauge theory:
  {D}uality cascades and {$\chi$SB}-resolution of naked singularities,  JHEP
  08 (2000) 052, hep-th/0007191.

\bibitem{Acharya:2000mu}
B.S. Acharya, J.P. Gauntlett,  N.~Kim,  Fivebranes wrapped on associative
  three-cycles,  Phys. Rev.  D 63 (2001) 106003, hep-th/0011190.

\bibitem{Nieder:2000kc}
H.~Nieder,  Y.~Oz,  Supergravity and {D}-branes wrapping special lagrangian
  cycles,  JHEP  03 (2001) 008, hep-th/0011288.

\bibitem{Gauntlett:2000ng}
J.P. Gauntlett, N.~Kim,  D.~Waldram,  M-fivebranes wrapped on supersymmetric
  cycles,  Phys. Rev.  D 63 (2001) 126001, hep-th/0012195.

\bibitem{Nunez:2001pt}
C.~N{\'u\~n}ez, I.Y. Park, M.~Schvellinger,  T.A. Tran,  Supergravity duals of
  gauge theories from {$F(4)$} gauged supergravity in six dimensions,  JHEP  04
  (2001) 025, hep-th/0103080.

\bibitem{Edelstein:2001pu}
J.D. Edelstein,  C.~N{\'u\~n}ez,  {D6}-branes and {M}-theory geometrical
  transitions from gauged supergravity,  JHEP  04 (2001) 028, hep-th/0103167.

\bibitem{Schvellinger:2001ib}
M.~Schvellinger,  T.A. Tran,  Supergravity duals of gauge field theories from
  {$SU(2)\times U(1)$} gauged supergravity in five dimensions,  JHEP  06 (2001)
  025, hep-th/0105019.

\bibitem{Maldacena:2001pb}
J.M. Maldacena,  H.~Nastase,  The supergravity dual of a theory with dynamical
  supersymmetry breaking, hep-th/0105049.

\bibitem{Gauntlett:2001qs}
J.P. Gauntlett, N.~Kim, S.~Pakis,  D.~Waldram,  Membranes wrapped on
  holomorphic curves, hep-th/0105250.

\bibitem{Hernandez:2001bh}
R.~Hernandez,  Branes wrapped on coassociative cycles, hep-th/0106055.

\bibitem{Gauntlett:2001ps}
J.P. Gauntlett, N.~Kim, D.~Martelli,  D.~Waldram,  Wrapped fivebranes and {$\cN
  = 2$} super {Y}ang-{M}ills theory, hep-th/0106117.

\bibitem{Gomis:2001vk}
J.~Gomis,  D-branes, holonomy and {M}-theory,  Nucl. Phys.  B 606 (2001) 3--17,
  hep-th/0103115.

\bibitem{Gomis:2001vg}
J.~Gomis,  T.~Mateos,  D6-branes wrapping {K\"a}hler four-cycles,
  hep-th/0108080.

\bibitem{Pernici:1984xx}
M.~Pernici, K.~Pilch,  P.~van Nieuwenhuizen,  Gauged maximally extended
  supergravity in seven dimensions,  Phys. Lett.  B 143 (1984) 103--107.

\bibitem{Chamseddine:1999uy}
A.H. Chamseddine,  W.A. Sabra,  {$D = 7$} {$SU(2)$} gauged supergravity from
  {$D = 10$} supergravity,  Phys. Lett.  B 476 (2000) 415--419, hep-th/9911180.

\bibitem{Nastase:2000tu}
H.~Nastase,  D.~Vaman,  On the nonlinear {KK} reductions on spheres of
  supergravity theories,  Nucl. Phys.  B 583 (2000) 211--236, hep-th/0002028.

\bibitem{Cvetic:1999pu}
M.~Cveti\v{c}, J.T. Liu, H.~L{\"u},  C.N. Pope,  Domain-wall supergravities
  from sphere reduction,  Nucl. Phys.  B 560 (1999) 230--256, hep-th/9905096.

\bibitem{Cvetic:2000dm}
M.~Cveti\v{c}, H.~L{\"u},  C.N. Pope,  Consistent {Kaluza-Klein} sphere
  reductions,  Phys. Rev.  D 62 (2000) 064028, hep-th/0003286.

\bibitem{Salam:1982xd}
A.~Salam,  J.~Strathdee,  On {K}aluza-{K}lein theory,  Annals Phys.  141 (1982)
  316--352.

\bibitem{deWit:1985nz}
B.~de~Wit, H.~Nicolai,  N.P. Warner,  The embedding of gauged {${\cal N}=8$}
  supergravity into {$D = 11$} supergravity,  Nucl. Phys.  B 255 (1985) 29--62.

\bibitem{Nilsson:1985dj}
B.E.W. Nilsson,  On the embedding of {$D = 4$} {${\cN} = 8$} gauged
  supergravity in {$D = 11$} {$\cN=1$} supergravity,  Phys. Lett.  B 155 (1985)
  54--58.

\bibitem{Khavaev:1998fb}
A.~Khavaev, K.~Pilch,  N.P. Warner,  New vacua of gauged {${\cal N} = 8$}
  supergravity in five dimensions,  Phys. Lett.  B487 (2000) 14--21,
  hep-th/9812035.

\bibitem{Townsend:1983kk}
P.K. Townsend,  P.~van Nieuwenhuizen,  Gauged seven-dimensional supergravity,
  Phys. Lett.  B 125 (1983) 41--46.

\bibitem{Salam:1983fa}
A.~Salam,  E.~Sezgin,  {$SO(4)$} gauging of {${\cal N}=2$} supergravity in
  seven-dimensions,  Phys. Lett.  B 126 (1983) 295--300.

\bibitem{Giani:1984dw}
F.~Giani, M.~Pernici,  P.~van Nieuwenhuizen,  Gauged {${\cal N}=4$} {$D = 6$}
  supergravity,  Phys. Rev.  D 30 (1984) 1680--1687.

\bibitem{Cowdall:1998rs}
P.M. Cowdall,  On gauged maximal supergravity in six dimensions,  JHEP  06
  (1999) 018, hep-th/9810041.

\bibitem{Cowdall:1998fn}
P.M. Cowdall,  Supersymmetric electrovacs in gauged supergravities,  Class.
  Quantum Grav.  15 (1998) 2937--2953, hep-th/9710214.

\bibitem{Rocek:1992ps}
M.~Ro{\v{c}}ek,  E.~Verlinde,  Duality, quotients, and currents,  Nucl. Phys.
  B 373 (1992) 630--646, hep-th/9110053.

\bibitem{delaOssa:1993vc}
X.C. de~la Ossa,  F.~Quevedo,  Duality symmetries from non-abelian isometries
  in string theory,  Nucl. Phys.  B 403 (1993) 377--394, hep-th/9210021.

\bibitem{Alvarez:1994qi}
E.~Alvarez, L.~Alvarez-Gaume, J.L.F. Barbon,  Y.~Lozano,  Some global aspects
  of duality in string theory,  Nucl. Phys.  B 415 (1994) 71--100,
  hep-th/9309039.

\bibitem{Alvarez:1994zr}
E.~Alvarez, L.~Alvarez-Gaume,  Y.~Lozano,  On non-abelian duality,  Nucl. Phys.
   B 424 (1994) 155--183, hep-th/9403155.

\bibitem{Alvarez:1994wj}
E.~Alvarez, L.~Alvarez-Gaume,  Y.~Lozano,  A canonical approach to duality
  transformations,  Phys. Lett.  B 336 (1994) 183--189, hep-th/9406206.

\bibitem{Alvarez:1995dn}
E.~Alvarez, L.~Alvarez-Gaume,  Y.~Lozano,  An introduction to {T} duality in
  string theory,  Nucl. Phys. Proc. Suppl.  41 (1995) 1--20, hep-th/9410237.

\bibitem{Bossard:2000xq}
A.~Bossard,  M.~Mohammedi,  Non-abelian duality in the string effective action,
   Nucl. Phys.  B 595 (2001) 93--118, hep-th/0008062.

\bibitem{Gunaydin:1986cu}
M.~G{\"u}naydin, L.J. Romans,  N.P. Warner,  Compact and noncompact gauged
  supergravity theories in five dimensions,  Nucl. Phys.  B 272 (1986)
  598--646.

\bibitem{Ferrara:1998bp}
S.~Ferrara, M.A. Lledo,  A.~Zaffaroni,  Born-{I}nfeld corrections to {D3} brane
  action in {$AdS_5\times S^5$} and {${\cal N} = 4$}, {$D = 4$} primary
  superfields,  Phys. Rev.  D 58 (1998) 105029, hep-th/9805082.

\bibitem{Das:1998ei}
S.R. Das,  S.P. Trivedi,  Three-brane action and the correspondence between
  {${\cal N} = 4$} {Y}ang {M}ills theory and anti de {S}itter space,  Phys.
  Lett.  B 445 (1998) 142--149, hep-th/9804149.

\bibitem{Das:1996wn}
S.R. Das,  S.D. Mathur,  Comparing decay rates for black holes and {D}-branes,
  Nucl. Phys.  B 478 (1996) 561--576, hep-th/9606185.

\bibitem{Callan:1997tv}
Jr. C.G.~Callan, S.S. Gubser, I.R. Klebanov,  A.A. Tseytlin,  Absorption of
  fixed scalars and the {D}-brane approach to black holes,  Nucl. Phys.  B 489
  (1997) 65--94, hep-th/9610172.

\bibitem{Ferrara:1998ej}
S.~Ferrara, C.~Fronsdal,  A.~Zaffaroni,  On {${\cal N} = 8$} supergravity on
  {$AdS_5$} and {${\cal N} = 4$} superconformal {Y}ang-{M}ills theory,  Nucl.
  Phys.  B 532 (1998) 153--162, hep-th/9802203.

\bibitem{Liu:1998bu}
H.~Liu,  A.A. Tseytlin,  {$D = 4$} super {Y}ang-{M}ills, {$D = 5$} gauged
  supergravity, and {$D = 4$} conformal supergravity,  Nucl. Phys.  B 533
  (1998) 88--108, hep-th/9804083.

\bibitem{Park:1999xz}
I.Y. Park,  Fundamental versus solitonic description of {D3}-branes,  Phys.
  Lett.  B 468 (1999) 213--218, hep-th/9907142.

\bibitem{Freedman:1998tz}
D.Z. Freedman, S.D. Mathur, A.~Matusis,  L.~Rastelli,  Correlation functions in
  the {$CFT_d/AdS_{d+1}$} correspondence,  Nucl. Phys.  B 546 (1999) 96--118,
  hep-th/9804058.

\bibitem{Muck:1998rr}
W.~M{\"u}ck,  K.S. Viswanathan,  Conformal field theory correlators from
  classical scalar field theory on {$AdS_{d+1}$},  Phys. Rev.  D 58 (1998)
  041901, hep-th/9804035.

\bibitem{Cederwall:1997pv}
M.~Cederwall, A.~von Gussich, B.E.W. Nilsson,  A.~Westerberg,  The {D}irichlet
  super-three-brane in ten-dimensional type {IIB} supergravity,  Nucl. Phys.  B
  490 (1997) 163--178, hep-th/9610148.

\bibitem{Bergshoeff:1997tu}
E.~Bergshoeff,  P.K. Townsend,  Super {D}-branes,  Nucl. Phys.  B 490 (1997)
  145--162, hep-th/9611173.

\bibitem{Aganagic:1997zk}
M.~Aganagic, J.~Park, C.~Popescu,  J.H. Schwarz,  Dual {D}-brane actions,
  Nucl. Phys.  B 496 (1997) 215--230, hep-th/9702133.

\bibitem{Metsaev:1998hf}
R.R. Metsaev,  A.A. Tseytlin,  Supersymmetric {D3}-brane action in
  {$AdS_5\times S^5$},  Phys. Lett.  B 436 (1998) 281--288, hep-th/9806095.

\bibitem{Metsaev:1998it}
R.R. Metsaev,  A.A. Tseytlin,  Type {IIB} superstring action in {$AdS_5\times
  S^5$} background,  Nucl. Phys.  B 533 (1998) 109--126, hep-th/9805028.

\bibitem{Kallosh:1998nx}
R.~Kallosh,  J.~Rahmfeld,  The {GS} string action on {$AdS_5\times S^5$},
  Phys. Lett.  B 443 (1998) 143--146, hep-th/9808038.

\bibitem{Hughes:1986fa}
J.~Hughes, J.~Liu,  J.~Polchinski,  Supermembranes,  Phys. Lett.  B 180 (1986)
  370--374.

\bibitem{Becker:1995kb}
K.~Becker, M.~Becker,  A.~Strominger,  Fivebranes, membranes and
  nonperturbative string theory,  Nucl. Phys.  B 456 (1995) 130--152,
  hep-th/9507158.

\bibitem{Intriligator:1997pq}
K.~Intriligator, D.R. Morrison,  N.~Seiberg,  Five-dimensional supersymmetric
  gauge theories and degenerations of {C}alabi-{Y}au spaces,  Nucl. Phys.  B
  497 (1997) 56--100, hep-th/9702198.

\bibitem{Ferrara:1998gv}
S.~Ferrara, A.~Kehagias, H.~Partouche,  A.~Zaffaroni,  {$AdS_6$} interpretation
  of {5D} superconformal field theories,  Phys. Lett.  B 431 (1998) 57--62,
  hep-th/9804006.

\bibitem{Brandhuber:1999np}
A.~Brandhuber,  Y.~Oz,  The {D4-D8} brane system and five-dimensional fixed
  points,  Phys. Lett.  B 460 (1999) 307--312, hep-th/9905148.

\bibitem{Aharony:1997bx}
O.~Aharony, A.~Hanany, K.~Intriligator, N.~Seiberg,  M.J. Strassler,  Aspects
  of {${\cal N} = 2$} supersymmetric gauge theories in three dimensions,  Nucl.
  Phys.  B 499 (1997) 67--99, hep-th/9703110.

\bibitem{deBoer:1997mp}
J.~de~Boer, K.~Hori, H.~Ooguri,  Y.~Oz,  Mirror symmetry in three-dimensional
  gauge theories, quivers and {D}-branes,  Nucl. Phys.  B 493 (1997) 101--147,
  hep-th/9611063.

\bibitem{deBoer:1997ck}
J.~de~Boer, K.~Hori, H.~Ooguri, Y.~Oz,  Z.~Yin,  Mirror symmetry in
  three-dimensional gauge theories, {$SL(2,Z)$} and {D}-brane moduli spaces,
  Nucl. Phys.  B 493 (1997) 148--176, hep-th/9612131.

\bibitem{deBoer:1997kr}
J.~de~Boer, K.~Hori,  Y.~Oz,  Dynamics of {$\cN = 2$} supersymmetric gauge
  theories in three dimensions,  Nucl. Phys.  B 500 (1997) 163--191,
  hep-th/9703100.

\bibitem{deBoer:1997ka}
J.~de~Boer, K.~Hori, Y.~Oz,  Z.~Yin,  Branes and mirror symmetry in {$\cN = 2$}
  supersymmetric gauge theories in three dimensions,  Nucl. Phys.  B 502 (1997)
  107--124, hep-th/9702154.

\bibitem{Witten:1996gx}
E.~Witten,  Small instantons in string theory,  Nucl. Phys.  B 460 (1996)
  541--559, hep-th/9511030.

\bibitem{Polchinski:1996df}
J.~Polchinski,  E.~Witten,  Evidence for heterotic - type {I} string duality,
  Nucl. Phys.  B 460 (1996) 525--540, hep-th/9510169.

\bibitem{Polchinski:1996fm}
J.~Polchinski, S.~Chaudhuri,  C.V. Johnson,  Notes on {D}-branes,
  hep-th/9602052.

\bibitem{Polchinski:1996na}
J.~Polchinski,  T{ASI} lectures on {D}-branes, hep-th/9611050.

\bibitem{Seiberg:1996bd}
N.~Seiberg,  Five-dimensional {SUSY} field theories, non-trivial fixed points
  and string dynamics,  Phys. Lett.  B 388 (1996) 753--760, hep-th/9608111.

\bibitem{Morrison:1997xf}
D.R. Morrison,  N.~Seiberg,  Extremal transitions and five-dimensional
  supersymmetric field theories,  Nucl. Phys.  B 483 (1997) 229--247,
  hep-th/9609070.

\bibitem{Janssen:1999sa}
B.~Janssen, P.~Meessen,  T.~Ortin,  The {D}8-brane tied up: {S}tring and brane
  solutions in massive type {IIA} supergravity,  Phys. Lett.  B 453 (1999)
  229--236, hep-th/9901078.

\bibitem{Massar:1999sb}
M.~Massar,  J.~Troost,  {D0-D8-F1} in massive {IIA SUGRA},  Phys. Lett.  B 458
  (1999) 283--287, hep-th/9901136.

\bibitem{Cachazo:2000ey}
F.A. Cachazo,  C.~Vafa,  Type {I$^\prime$} and real algebraic geometry,
  hep-th/0001029.

\bibitem{Balasubramanian:1998sn}
V.~Balasubramanian, P.~Kraus,  A.E. Lawrence,  Bulk vs. boundary dynamics in
  anti-de {S}itter spacetime,  Phys. Rev.  D 59 (1999) 046003, hep-th/9805171.

\bibitem{Gubser:2000nd}
S.S. Gubser,  Curvature singularities: {T}he good, the bad, and the naked,
  hep-th/0002160.

\bibitem{Lavrinenko:1999xi}
I.V. Lavrinenko, H.~L{\"u}, C.N. Pope,  K.S. Stelle,  Superdualities, brane
  tensions and massive {IIA/IIB} duality,  Nucl. Phys.  B 555 (1999) 201--227,
  hep-th/9903057.

\bibitem{Buchel:2001gw}
A.~Buchel, C.P. Herzog, I.R. Klebanov, L.~Pando Zayas,  A.A. Tseytlin,
  Non-extremal gravity duals for fractional {D3}-branes on the conifold,  JHEP
  04 (2001) 033, hep-th/0102105.

\bibitem{Gubser:2001ri}
S.S. Gubser, C.P. Herzog, I.R. Klebanov,  A.A. Tseytlin,  Restoration of chiral
  symmetry: {A} supergravity perspective,  JHEP  05 (2001) 028, hep-th/0102172.

\bibitem{Buchel:2001qi}
A.~Buchel,  A.~Frey,  Comments on supergravity dual of pure {$\cN=1$} super
  {Y}ang {M}ills theory with unbroken chiral symmetry, hep-th/0103022.

\bibitem{Chamseddine:1997nm}
A.H. Chamseddine,  M.S. Volkov,  Non-{A}belian {BPS} monopoles in {${\cal N} =
  4$} gauged supergravity,  Phys. Rev. Lett.  79 (1997) 3343--3346,
  hep-th/9707176.

\bibitem{Chamseddine:1998mc}
A.H. Chamseddine,  M.S. Volkov,  Non-{A}belian solitons in {${\cal N} = 4$}
  gauged supergravity and leading order string theory,  Phys. Rev.  D 57 (1998)
  6242--6254, hep-th/9711181.

\bibitem{Klemm:1998in}
D.~Klemm,  {BPS} black holes in gauged {$\cN = 4$}, {$D = 4$} supergravity,
  Nucl. Phys.  B 545 (1999) 461--478, hep-th/9810090.

\bibitem{Itzhaki:1998dd}
N.~Itzhaki, J.M. Maldacena, J.~Sonnenschein,  S.~Yankielowicz,  Supergravity
  and the large {$N$} limit of theories with sixteen supercharges,  Phys. Rev.
  D 58 (1998) 046004, hep-th/9802042.

\bibitem{Chamseddine:2001hk}
A.H. Chamseddine,  M.S. Volkov,  Non-abelian vacua in {$D = 5$}, {${\cal N} =
  4$} gauged supergravity,  JHEP  04 (2001) 023, hep-th/0101202.

\bibitem{Romans:1986ps}
L.J. Romans,  Gauged {${\cal N}=4$} supergravities in five dimensions and their
  magnetovac backgrounds,  Nucl. Phys.  B 267 (1986) 433--447.

\bibitem{Maldacena:2000dr}
J.M. Maldacena,  L.~Maoz,  De-singularization by rotation, hep-th/0012025.

\bibitem{Kinar:1998vq}
Y.~Kinar, E.~Schreiber,  J.~Sonnenschein,  {$Q\overline{Q}$} potential from
  strings in curved spacetime{:} {C}lassical results,  Nucl. Phys.  B 566
  (2000) 103--125, hep-th/9811192.

\bibitem{Sonnenschein:1999if}
J.~Sonnenschein,  What does the string/gauge correspondence teach us about
  {W}ilson loops?, hep-th/0003032.

\bibitem{Pilch:2000ej}
K.~Pilch,  N.P. Warner,  A new supersymmetric compactification of chiral {IIB}
  supergravity,  Phys. Lett.  B 487 (2000) 22--29, hep-th/0002192.

\bibitem{Nahm:1978tg}
W.~Nahm,  Supersymmetries and their representations,  Nucl. Phys.  B 135 (1978)
  149--166.

\bibitem{Huq:1985im}
M.~Huq,  M.A. Namazie,  Kaluza-{K}lein supergravity in ten dimensions,  Class.
  Quantum Grav.  2 (1985) 293--308.

\bibitem{Bergshoeff:1995as}
E.~Bergshoeff, C.~Hull,  T.~Ortin,  Duality in the type {II} superstring
  effective action,  Nucl. Phys.  B 451 (1995) 547--578, hep-th/9504081.

\bibitem{DeWitt:1982wm}
B.S. DeWitt,  P.~van Nieuwenhuizen,  Explicit construction of the exceptional
  superalgebras {$F(4)$} and {$G(3)$},  J. Math. Phys.  23 (1982) 1953--1963.

\bibitem{D'Auria:2000ad}
R.~D'Auria, S.~Ferrara,  S.~Vaul{\`a},  Matter coupled {$F(4)$} supergravity
  and the {$AdS_6/CFT_5$} correspondence,  JHEP  10 (2000) 013, hep-th/0006107.

\bibitem{Andrianopoli:2001rs}
L.~Andrianopoli, R.~D'Auria,  S.~Vaul{\`a},  Matter coupled {$F(4)$} gauged
  supergravity lagrangian,  JHEP  05 (2001) 065, hep-th/0104155.

\bibitem{Townsend:1984xs}
P.K. Townsend, K.~Pilch,  P.~van Nieuwenhuizen,  Selfduality in odd dimensions,
   Phys. Lett.  B 136 (1984) 38--42.

\end{thebibliography}
